\documentclass[a4paper,11pt]{article}
\pdfoutput=1 
\usepackage{amssymb,mathrsfs}
\usepackage{bbm}
\usepackage{siunitx}
\usepackage{amsmath}
\usepackage{braket}
\usepackage{mathtools}
\usepackage[usenames,dvipsnames,svgnames,table]{xcolor}
\usepackage [utf8] {inputenc}
\usepackage{color}
\usepackage{lmodern}
\usepackage{footnote}
\usepackage{slashed}
 \usepackage{mathrsfs}
 \usepackage[pdftex] {graphicx}
\usepackage{multirow}
\usepackage{jheppub} 
\usepackage{here}
\usepackage{hyperref}
\usepackage[justification=centering, singlelinecheck=false]{caption}
\usepackage[list=true, labelfont=bf, labelformat=brace, position=top]{subcaption}

\usepackage[colorlinks=true,linkcolor=blue,urlcolor=blue,citecolor=blue,anchorcolor=blue]{hyperref}
\usepackage[capitalise]{cleveref}
\Crefname{section}{Sec.}{Secs.}
\crefrangelabelformat{equation}{(#3#1#4--#5#2#6)}
\usepackage{tikz}
\usetikzlibrary{calc,decorations.markings,decorations.pathmorphing}
\usepackage{multirow}
\usepackage{amsfonts}
\usepackage{ulem}
\usepackage{diagbox}
\newcommand{\eqq}{\begin{eqnarray}}
\newcommand{\en}{\end{eqnarray}}

\newcommand{\db}{\mathbf{d}}
\newcommand{\kb}{\mathbf{k}}
\newcommand{\lb}{\mathbf{l}}
\newcommand{\nb}{\mathbf{n}}

\newcommand{\Pb}{\mathbf{P}}
\newcommand{\JJ}{\mathcal{J}}
\newcommand{\Deltab}{\boldsymbol{\Delta}}

\newcommand{\nn}{\nonumber}
\newcommand{\dkv}[1]{\int \frac{d^3#1}{(2\pi)^3}}
\newcommand{\skv}[1]{L^{-3}\sum_{#1=\frac{2\pi}{L}\mathbf{n},\mathbf{n}\in\mathbb{Z}^3}}

\title{Generalized boost transformations in finite volumes and application to Hamiltonian methods}

\author[1,2]{Yan Li, }
\affiliation[1]{University of Chinese Academy of Sciences (UCAS), Beijing 100049, China}
\affiliation[2]{Department of Physics, University of Cyprus, 20537 Nicosia, Cyprus}
\emailAdd{li.yan@ucy.ac.cy}
\author[1,3]{Jia-Jun Wu, }
\affiliation[3]{Southern Center for Nuclear-Science Theory (SCNT), Institute of Modern Physics,
Chinese Academy of Sciences, Huizhou 516000, Guangdong Province, China}
\emailAdd{wujiajun@ucas.ac.cn}
\author[4]{T.-S. H. Lee, }
\affiliation[4]{Physics Division, Argonne National Laboratory, Argonne, Illinois 60439, USA}
\emailAdd{tshlee@anl.gov}
\author[5]{and R. D. Young}
\affiliation[5]{Special Research Center for the Subatomic Structure of Matter (CSSM),
Department of Physics,  The University of Adelaide, Adelaide SA 5005, Australia}
\emailAdd{ross.young@adelaide.edu.au}

\abstract{

\noindent
The investigation of hadron interactions within lattice QCD has been facilitated by the well-known quantisation condition, linking scattering phase shifts to finite-volume energies.
Additionally, the ability to utilise systems at finite total boosts has been pivotal in smoothly charting the energy-dependent behaviour of these phase shifts. 
The existing implementations of the quantization condition at finite boosts rely on momentum
transformations between rest and moving frames, defined directly in terms of the energy eigenvalues. 
This energy dependence is unsuitable in the formulation of a Hamiltonian.
In this work, we introduce a novel approach to generalise the three-momentum boost prescription, enabling the incorporation of energy-independent finite-volume Hamiltonians within moving frames. 
We demonstrate the application of our method through numerical comparisons, employing a phenomenological $\pi\pi$ scattering example.

}
\allowdisplaybreaks
\begin{document}
\maketitle
\flushbottom

\section{Introduction}\label{sec:int}

The successful description of hadronic resonances in terms of the quark model was instrumental in the establishment of QCD as
the fundamental theory of the strong interactions. 
Beyond the quark model however, it remains an ongoing challenge to
resolve the nature of hadron resonances in terms of the fundamental
quark and gluon degrees of freedom \cite{Shepherd:2016dni}.
Of particular interest, at present, is the demand to explain a number of
exotic states being identified in collider experiments that challenge
the quark model paradigm \cite{Guo:2017jvc}.
Lattice QCD studies are rapidly advancing towards this goal
(for reviews, please see e.g.~Refs.~\cite{Liu:2016kbb,
  Briceno:2017max,Padmanath:2019wid}).
However, states involving many-body decay channels represents a particular challenge, and hence there is ongoing effort dedicated to describing 3-body systems and beyond on the lattice
\cite{
  Detmold:2008gh,
  Detmold:2011kw,
  Briceno:2012rv,
  Polejaeva:2012ut,
  Hansen:2014eka,
  Hansen:2015zga,
  Briceno:2017tce,
  Hammer:2017uqm,
  Hammer:2017kms,
  Mai:2017bge,
  Mai:2018djl,
  Briceno:2018aml,
  Jackura:2019bmu,
  Horz:2019rrn,
  Blanton:2019vdk,
  Blanton:2019igq,
  Mai:2019fba,
  Culver:2019vvu,
  Pang:2019dfe,
  Jackura:2019bmu,
  Romero-Lopez:2019qrt,
  Blanton:2020gmf,
  Blanton:2020gha,
  Blanton:2020jnm,
  Muller:2020wjo,
  Fischer:2020jzp,
  Hansen:2020otl,
  Hansen:2020zhy,
  Alexandru:2020xqf,
  Muller:2021uur,
  Brett:2021wyd,
  Mai:2021lwb,
  Blanton:2021llb,
  Mai:2021nul}
(see Ref.~\cite{Hansen:2019nir} for a review).

The fundamental framework for the study of hadron interactions in
lattice QCD is built upon the quantisation condition introduced by
L\"uscher \cite{Luscher:1986pf,Luscher:1990ux}, and extended to a
describe range of different systems over recent years, please see a review paper, Ref.~\cite{Briceno:2017max}.
From the point of view of numerical implementation, the adaptation of
the quantisation condition to consider moving frames
\cite{Rummukainen:1995vs} has
proven particularly powerful.
In particular, the extension to study boosted systems has enabled an
almost-smooth mapping of the energy dependence of
scattering phase shifts.
These smooth phase shifts permit a clear identification of
resonant structures \cite{Dudek:2012xn}, even in intricate coupled
channels \cite{Dudek:2014qha} where
simple level identifications are destined to break down \cite{Briceno:2017max}.
The treatment of moving frames requires a prescription
to translate between the boosted boundary condition on the lattice and
the conventional (continuum) partial-wave decomposition in the centre of mass (CM).
In this article we present a general framework that connects different
finite-volume boost prescriptions that underpin the quantisation
condition in moving frames.
We consider three distinct boost transformations
which connect the discrete momentum sums of a boosted frame to the
corresponding centre-of-mass system, where one can match onto the
familiar partial wave decomposition of the continuum.
In particular, we consider the kinematic boost
originally introduced by
Rummukainen and Gottlieb (RG) \cite{Rummukainen:1995vs}, the 2-body
Bethe-Salpeter prescription of Kim, Sachrajda and Sharpe (KSS)
\cite{Kim:2005gf}, and we introduce a 
new method in terms of the on-shell non-interacting states ---
this transformation has previously been identified as the 
``Wu boost'' \cite{Blanton:2020gha}.
All formulations are demonstrated to describe precisely the same power-law
behavior of the finite-volume energy spectra, with differences that
are exponentially suppressed in $m_\pi L$.
By the introduction of the new formulation, we establish the framework for
applications of the finite-volume Hamiltonian technique
\cite{Hall:2013qba, Wu:2014vma, Abell:2021awi} at finite boost --- such as appearing in
Ref.~\cite{Li:2021mob}.
In this new method, an important feature is that the boost definition
is defined independently of the energy eigenvalue, which makes it
suitable for direct application to a Hamiltonian-based formulation.
The eigenvectors from an energy-independent Hamiltonian form a complete orthonormal basis in the Hilbert space of the Hamiltonian, and are useful in many previous studies \cite{Hall:2013qba, Wu:2014vma, Wu:2016ixr, Li:2019qvh}.
The energy-dependence of a Hamiltonian usually comes from the incompleteness of the Hilbert space \cite{Bloch:1958determination,Luu:2005qr,Li:2022aru}, and should not come from the boost transformation.
Furthermore, as pointed out in Refs.~\cite{Blanton:2020gha, Muller:2021uur}, in the 3-body system this transformation method ensures that the velocity of any two-body system is smaller than speed of light for all choices of spectator momenta.
Here we summarise the main features of the present work.
In general, L\"uscher's quantization condition can be summarized by
\begin{equation}
  \det\left[F^{-1}(E,\mathbf{P};L)-T(E,\mathbf{P})\right]=0,
\end{equation}
where $T$ describes the infinite-volume $2\to 2$ scattering matrix, $E$ and $\mathbf{P}$ are the total energy and momentum of the system, respectively, and $F$ encodes the kinematics associated with the finite-volume boundary conditions on a lattice of finite spatial
extent, $L$. 
In the general evaluation of $F^{-1}$, we require finite volume summations of the form
\begin{equation}
\frac{1}{L^3}\sum_{\kb}\JJ\frac{f(\kb^*)}{{q^*}^2-{\mathbf{k}^*}^2}.
\label{eq:Sintro}
\end{equation}
The summation is over the discrete momenta of the moving frame $\kb=(2\pi/L)\nb$, whereas $\kb^*$ denotes the transformation of $\kb$ to the CM frame. 
The factor $\JJ$ describes a Jacobian for the boost transformation, and $q^*$ is the magnitude of the on-shell momentum in the centre-of-mass frame, for total energy given by $E^*=\sqrt{E^2-\mathbf{P}^2}$.
As detailed in Sect.~\ref{sec:qd}, the amplitude $f$ is expanded in terms of spherical harmonics in the CM frame, where it is noted that the form of $f$ does not itself have any dependence on the boost transformation. 
The regularization of $S$ in Eq.~(\ref{eq:Sintro}) is implicitly assumed, the details of which appear in the main body of the paper.

The three different boost schemes we consider in this paper are each realised in the particular form of the centre-of-mass momenta $\kb^*$ and Jacobian $\JJ$. 
The primary objective of the present manuscript is to introduce a new form for $\kb^*$ and $\JJ$ that is suitable for a relativistic lattice Hamiltonian. 
A key feature of the boost is that the transformation is functionally defined in terms of
$\kb$ and $\Pb$, $\kb^*=\kb^*(\kb,\Pb)$ and $\JJ=\JJ(\kb,\Pb)$. 
This is in contrast to the existing formulations which manifestly depend on knowledge of $E^*$, $\kb^*=\kb^*(\kb,\Pb,E^*)$ and $\JJ=\JJ(\kb,\Pb,E^*)$. 
Here we prove that the new Hamiltonian scheme provides a quantisation scheme that is equivalent to the established schemes\footnote{By equivalent, we take the usual interpretataion that any differences between the quantisation condition are suppressed by the characteristic factor,
$e^{-mL}$. 
The equivalence of the first two schemes, RG and KSS, was originally established in Ref.~\cite{Kim:2005gf}.
In \cref{app:ABmodel} we show that RG and KSS are not only equivalent up to exponentially-supressed terms, but are in fact identical. 
Given the vastly different form of the transformations, this feature is not obvious at the outset.}.
This paper is organized as follows.
In \cref{sec:qd}, we give a detailed derivation of the quantization
condition in the moving-frame finite volume.
In \cref{sec:comparing}, we introduce a generalised form for the three-momentum transformation required to evaluate the lattice eigenenergy quantisation condition.
This general form is then used to introduce the new scheme, and also encode the schemes introduced by KSS \cite{Kim:2005gf} and RG \cite{Rummukainen:1995vs}.
Using a phenomenological model for the $\pi\pi$ interaction, a numerical demonstration of the equivalence of the three approaches is presented in \cref{sec:NCC}.
Finally, a brief summary in given in \cref{sec:Sum}.

\section{Derivation of a moving-frame finite volume quantization condition}
\label{sec:qd}

It is well-known that the finite-volume energy levels calculated from lattice QCD (LQCD) appear as poles of the finite-volume Green function which is related to the Bethe-Salpeter (BS) equation.
It was demonstrated in Refs.~\cite{Kim:2005gf,Doring:2011vk,Bernard:2008ax} that the moving-frame finite-volume quantization condition, which relates the phase shifts to the spectrum predicted by LQCD, can be derived from examining the differences in the Bethe-Salpeter equations in finite volume and in infinite space. 

In  this section, we will follow the same approach, but will give a different derivation which provides a generalized quantization condition in the moving frame.

\subsection{The BS Equation in the Infinite Volume}

\begin{figure}[tbp]
	\centering
	\includegraphics[width=\columnwidth]{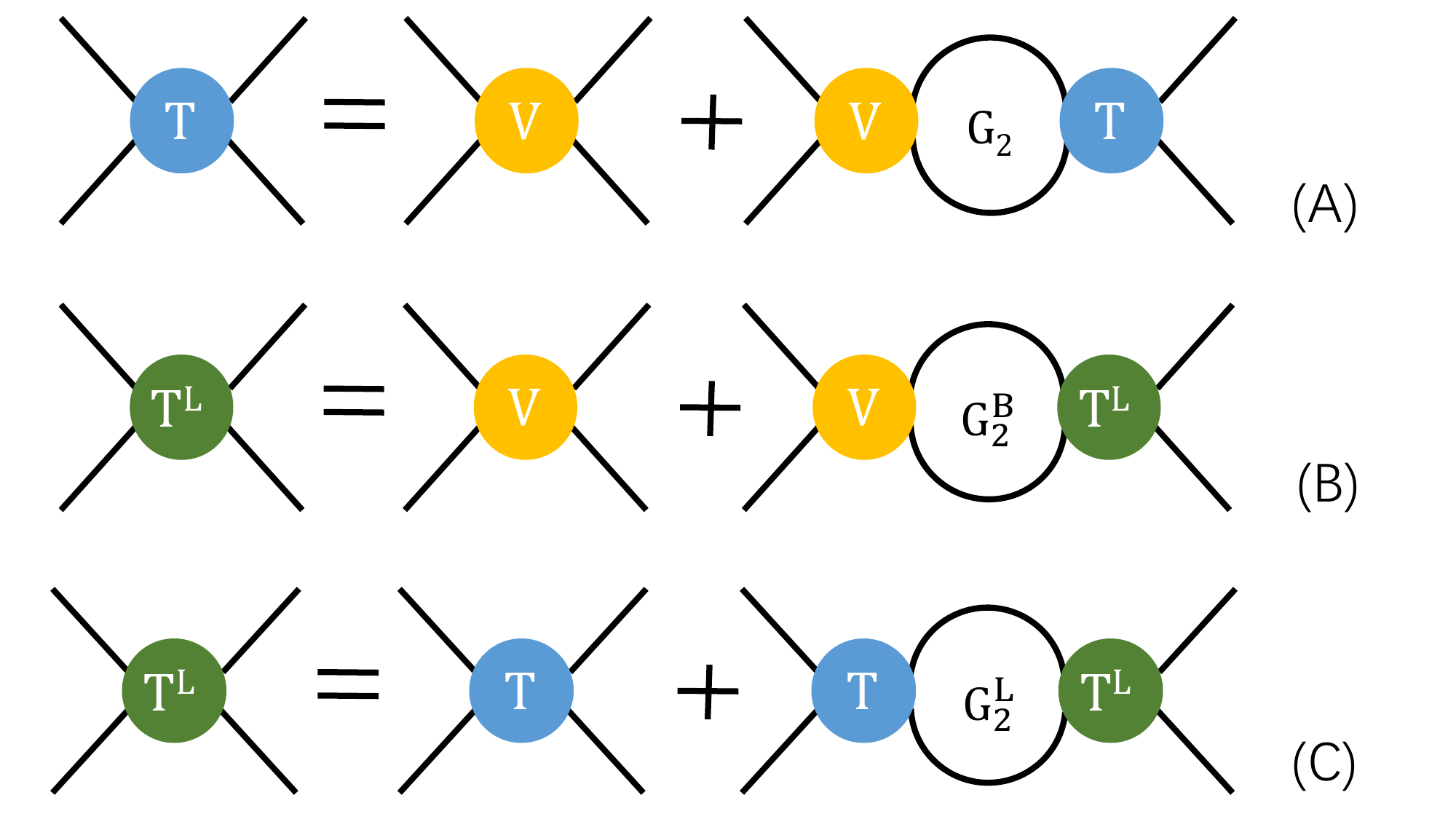}
	\caption{Diagram (A) indicates the usual
          Bethe-Salpeter equation in the continuum, relating the $T$
          matrix to the scattering kernel, $V$.
          Diagram (B) shows the analogous finite volume amplitude $T^L$ can be expressed
          in the same form, with the 2-body propagator replaced with a
          box-quantised representation, $G_2^B$.
          Diagram (C) represents Eq.~(\ref{eq:TGLTL}), demonstrating
        that $T^L$ can similarly be expressed in terms of infinite
        volume $T$-matrix, with the 2-body propagator replaced with
        $G_2^L=G_2^B-G_2$.
        \label{fig:BSE}}
\end{figure}

We start from the BS equation for the two-particle scattering $1(\mathbf{p}_i)+2(\mathbf{P}-\mathbf{p}_i)\rightarrow 1(\mathbf{p}_f)+2(\mathbf{P}-\mathbf{p}_f)$.
In the rest frame with total momentum $P^*=(P^{*\,0},\mathbf{0})$ ($^*$ indicates this variable in the rest frame), we have
\begin{eqnarray}\label{eq:BSEifvrest}
   T(p_f^*,p_i^*;P^*) &=& V(p_f^*,p_i^*;P^*)
+ \int \frac{d^4k^*}{(2\pi)^4} V(p_f^*,k^*;P^*) 
G_2(k^*;P^*) T(k^*,p_i^*;P^*) \,,
\label{eq:bs-0}
\end{eqnarray}
where $V(p_f^*,p_i^*;P^*)$ is the two-particle potential;
$k^*$ is the four-momentum of the particle-$1$ in the intermediate state and we thus have 
\begin{eqnarray}
G_2(k;P)=\frac{1}{(k^{2}-m^2_1+i\epsilon)\,}
\frac{1}{((P-k)^2-m^2_2+i\epsilon)}\,.
\label{eq:g2}
\end{eqnarray}
Graphically, Eq.~(\ref{eq:bs-0}) can be represented as \cref{fig:BSE}(A).

To proceed, it is necessary to make the partial-wave expansion of the scattering amplitude. 
We choose the normalization 
$\braket{\mathbf{k}|\mathbf{k}^{\,'}}=(2\pi)^3\delta^{(3)}(\mathbf{k}-\mathbf{k}^{\,'})$ for plane-wave states and the S-matrix is related to the amplitude $T$ of Eq.~(\ref{eq:bs-0}) by 
\begin{eqnarray}
\braket{\mathbf{p}^*_f|S|\mathbf{p}^*_i}&=& (2\pi)^3\delta^{3}(\mathbf{p}^*_f-\mathbf{p}^*_i)
-i2\pi\delta( E^*(\mathbf{p}^*_f)-E^*(\mathbf{p}^*_i))\,\left[N(\mathbf{p}^*_f)T(p^*_f,p^*_i;P^*)N(\mathbf{p}^*_i)\right]\,,
\end{eqnarray}
where
\begin{eqnarray}
E^*(\mathbf{k})&=&\omega_1(\mathbf{k})+\omega_2(\mathbf{k}),\nonumber \\
N(\mathbf{k})&=&\frac{i}{\sqrt{4\omega_1(\mathbf{k})\omega_2(\mathbf{k})}}\,,
\end{eqnarray}
where $\omega_i(\mathbf{q})=\sqrt{\mathbf{q}^2+m^2_i}$.

At the on-shell point, we have
\begin{align}
    &|\mathbf{p}_f^*|=|\mathbf{p}_i^*|\equiv q\,,\; p_{f}^{*\,0}=p_{i}^{*\,0}=\omega_1(q) \label{eq:on-e-1}\,,\\
    &P^{*\,0}=E^*(q) = \omega_1(q)+\omega_2(q) \,. \label{eq:on-e-2}
\end{align}
Then we are able to define
\begin{align}
    T(p^*_f,p^*_i;P^*)&=\sum_{lm} T_{l}(q)Y_{lm}(\hat{\mathbf{p}}^*_f)Y^*_{lm}(\hat{\mathbf{p}}^*_i) \,,\label{eq:t-par}\\
    T_l(q) &=  
 i \frac{32\pi^2 E^*(q)}{ \, q} e^{i\delta_l(q)}\sin\delta_l(q)  \,,\label{eq:G4IFVpw}
\end{align}
where $\delta_l(q)$ is the scattering phase shift defined by the partial-wave S-matrix $S_l(E)=e^{2\,i\delta_l}$.

In practice, it is helpful to reduce BS equation to a 3-dimensional representation.
In defining the scattering equation, only the poles of $G_2(k^*;P)$ in the upper half of $k^*_0$ plane are kept:
(1) $k^*_0=-\omega_1(\mathbf{k}^*)+i\epsilon$, (2) $k^*_0 =  P^*_0-\omega_2(\mathbf{k}^*)+i\epsilon$.
Keeping both of the poles, Eq.~(\ref{eq:g2}) can be written as
\footnote{Here, we note that $\int \frac{f(x)}{x-a+i\epsilon} dx=2\pi i f(a)$, so $\frac{1}{x-a+i \epsilon}$ can be replaced as $2\pi i\, \delta(x-a)$ in the integration.}
\begin{align}
G_2(k^*;P^*) \to &~~
\frac{1}{-2\omega_1(\mathbf{P}^*-\mathbf{k}^*)}
\frac{1}{P^*_0+\omega_1(\mathbf{k}^*)-\omega_2(\mathbf{k}^*)}
\frac{(2\pi)i\,\delta(k_0^*+\omega_1(\mathbf{k}^*))}
{P_0^*+\omega_1(\mathbf{k}^*)+\omega_2(\mathbf{k}^*)} 
\nonumber\\
&+
 \frac{1}{2\omega_2(\mathbf{k}^*)}
\frac{1}{P^*_0+\omega_1(\mathbf{k}^*)-\omega_2(\mathbf{k}^*)}
\frac{(2\pi)i\,\delta(k_0^*-(P^*_0-\omega_2(\mathbf{k}^*)))}{P_0^*-\omega_1(\mathbf{k}^*)-\omega_2(\mathbf{k}^*)+i\epsilon}. %
\label{eq:g2-2} 
\end{align}
The first pole corresponds to setting the four-momentum of particle 1 to $p^*_1=(-E_1(\mathbf{k}), \mathbf{k}^*)$ which is in the anti-particle space, thus, it can be neglected directly, i.e., the first term in Eq.~(\ref{eq:g2-2}) is removed.   
Based on Relativistic Quantum Mechanics formulated by Dirac~\cite{Dirac:1936tg}, $P^{*\,0}$ in the above equation can be defined by the integration variable $\mathbf{k}^*$ as $P^{*\,0}=\omega_1(\mathbf{k}^*)+\omega_2(\mathbf{k}^*)$, which can be used for the quantity in $\delta$ function and the second denominator in the second term of Eq.~(\ref{eq:g2-2}).
Furthermore, the third denominator of this term will generate the right hand singularity, where $P^{*\,0}$ can be defined by the on-shell momentum $q$ of Eq.~(\ref{eq:on-e-2}) as $P^{*\,0}=\omega_1(q)+\omega_2(q)=E^*(q)$.

Then we can obtain the familiar form: 
\begin{align}
G_2(k^*;P^*)\to &~  
\frac{i}{2\omega_2(\mathbf{k}^*)}
\frac{1}{2\omega_1(\mathbf{k}^*)}
\frac{(2\pi)\,\delta(k_0^*-\omega_1(\mathbf{k}^*))}{E^*(q)-\omega_1(\mathbf{k}^*)-\omega_2(\mathbf{k}^*)+i\epsilon}.
\label{eq:g2-3}
\end{align}
By using Eq.~(\ref{eq:g2-3}) in the evaluation of Eq.~(\ref{eq:bs-0}), we then
obtain the following three-dimensional equation in the rest frame,
\begin{eqnarray}
   T(\mathbf{p}_f^{*},\mathbf{p}_i^{*};E^{*}(q)) &=& V(\mathbf{p}_f^*,\mathbf{p}_i^*;E^*(q)) \nonumber \\
&&+
i\int \frac{d^3 k^*}{(2\pi)^3} 
\frac{V(\mathbf{p}_f^{*},\mathbf{k}^{*};E^{*}(q)}{4\omega_1(\mathbf{k}^*)\omega_2(\mathbf{k}^*)}
\frac{ T(\mathbf{k}^*,\mathbf{p}_f^{*};E^{*}(q))}{E^*(q)-\omega_1(\mathbf{k}^*)-\omega_2(\mathbf{k}^*)+i\epsilon}
 \,.\label{eq:bs-1}
\end{eqnarray}
Rigorously speaking, dropping the anti-particle-pole term $\delta G$ in \cref{eq:g2-2} to get \cref{eq:g2-3} will modify the potential $V$ to $\tilde{V}$ defined through $\tilde{V}=V + V\,\delta G \,\tilde{V}$.
Using $\tilde{V}$ will not change the following discussion as long as
$\delta G$ is regular in the energy region of interest.
To simplify the notation, we proceed with $V$ instead of $\tilde{V}$.
Here we note that the three-dimensional reduction of the Bethe-Salpeter Equation we have chosen  is one of the possible approaches as reviewed in Ref.~\cite{Klein:1974aa}.
It is instructive to write this equation in matrix form as follows,
\begin{eqnarray}
   T=V+V G_2 T,\label{eq:bsmatrix-1}
\end{eqnarray}
which is easily solved by 
\begin{eqnarray}
   T=(1-V G_2)^{-1}V.\label{eq:bsmatrix-2}
\end{eqnarray}

The above derivation is considered to be in the center mass of system
(CM) since the 
potential $V$ matrix is diagonal in $lm$, which leads to Eqs.~(\ref{eq:t-par}-\ref{eq:G4IFVpw}).
In contrast, the BS equation directly boosted to the moving frame from \cref{eq:bs-0} will reduce to a boosted version of \cref{eq:g2-3,eq:bs-1} with both the potential $V$ and the propagator dependent on the total momentum, which is not diagonalizable in the partial-wave basis.
In this work, we introduce a prescription that boosts the interaction kernel after the center-mass-system reduction of the BS equation.


\subsection{The Quantization Condition in the Finite Volume}
\label{subsec:qd}

The quantization in a finite volume has been presented numerous times in the past in various forms.
Here we restate a similar version that allows us to more directly connect to the BS language described above and connect to a generalization of the implementation of the quantization condition at finite boost.

The 2-point correlation function can be expressed in terms of the BS kernel through the series shown in Fig.~\ref{fig:BSE} in Ref.~\cite{Kim:2005gf}.
Similarly, in the finite volume, based on the correlation functions
computed on a lattice, we can define the finite-volume T-matrix, $T^L$, as shown in Fig.~\ref{fig:BSE}(B).

It is the poles of $T^L$ that correspond to the energy eigenvalues resolved in a box, up to the familiar caveats that we only consider energies below any inelastic thresholds, and exponentially-suppressed finite-volume effects have been neglected.
In particular, for states with total energy below any inelastic threshold, it is only the 2-body propagator which will experience power corrections associated with the finite boundary conditions.
The potential, $V^L$, is only sensitive to the volume through virtual loop effects that are exponentially-suppressed with the box size, which leads to an important approximation, $V^L \sim V$. 
Similarly to Eq.~(\ref{eq:bsmatrix-1}), the finite-volume T-matrix is then given by
\begin{align}\label{eq:bsfv-1}
   T^L=V+VG_2^B T^L \,,
\end{align}
where $G_2^B$ is the 2-body propagator matrix, subject to the box quantization conditions. 
We will soon come to the details of the propagator $G_2^B$, but first we demonstrate how $T^L$ is related to the physical T-matrix, $T$.

From Eq.~(\ref{eq:bsfv-1}), we can derive,
\begin{eqnarray}
   T^L=V+V\left(G_2+G_2^B-G_2\right)T^L = V+V\left(G_2+G_2^L\right)T^L,\label{eq:VGBTL}
\end{eqnarray}
where we identify the familiar finite-volume propagator written as a sum minus integral, $G_2^L\equiv G_2^B-G_2$.
Then compare Eq.~(\ref{eq:bsmatrix-1}) and (\ref{eq:VGBTL}), we can immediately obtain 
\begin{eqnarray}
   T-T^L=-(1-VG_2)^{-1}VG_2^L T^L.\label{eq:TmTL}
\end{eqnarray}
By using Eq.~(\ref{eq:bsmatrix-2}), we this have
\begin{eqnarray}
   T^L=T+TG_2^L T^L ,\label{eq:TGLTL}
\end{eqnarray}
which graphically is given by Fig.\ref{fig:BSE}(C) --- an equivalent proof can be found in Ref.~\cite{Luscher:1986pf}.
One can then identify that the finite-volume BS equation can be expressed with the infinite volume T-matrix as the interaction kernel,
see also Refs.~\cite{Kim:2005gf,Briceno:2017max,Doring:2011vk}.

Now we write down the detailed version of \cref{eq:bsfv-1}:
\begin{align}
   T^{L}(p_f^*,p_i^*;P^*) &= V(p_f^*,p_i^*;P^*) 
   + \int \frac{d^3 \mathbf{k}^* }{(2\pi)^3}
V(p_f^*,k^*;P^*) G_2^B(k^*,P^*) T^{L}(k^*,p_i^*;P^*) \,,
\end{align}
where
\begin{align}\label{eq:repks}
    G_2^B(k^*,P^*):=\sum_{\mathbf{k}=(2\pi/L)\mathbf{n}\ (\mathbf{n}\in {\cal Z}^3)}(2\pi)^3\delta^3(\mathbf{k}^*-\mathbf{k})G_2(k,P^*) \,.
\end{align}
Performing the integral over these delta functions, one gets
\begin{align}
   T^{L}(p_f^*,p_i^*;P^*) &= V(p_f^*,p_i^*;P^*) 
   + \sum_{\mathbf{k}^*=\frac{2\pi\mathbf{n}}{L},\,\mathbf{n}\in {\cal Z}^3}
V(p_f^*,k^*;P^*) G_2(k^*,P^*) T^{L}(k^*,p_i^*;P^*) \,. 
\end{align}
Furthermore, when the total momentum of the system is non-vanishing in the box, the momentum modes of the field quanta are more naturally expressed with respect to the lattice rest frame -- we denote these momenta by a superscript r, e.g. $\mathbf{k}^r$.
To connect the partial wave amplitude, the scattering amplitude at infinite volume of CM system is necessary.
A boost transformation $\mathbf{k}^r \to \mathbf{k}^*$ is introduced
in the finite volume, and the
Jacobian factor also appears in the summation as follows,
\begin{eqnarray}
\int \frac{d^3k^*}{(2\pi)^3}\,\to\, \int \frac{d^3k^r}{(2\pi)^3}\JJ^r\,  \to\, \frac{1}{L^3}\sum_{\mathbf{k}^r}\JJ^r,\label{eq:summJ}
\end{eqnarray}
where $\JJ^r$ is the Jacobi factor and is discussed in the next section.
Hence the finite volume BS equation becomes,
\begin{align}\label{eq:BSErL2}
   T^{r,L}(p_f^*,p_i^*;P^*) &= V(p_f^*,p_i^*;P^*) 
   + \int \frac{dk_0^* }{2\pi} \frac{1}{L^3}\sum_{\mathbf{k}^r}
\JJ^r
V(p_f^*,k^*;P^*) G_2(k^*,P^*) T^{r,L}(k^*,p_i^*;P^*) \,. 
\end{align}
It is worth emphasizing that although the arguments are written in terms of $k^*$, the spatial components should be implicitly understood to be functions of the lattice rest-frame vectors, $\mathbf{k}^r$.

To exploit the connection to the usual T-matrix highlighted by Eq.~(\ref{eq:TGLTL}), we again add and subtract the integral form,
\begin{align}
    \int \frac{dk_0^* }{2\pi}  \left(\frac{1}{L^{3}}\sum_{\mathbf{k}^r}\right) \JJ^r
        \to   \int \frac{dk_0^* }{2\pi}  \int \frac{d^3k^r}{(2\pi)^3}\JJ^r 
        + \int \frac{dk_0^* }{2\pi}  \left(\frac{1}{L^{3}}\sum_{\mathbf{k}^r} - 
\int \frac{d^3k^r}{(2\pi)^3} \right) \JJ^r
\end{align}
and hence, as sketched above in Eqs.~(\ref{eq:VGBTL},\ref{eq:TGLTL}), we have  
\begin{align}
    T^{r,L}(p_f^*,p_i^*;P^*) &= T(p_f^*,p_i^*;P^*) 
+ \int \frac{dk_0^* }{2\pi}  \left(\frac{1}{L^{3}}\sum_{\mathbf{k}^r} - \int \frac{d^3k^r}{(2\pi)^3} \right) 
\nonumber\\&\qquad\times
\JJ^r  T(p_f^*,k^*;P^*) G_2(k^*,P^*) T^{r,L}(k^*,p_i^*;P^*) \,.
\label{eq:qc-1}
\end{align}

Furthermore, if we perform the $k_0^*$ integral as specified in the equations leading to Eq.~(\ref{eq:bs-1}), we have the corresponding 3-D reduction of the finite-volume T-matrix equation as follows,
\begin{align}
    T^{r,L}(\mathbf{p}_f^*,\mathbf{p}_i^*;E^*(q)) &= T(\mathbf{p}_f^*,\mathbf{p}_i^*;E^*(q)) 
+ i\left(\frac{1}{L^{3}}\sum_{\mathbf{k}^r} - \int \frac{d^3 k^r}{(2\pi)^3} \right) 
\nonumber\\&\qquad\times
\JJ^r  \frac{T(\mathbf{p}_f^*,\mathbf{k}^*;E^*(q))}{4\omega_1(\mathbf{k}^*)\omega_2(\mathbf{k}^*)}
\frac{ T^{r,\,L}(\mathbf{k}^*,\mathbf{p}_i^*;E^*(q))}{E^*(q)-\omega_1(\mathbf{k}^*)-\omega_2(\mathbf{k}^*)+i\epsilon} \,.
\label{eq:qc-2}
\end{align}
where, as above, we note that $E^*(q)=\omega_1(q)+\omega_2(q)$.

Here we have a brief diversion to note that the Poisson summation formula will
ensure that only the singularity term will contribute the subtraction between summation and integration.
Thus, to obtain 3-D dimensional finite-volume T-matrix equation is not
necessary to use the same reduction as in the infinite volume case, as
used to arrive at Eq.~(\ref{eq:g2-3}).
This comes as a consequences of the Poisson summation formula (see e.g.~\cite{Kim:2005gf}),
\begin{eqnarray}
     \left(\frac{1}{L^{3}}\sum_{\mathbf{k}^r} - \int \frac{d^3k^r}{(2\pi)^3} \right)g(\mathbf{k}) = \sum_{\mathbf{l}\neq\vec{0}}\int\frac{d^3k^r}{(2\pi)^3}e^{iL\lb\cdot \mathbf{k}}g(\mathbf{k})  \propto e^{-m_{min} L} \,.
\label{eq:poisson}
\end{eqnarray}
where the $m_{min}$ should be the lowest single-particle mass in the physical system.
Eq.~(\ref{eq:poisson}) is valid for any amplitude function $g(\mathbf{k})$ without singularities for real $\mathbf{k}$ and falling off fast enough at $|\mathbf{k}|\to \infty$ avoiding ultraviolet divergence.
As $L\to \infty$, the right hand side of
Eq.~(\ref{eq:poisson}) will be suppressed faster than any power of
$L^{-1}$, and can be safely neglected.
For example, as calculated in Ref.~\cite{Beane:2004tw}, the $g(\mathbf{k})=(\mathbf{k}^2+m)^{3/2}$, the right hand side will be $\propto e^{-mL}$.
Thus, in the calculation of Eq.~(\ref{eq:qc-1}), we will just keep the singularity terms and neglect the regular terms, i.e., we also can remove the first regular term in Eq.~(\ref{eq:g2-2}), and then use the pole function $P^*_0-\omega_2(\mathbf{k}^*)=\omega_1(\mathbf{k}^*)$ to perform the derivation in Eq.~(\ref{eq:g2-3}).  
Clearly, we could have,  
\begin{align}
    &\int \frac{dk_0^*}{2\pi}  \left(L^{-3}\sum_{\mathbf{k}} - \int \frac{d^3k}{(2\pi)^4} \right) G_2(k^*,P^*)f(k_0^*,|\mathbf{k}|^*,\hat{\mathbf{k}}^*) \nonumber\\
    \sim&  i\left(L^{-3}\sum_{\mathbf{k}} - \int \frac{d^3k}{(2\pi)^4} \right)\frac{f(\omega_1(q),q,\hat{\mathbf{k}}^*)}{4\omega_1(\mathbf{k}^*)\omega_2(\mathbf{k}^*)}
\frac{ 1}{E^*(q)-\omega_1(\mathbf{k}^*)-\omega_2(\mathbf{k}^*)+i\epsilon} \,.\label{eq:simply} 
\end{align}

To apply Eq.~(\ref{eq:simply}) to the integral of Eq.~(\ref{eq:qc-2}), one can add
  and subtract the numerator evaluated at the pole location,
  \begin{align}
\frac{T(\mathbf{p}_f^*,\mathbf{k}^*)T^L(\mathbf{k}^*,\mathbf{p}_i^*)}{E^*(q)-\omega_1(\mathbf{k}^*)-\omega_2(\mathbf{k}^*)+i\epsilon}\rightarrow&
\underbrace{
\frac{T(\mathbf{p}_f^*,\mathbf{k}^*)T^L(\mathbf{k}^*,\mathbf{p}_i^*)-T(\mathbf{p}_f^*,q\hat{\mathbf{k}}^*)T^L(q\hat{\mathbf{k}}^*,\mathbf{p}_i^*)}{E^*(q)-\omega_1(\mathbf{k}^*)-\omega_2(\mathbf{k}^*)+i\epsilon}}_{\textrm{residue
vanishes at
$|\mathbf{k}^*|=q$}}\nonumber\\
&+\frac{T(\mathbf{p}_f^*,q\hat{\mathbf{k}}^*)T^L(q\hat{\mathbf{k}}^*,\mathbf{p}_i^*)}{E^*(q)-\omega_1(\mathbf{k}^*)-\omega_2(\mathbf{k}^*)+i\epsilon}, \label{eq:treduce}
  \end{align}
where $\hat{\mathbf{k}}$ indicates the unit vector in the direction of $\mathbf{k}$, and the explicit $E^*(q)$ dependence of $T$ and $T^L$ has been suppressed. 
Importantly, the first two terms combine to give vanishing residue
such that the integrand is free of any singularities and hence will give a contribution that is exponentially suppressed in the volume according to Eq.~(\ref{eq:poisson}).
\footnote{A similar transformation can, optionally, be performed for the the scale factors $\JJ^r$ and $(\omega_1(\mathbf{k}^*)\omega_2(\mathbf{k}^*))$.} 
As a consequence, the dominant effects of the finite-volume
quantisation described by Eq.~(\ref{eq:qc-2}) can be evaluated using only the
on-shell values of $T$; and with the $\mathbf{k}$ values in the
sum/integral only depending on the on-shell magnitude of $\mathbf{k}$,
we are reduced to determining the summation over the rotations of
$\hat{\mathbf{k}}$ for $T$ and $T^L$.
Furthermore, it is notable that the above discussion is based on the
assumption that the first term in Eq.~(\ref{eq:treduce}) has no
singularities (or rapid momentum dependence), i.e., 
$\left(T(\mathbf{p}_f^*,\mathbf{k}^*)T^L(\mathbf{k}^*,\mathbf{p}_i^*)-T(\mathbf{p}_f^*,q\hat{\mathbf{k}}^*)T^L(q\hat{\mathbf{k}}^*,\mathbf{p}_i^*)\right)\propto \left(q^2-\mathbf{k}^{*\,2}\right)$.
However, for including an effective long-range interaction, such
assumption will be limited in applicability, please check recent papers~\cite{Hansen:2024ffk, Raposo:2023oru, Meng:2023bmz,Bubna:2024izx}. 
The general quantization condition in the moving frame for such
long-range effects remains the focus of ongoing work.
The rotational symmetry of the system is broken by the box -- and further broken at finite $\mathbf{P}$.
As a consequence, $T^{r,L}$ is no longer diagonal in $lm$, as compared to the usual case, Eq.~(\ref{eq:t-par}),
\begin{align}
    T^{r,L}(q\mathbf{\hat{p}}_f^*,q\mathbf{\hat{p}}_i^*;E^*(q))&=\sum_{lm;l'm'} 
[T^{r,L}(q;\mathbf{P})]_{lm,l'm'}Y_{lm}(\hat{\mathbf{p}}^*_f)Y^*_{l'm'}(\hat{\mathbf{p}}^*_i) \,,
\label{eq:tr-0}
\end{align}
where $\mathbf{P}$-dependence is introduced as the moving-frame total momentum
\footnote{Although the left hand side of \cref{eq:tr-0} does not display explicit dependence on $\mathbf{P}$, it is implied by the transformation denoted by $r$.}.
Similarly, we will have the partial-wave expansion as follows, 
\begin{eqnarray}
    T(q\mathbf{\hat{p}}_i^*, q\mathbf{\hat{k}}^*;E^*(q))&=&\sum_{lm;l'm'} 
[T(q;\mathbf{P})]_{lm,l'm'}Y_{lm}(\mathbf{\hat{p}}^*_f)Y^*_{l'm'}(\mathbf{\hat{k}}^*)\,,
\label{eq:tr-1}\\
   T^{r,L}(q\mathbf{\hat{k}}^*,q\mathbf{\hat{p}}_i^*;E^*(q))&=&\sum_{lm;l'm'} 
[T^{r,L}(q;\mathbf{P})]_{lm,l'm'}Y_{lm}(\mathbf{\hat{k}}^*)Y^*_{l'm'}(\mathbf{\hat{p}}^*_i).
\label{eq:tr-2}
\end{eqnarray}
Given the spherical symmetry of the infinite-volume theory, the components of $T$ are diagonal in $l$ and $m$. The angular components are mixed in the finite box, where by combining Eqs.~(\ref{eq:tr-0}), (\ref{eq:tr-1}) and (\ref{eq:tr-2}), Eq.~(\ref{eq:qc-2}) gives
\begin{eqnarray}
    [T^{r,L}(q;\mathbf{P})]_{lm,l'm'} &=& [T(q)]_{lm,l'm'}\nonumber \\
&& + 
\sum_{\bar{l}\bar{m},\bar{l}'\bar{m}'} [T(q)]_{lm,\bar{l}\bar{m}}\,\,
[F(q;\mathbf{P})]_{\bar{l}\bar{m},\bar{l}'\bar{m}'}\,\,
[T^{r,L}(q;\mathbf{P})]_{\bar{l}'\bar{m}',l'm'} \,,
\label{eq:qc-mx}
\end{eqnarray}
with
\begin{align}
    [F(q;\mathbf{P})]_{lm,l'm'}&= \left( L^{-3}\sum_{\mathbf{k}} - \int \frac{d^3k}{(2\pi)^3}  \right)\frac{i \JJ^r}{4\omega_1(\mathbf{k}^*)\,\omega_2(\mathbf{k}^*)} \frac{Y_{lm}(\hat{k}^*)Y^*_{l'm'}(\hat{k}^*) \left(\frac{|\mathbf{k}^*|}{q}\right)^{l+l'}}{P_0^*-(\omega_1(\mathbf{k}^*)+\omega_2(\mathbf{k})^*)+i \varepsilon} \,.
\end{align}
The factor $\left(\frac{|\mathbf{k}^*|}{q}\right)^{l+l'}$ is introduced to make $Y_{lm}(\hat{k}^*)$ well-defined at $|\mathbf{k}^*|=0$ and will be $1$ when $|\mathbf{k}^*|\to q$.

The matrix equation Eq.~(\ref{eq:qc-mx}) can be solved as
\begin{align}
    [T^{r,L}(q;\mathbf{P})] 
= \left( [T(q)]^{-1} - [F(q;\mathbf{P})] \right)^{-1} \,,
\end{align}
Since the finite-volume eigenenergies correspond to the poles of  $T^{r,L}(q;\mathbf{P})$,
the quantization condition can be expressed as:
\begin{align}
    \det\left( [T(q)]^{-1} - [F(q;\mathbf{P})] \right)=0 \,.
\label{eq:qc-mx-1}
\end{align}

For the numerical implementation, we have used another notation commonly seen in the literature (see e.g. Ref.~\cite{Gockeler:2012yj}).
By noticing $\text{Re}[(e^{i\delta}\sin\delta)^{-1}]=\cot\delta$ based on Eq.~(\ref{eq:G4IFVpw}), Eq.~(\ref{eq:qc-mx-1}) can be transformed into an alternative form:
\begin{align}\label{eq:qc-mx-2}
    \det\left( [\cot\delta(q)] + [M(q;\mathbf{P})] \right)=0 \,,
\end{align}
where
\begin{align}
    [\cot\delta(q)]_{lm,l'm'}=\cot\delta_l(q) \delta_{l,l'}\delta_{m,m'}\,,
\end{align}
and
\begin{align}
    [M(q;\mathbf{P})]_{lm,l'm'} &= \frac{32\pi^2E^*(q)}{\,q}  \text{Re}\left[ -i\,F(q;P) \right] \nonumber\\
    &= \frac{1}{q} \left(\frac{1}{ L^{3}}\sum_{\mathbf{k}} - \mathcal{P}\int \frac{d^3k^r}{(2\pi)^3}  \right) \frac{32\pi^2E^*(q)\JJ^r}{4\omega_1(\mathbf{k}^*)\,\omega_2(\mathbf{k}^*)} 
    \frac{Y_{lm}(\hat{\mathbf{k}}^*)Y^*_{l'm'}(\hat{\mathbf{k}}^*) \left(\frac{|\mathbf{k}^*|}{q}\right)^{l+l'}}
    {E^*(q)-(\omega_1(\mathbf{k}^*)+\omega_2(\mathbf{k}^*))}\,. \label{eq:Mlmlm}
\end{align}
We remind the readers that $\mathbf{k}^*$ is a function of
$\mathbf{k}^r$, to be discussed in detail in the following section.

Furthermore, to compare with common implementations,
Eq.~(\ref{eq:Mlmlm}) can be transformed as
\begin{align}
		\frac{1}{4\omega_1(\mathbf{k}^*)\omega_2(\mathbf{k}^*)} \frac{1}{E^*(q)-(\omega_1(\mathbf{k}^*)+\omega_2(\mathbf{k}^*))}
		\to \frac{1}{2E^*(q)}\frac{1}{q^2-\mathbf{k}^{*\,2}} \,,
	\end{align}
such that the pole location and residue is preserved at the on-shell point.
Hence we have an equivalent form for $M(q;\mathbf{P})$:
\begin{align}
    [M(q;\mathbf{P})]_{lm,l'm'}
    &= \frac{16\pi^2}{q} \left(\frac{1}{ L^{3}}\sum_{\mathbf{k}} - \mathcal{P}\int \frac{d^3k^r}{(2\pi)^3}  \right) {\cal J}^r \frac{Y_{lm}(\hat{\mathbf{k}}^*)Y^*_{l'm'}(\hat{\mathbf{k}}^*) \left(\frac{|\mathbf{k}^*|}{q}\right)^{l+l'}}
    {q^2-{k}^{*\,2}}\,. \label{eq:Mlmlm2}
\end{align}
This form matches the $M$ used in Ref.~\cite{Gockeler:2012yj}, up to the generalization of the boost transformation $\mathbf{k}^r\to \mathbf{k}^*$ and the corresponding Jacobian.
Further details on the numerical implementation of $M$ are presented in \cref{app:deriM}.

\section{Three-momentum transformation}
\label{sec:comparing}

Now let us specify the exact three-momentum transformation $\kb^*\to\kb^r$ as an imitation of the four-momentum Lorentz boost.
Conceptually, this describes a system in the rest of two-body system transformed to a moving frame.
The key problem is that the energy components are missing for these three-momenta. 
Thus, we will introduce two variables $a^*$ and $b^*$ for the energy parts of $\mathbf{P}^*$ and $\mathbf{k}^*$, respectively.
Clearly, such transformation should have some freedom because of the different choices of $a^*$ and $b^*$. 
In addition, the choice of $a^*$ and $b^*$ must satisfy the on-shell condition.
The on-shell condition means when $|\mathbf{k}^*|=q$, we should have $a^*=\omega_1(q)+\omega_2(q)=E^*(q)$ and $b^*=\omega_1(q)=\frac{{E^*}^2(q)+m_1^2-m^2_2}{2E^*(q)}$.
Then we have several typical choices of $a^*$ and $b^*$ as follows,
\begin{align}\label{eq:asbs}
a^*&=E^*(q) \,\,\text{ or }\,\, \omega_1(\mathbf{k}^*)+\omega_2(\mathbf{k}^*)\,, \\
b^*&=\omega_1(q) \,\text{ or }\,\,  \omega_1(\mathbf{k}^*)\,.
\end{align}
In \cref{app:proof}, it is proven that any combinations of these choices for $\left(a^*, b^*\right)$ will give equivalent schemes up to the correction suppressed by factor $e^{-mL}$. 
Then it will be easy to obtain the transformation based on standard four-momentum Lorentz transformation as follows,
\begin{align}
\mathbf{k}^r &= (k^r_{\parallel},\mathbf{k}^r_{\perp}) = (\gamma\,\beta\,b^* + \gamma\,k_{\parallel}^*,\mathbf{k}^*_{\perp}) \equiv \mathcal{A}\,\mathbf{k}^*_{\parallel} + \mathcal{B}\,\mathbf{P} + \mathbf{k}^*_{\perp} \,,\label{eq:repa2}\\
\beta&=\frac{|\mathbf{P}|}{\sqrt{a^{*\,2}+\mathbf{P}^2}}\,, \qquad
		\mathcal{A} = \gamma= \frac{\sqrt{a^{*\,2}+\mathbf{P}^2}}{a^*} \,, \qquad \mathcal{B} = \frac{b^*}{a^*} \label{eq:ABdef}\,.
\end{align}
Here, the function $\mathbf{k}^*(\mathbf{k}^r)$ can be obtained from \cref{eq:repa2} easily.

In the next three subsections, we will give the quantization conditions based on the above momentum transformation with three choices of $a^*$ and $b^*$.
First, we will give a new form and refer to it as \textbf{LWLY}, while the other two are named as \textbf{KSS} and \textbf{RG} using the same three-momentum transformation defined in Refs.~\cite{Kim:2005gf} and \cite{Rummukainen:1995vs}, respectively.
In \cref{app:proof}, we will show that all approaches (\textbf{LWLY}, \textbf{KSS}, and \textbf{RG}) are equivalent up to the exponentially suppressed correction ($\propto e^{-mL}$).

\subsection{The new transformation and quantization implementation \textbf{LWLY}}

Based on the derivation of the two-body propagator, $G_2$, we observe that the form is consistent with that in relativistic quantum mechanics formulated by Dirac~\cite{Dirac:1936tg}.
Actually, in the instant form of relativistic quantum mechanics of Dirac within which both particles are on mass-shell with
\begin{eqnarray}
p^*_1&=&(\omega_1(\mathbf{k}^*),\mathbf{k}^*) \label{eq:p1p2-a}\\
p^*_2&=&(\omega_2(\mathbf{k}^*),-\mathbf{k}^*)  \label{eq:p1p2-b}
\end{eqnarray}
Here we also note that the Hamiltonian approach to finite-volume
effective field theory calculations of two-particle scattering also has the scattering equation similar to that defined by Eqs.~(\ref{eq:bs-1}). 
Thus what will be developed below is most suitable to the Hamiltonian
approach of finite-volume spectra.
 
The transformation from $(p^*_1,\,p^*_2)$ to $(p_1,\,p_2)$ in the moving frame with 
$P=(P_0,\,\mathbf{P})$ is well defined within relativistic quantum mechanics based on the mass-shell condition. 
They are
\begin{eqnarray}
p_1&=&(\omega_1(\mathbf{k}^r), \mathbf{k}^r)\nonumber \\
p_2&=&(\omega_2(\mathbf{P}-\mathbf{k}^r),\mathbf{P}-\mathbf{k}^r)
\end{eqnarray}
where $\mathbf{k}^r$ is expressed in terms of $\mathbf{k}^*$.
It is rather straightforward that this transformation is just the case we choose $ a^*=\omega_1(\mathbf{k}^*)+\omega_2(\mathbf{k}^*)$ and $b^*=\omega_1(\mathbf{k}^*)$, which will be named \textbf{LWLY}, i.e., $r=\textbf{LWLY}$.
Accordingly, we find that
\begin{align}
\mathbf{k}^r &= \frac{\sqrt{\left(\omega_1(\mathbf{k}^*)+\omega_2(\mathbf{k}^*)\right)^2+\mathbf{P}^2}}{\omega_1(\mathbf{k}^*)+\omega_2(\mathbf{k}^*)}\mathbf{k}^*_{\parallel} + \frac{\omega_1(\mathbf{k}^*)}{\omega_1(\mathbf{k}^*)+\omega_2(\mathbf{k}^*)}\mathbf{P} + \mathbf{k}^*_{\perp} \,, \label{eq:kksSC}\\
\mathbf{k}^* &= \frac{\omega_1(\mathbf{k}^r)+\omega_2(\mathbf{P}-\mathbf{k}^r)}{\sqrt{\left(\omega_1(\mathbf{k}^r)+\omega_2(\mathbf{P}-\mathbf{k}^r)\right)^2-\mathbf{P}^2}}\mathbf{k}^r_{\parallel} - \frac{\omega_1(\mathbf{k}^r)}{\sqrt{\left(\omega_1(\mathbf{k}^r)+\omega_2(\mathbf{P}-\mathbf{k}^r)\right)^2-\mathbf{P}^2}}\,\mathbf{P} + \mathbf{k}^r_{\perp}\,, \label{eq:kskSC} \\
{\mathcal J}^r&=\left|\frac{\partial
\mathbf{k}^*}{\partial \mathbf{k}^r}\right|
=\frac{\omega_1(\mathbf{k}^*)\omega_2(\mathbf{k}^*)}
{\omega_1(\mathbf{k}^*)+\omega_2(\mathbf{k}^*)}
\frac{\omega_1(\mathbf{k}^r)+\omega_2(\mathbf{P}-\mathbf{k}^r)}
{\omega_1(\mathbf{k}^r)\omega_2(\mathbf{P}-\mathbf{k}^r)}.\label{eq:SC}
\end{align}

The evaluation of $M$ proceeds with these forms via equations provided in Appendix \ref{app:deriM}. 
The general equivalence between this new scheme and others, including RG
and KSS, in presented in \cref{app:proof}.

\subsection{ Boost form of \textbf{KSS}}

In this case, we take the three-momentum transformation as follows ($r=\textbf{KSS}$),
\begin{align}\label{eq:SA}
		& a^*=E^*(q)\,,\quad b^*=\omega_1(\mathbf{k}^*)\,,\quad \mathcal{J}^{r}= \frac{\omega_1(\mathbf{k}^*)}{\omega_1(\mathbf{k}^r)} \nn\\
		& \mathbf{k}^r = \frac{E(q)}{E^*(q)}\mathbf{k}^*_{\parallel} + \frac{\omega_1(\mathbf{k}^*)}{E^*(q)}\mathbf{P} + \mathbf{k}^*_{\perp} \,, \quad \mathbf{k}^* = \frac{E(q)}{E^*(q)}\mathbf{k}^r_{\parallel} - \frac{\omega_1(\mathbf{k}^r)}{E^*(q)}\mathbf{P} + \mathbf{k}^r_{\perp} \,,
\end{align}
where the energies for the two particles are $\omega_1(\mathbf{k}^r)$
and $E(q)-\omega_1(\mathbf{k}^r)$ in the moving frame, respectively,
so only the first particle is taken to be on the mass shell.
It has the similar form as Ref.~\cite{Kim:2005gf}.
To show the comparison, we explicitly provide $\bar{M}_{00}(q,\,\mathbf{P})$ defined in Eq.~(\ref{eq:mbar}) for the $S$-wave case,
\begin{align}\label{eq:MMA} 
\bar{M}^{\textbf{KSS}}_{00}(q,\,\mathbf{P}) &=
\frac{4\pi}{q}\left[
\frac{1}{L^3}
		\sum_{\mathbf{k}=\frac{2\pi}{L}\mathbf{n},\,\mathbf{n} \in \mathbb{Z}^3}
		\frac{\omega_1(\mathbf{k}^*)}{\omega_1(\mathbf{k})}
		\frac{e^{\alpha(q^2-\mathbf{k}^{*\,2})}}{q^2-\mathbf{k}^{*\,2}}
		- 
		\mathcal{P} \int \frac{d^3k^*}{(2\pi)^3}
		\frac{e^{\alpha(q^2-\mathbf{k}^{*\,2})}}{q^2-\mathbf{k}^{*\,2}}
		\right]\nonumber\\
		&-\frac{1}{\pi qL}\sum_{\mathbf{n} \in \mathbb{Z}^3, \mathbf{n}\neq 0}
		\int^{\alpha}_{0} dt\, e^{tq^2}\int dk^*\, e^{-tk^{*\,2}}\nn\\
		&\times \cos \left[ L\frac{\omega_1(\mathbf{k}^*)}{E^*(q)}\mathbf{n}\cdot\Pb \right]  \frac{2k^*\sin \left[L\,D_{\textbf{KSS}}\,k^* \right]}{D_{\textbf{KSS}}} \,,\nn\\
		D_{\textbf{KSS}}&=\sqrt{\mathbf{n}^2  +\left(\frac{\mathbf{n}\cdot \mathbf{P}}{E^*(q)}\right)^2} \,. 
	\end{align}
In the first line of which the expression in the square brackets are exactly the same as Eq.~(3) of Ref.~\cite{Kim:2005gf} in the equal-mass case.
And Ref.~\cite{Kim:2005gf} emphasized that $\alpha$ is not an additional parameter and it should be close to zero.
In fact, when $\alpha \to 0^+$, the other two terms in our expression vanishes since $\int_0^{\alpha\to0^+}dt\to0$.
Furthermore, when that term is included, the expression will be independent of $\alpha$ exactly. 
We will show this in the numerical calculation in \cref{sec:NCC}.
Therefore, our formalism with this three-momentum transformation encodes the quantization condition of Ref.~\cite{Kim:2005gf} successfully.

\subsection{Boost form of \textbf{RG}}

Here we take the three-momentum transformation as follows ($r=\textbf{RG}$),
\begin{align}\label{eq:SB}
& a^*=E^*(q)\,,\quad b^*=\frac{E^*(q)}{2}+\frac{m^2_1-m^2_2}{2E^*(q)}=\omega_1(q)\,,\quad \mathcal{J}^r= \frac{E^*(q)}{E(q)}\,, \nn\\
& \mathbf{k}^r = \frac{E(q)}{E^*(q)}\mathbf{k}^*_{\parallel} + \frac{1}{2}\left(1+\frac{m^2_1-m^2_2}{E^{*\,2}(q)}\right)\mathbf{P} + \mathbf{k}^*_{\perp} \,, \nn\\
& \mathbf{k}^* = \frac{E^*(q)}{E(q)}\left(\mathbf{k}^r_{\parallel} - \frac{1}{2}\left(1+\frac{m^2_1-m^2_2}{E^{*\,2}(q)}\right)\mathbf{P}\right) + \mathbf{k}^r_{\perp} \,,
	\end{align}
where the total rest energy and the energy of particle 1 are both dependent only on the on-shell momentum $q$.
Neither of the two particles are on shell.
This method is used in various references, such as Refs.~\cite{Rummukainen:1995vs,Gockeler:2012yj,Leskovec:2012gb}.
$\bar{M}_{00}(q,\,\mathbf{P})$ is defined in Eq.~(\ref{eq:mbar}) for the $S$-wave case.
It shows that the usual quantization condition can be derived explicitly.
In the S-wave case, we have
\begin{align}\label{eq:MMB}
\bar{M}^{\textbf{RG}}_{00}(q,\,\mathbf{P}) &=
	\frac{4\pi}{q}\left[\frac{1}{L^3}\sum_{\mathbf{k}=\frac{2\pi}{L}\mathbf{n},\,\mathbf{n} \in \mathbb{Z}^3}\frac{E^*(q)}{E(q)}\frac{e^{\alpha(q^2-\mathbf{k}^{*\,2})}}{q^2-\mathbf{k}^{*\,2}}
		- 
\mathcal{P} \int \frac{d^3k^*}{(2\pi)^3}\frac{e^{\alpha(q^2-\mathbf{k}^{*\,2})}}{q^2-\mathbf{k}^{*\,2}}\right]\nonumber\\
&-\frac{1}{\pi qL}\sum_{\mathbf{n} \in \mathbb{Z}^3, \mathbf{n}\neq 0}
\cos \left[ \frac{L\,\mathbf{n}\cdot\Pb}{2}\left(1+\frac{m_1^2-m^2_2}{E^{*\,2}(q)}\right)  \right]
\int^{\alpha}_{0} dt\, e^{tq^2}\nn\\
&\times\int dk^*\, e^{-tk^{*\,2}} 
\frac{2k^*\sin \left[L\,D_{\textbf{RG}}\,k^* \right]}{D_{\textbf{RG}}} \,,\nn\\
D_{\textbf{RG}}&=D_{\textbf{KSS}}=\sqrt{\mathbf{n}^2  +\left(\frac{\mathbf{n}\cdot \mathbf{P}}{E^*(q)}\right)^2} \,.
\end{align}
By using the following equations:
\begin{align}
&\int dk^*\,e^{-tk^{*\,2}}
\frac{2k^*\sin \left[ L\,D_{\textbf{RG}}\,k^* \right]}{D_{\textbf{RG}}}
	=\frac{\sqrt{\pi} L}{2t^{3/2}}e^{-\frac{D^2_{\textbf{RG}} L^2}{4t}},
	\nn\\
	&-\frac{4\pi}{q}\mathcal{P} \int \frac{d^3k^*}{(2\pi)^3}
	\frac{e^{\alpha(q^2-k^{*\,2})}}{q^2-k^{*\,2}}
	=\frac{1}{\pi qL}\left[\sqrt{\frac{\pi L^2}{\alpha}}-
	\int_0^{\alpha\left(\frac{2\pi}{L}\right)^2}dt \left(e^{t\left(\frac{qL}{2\pi}\right)^2}-1\right)
	\left(\frac{\pi}{t}\right)^{\frac{3}{2}}\right]\,,
	\end{align}
and choosing $\alpha = \left(\frac{L}{2\pi}\right)^2$, $\bar{M}^B_{00}$ can be expressed as
\begin{align}\label{eq:MB00}
		\bar{M}^B_{00}(q,\mathbf{P}) &=
		\frac{1}{\pi qL}\frac{E^*(q)}{E(q)}
		\sum_{\mathbf{k}=\frac{2\pi}{L}\mathbf{n},\,\mathbf{n} \in \mathbb{Z}^3}
		\frac{e^{\left(\frac{L\kb^*}{2\pi}\right)^2-\left(\frac{Lq}{2\pi}\right)^2}}{\left(\frac{L\kb^*}{2\pi}\right)^2-\left(\frac{Lq}{2\pi}\right)^2}
		\nn\\
		&+
		\frac{1}{\pi qL}\left[2\pi^{\frac{3}{2}}-
		\int_0^1dt \left(e^{t\left(\frac{Lq}{2\pi}\right)^2}-1\right)
		\left(\frac{\pi}{t}\right)^{\frac{3}{2}}\right]
		\nonumber\\
		&-\frac{1}{\pi qL}\sum_{\mathbf{n} \in \mathbb{Z}^3, \mathbf{n}\neq 0}
		\cos (\pi \mathbf{n}\cdot\Deltab)
		\int^{1}_{0} dt\,e^{t\left(\frac{Lq}{2\pi}\right)^2}
		\left(\frac{\pi}{t}\right)^{\frac{3}{2}}e^{-\frac{\pi^2
		D_{\textbf{RG}}^2}{t}}\nn\\
		&=-\frac{1}{\pi qL} \frac{E^*(q)}{E(q)} \sqrt{4\pi} \mathcal{Z}^{\mathbf{\Delta}}_{00}(1;\left(\frac{Lq}{2\pi}\right)^2) \,,
\end{align}
where $\mathbf{\Delta}=\left[ \frac{\mathbf{P}\,L}{2\pi}\left(1+\frac{m_1^2-m^2_2}{E^{*\,2}(q)}\right)  \right]$ and the zeta function $\mathcal{Z}^{\mathbf{\Delta}}_{00}$ is defined in Ref.~\cite{Gockeler:2012yj} (also defined in Ref.~\cite{Leskovec:2012gb} but with a different notation). 
It will reduce into the zeta function defined in Ref.~\cite{Rummukainen:1995vs} in the equal-mass case for moving frame.
It also reduces into L\"uscher's zeta function \cite{Luscher:1986pf,Luscher:1990ux} for the rest frame, i.e., $\mathbf{P}=\mathbf{\Delta} = (0,0,0)$.
Furthermore, the pure S-wave quantization condition will be simplified as
\begin{align}
	\tan \delta_0(q)&=\frac{E(q)}{E^*(q)}\,\frac{Lq\sqrt{\pi}}{2}\frac{1}
	{\mathcal{Z}^{\mathbf{\Delta}}_{00}(1;\left(\frac{Lq}{2\pi}\right)^2)} \,,
	\end{align}
which reproduces the results of Refs.~\cite{Rummukainen:1995vs,Gockeler:2012yj,Leskovec:2012gb} exactly. 
It is straightforward to extend \cref{eq:MB00} to any partial-wave case as 
\begin{align}
	\bar{M}^{\textbf{RG}}_{lm}(q,\,\mathbf{P}) &= -\frac{1}{\pi qL} \frac{E^*(q)}{E(q)} \sqrt{4\pi} \mathcal{Z}^{\mathbf{\Delta}}_{lm}(1;\left(\frac{Lq}{2\pi}\right)^2) \,.
\end{align}
Thus, from \textbf{RG}, we can reproduce the  quantization condition of Refs.~\cite{Rummukainen:1995vs,Gockeler:2012yj,Leskovec:2012gb}.
%


\section{Numerical Comparison and Discussion}
\label{sec:NCC}

To numerically confirm the equivalence of all three approaches, and compare the differences, we build a simple model for the $\pi\pi \to \pi\pi$ scattering. 
It includes two-pion states and one bare state.
	The partial-wave amplitude $t_l$ is related to the phase shift by
	\begin{align}
		e^{2i\delta_l(q)}&=
		1-i\frac{\pi q E^*(q)}{2}t_{l}(q,q,E^*(q))\,,
	\end{align}	
	and can be solved from
	\begin{align}
		t_l(q,\,q',\,E)&=\frac{f_l(q)f^*_l(q')}{E-m_B}+\int k^2dk \frac{f_l(q)f^*_l(k)}{E-m_B}
		\frac{1}{E-2\sqrt{m_\pi^2+k^2}+i\epsilon}
		t_l(k,\,q',\,E)\,,
	\end{align}
	to give
	\begin{align}
		t_l(q,\,q',\,E)&=f_{l}(q)f_l(q')\left(
		E-m_B-\int k^2dk\frac{f_l(k)f^*_l(k)}{E-2\sqrt{m_\pi^2+k^2}+i\epsilon}
		\right)^{-1}\,,
	\end{align}
	where $m_\pi=138.5$ MeV and $m_B$ are the masses of the pion and the bare state respectively.
	We choose the parameterization of $f_l$ as
	\begin{align}
		f_l(q)&=\frac{g_l}{m_\pi^{l+1/2}}\frac{q^l}{\left(1+(c_{l}\,q)^2\right)^{l/2+1}}\,.
	\end{align}

\begin{figure}[htbp] \vspace{-0.cm}
	\begin{center}
	\includegraphics[width=0.9\columnwidth]{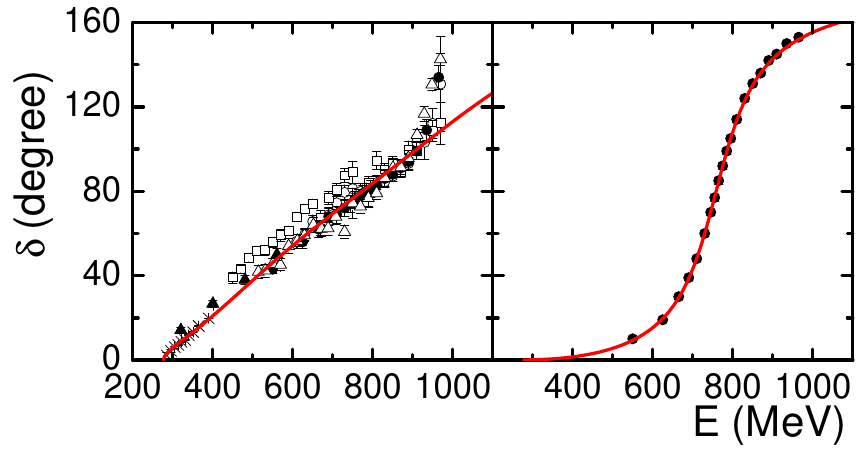}
	\caption{S(left)- and P(right)-wave phase shifts solved with parameters from \cref{tab:para} are compared with S- and P-wave data taken from Refs.~\cite{Batley:2007zz,Hyams:1973zf,Estabrooks:1973dya,Protopopescu:1973sh,doi:10.1063/1.2948709,Manner:1974ak,Froggatt:1977hu} and Ref.~\cite{Protopopescu:1973sh} respectively.}
	\label{fig:phase}
	\end{center}
	\end{figure}

	Then we fit this model to phase shifts of the S-wave taken from Refs.~\cite{Batley:2007zz,Hyams:1973zf,Estabrooks:1973dya,Protopopescu:1973sh,doi:10.1063/1.2948709,Manner:1974ak,Froggatt:1977hu} and the P-wave taken from Ref.~\cite{Protopopescu:1973sh}, the resulting parameters and phase shifts are shown in \cref{tab:para} and \cref{fig:phase} respectively.

	\begin{table}[htbp]
	     \setlength{\tabcolsep}{0.15cm}
	\begin{center}
	\caption{The model parameters used for the numerical investigation.}
	\begin{tabular}{ccccc}\hline
	                           &l=0     &     l=1         \\
	\hline
	 $m_B$(MeV)      &948.96   &  852.50   \\
	 $g_{l}$         &0.64698  &  0.095626 \\
	 $c_{l}$(fm)     &0.43979  & 0.48477 \\
	\hline\end{tabular}  \label{tab:para}
	\end{center}
	\end{table}

	\begin{table}[htbp]
	     \setlength{\tabcolsep}{0.15cm}
	\begin{center}
	\caption{Numerical results of $\bar{M}_i$ (i=a, b, c) and $\bar{M}_{00}$ with $L=1$ fm, $E^*(q) = 500$ MeV, $\mathbf{P} = \frac{2\pi}{L}(1,1,1)$, $\alpha_1 =$ $0.01/(100)^2$ MeV$^{-2}$ and $\alpha_2 =$ $0.1/(100)^2$ MeV$^{-2}$.}
	\begin{tabular}{cccccc}\hline
	  Scheme     &    $\alpha$   & $\bar{M}_a$    & $\bar{M}_b$  & $\bar{M}_c$  & $\bar{M}_{00}$     \\
	\hline
	 $\textbf{KSS}$         &$\alpha_1$     & $-12.59492$     & $2.59249$   & $-1.37813$  & $-11.3806$\\
	             &$\alpha_2$     & $-6.57847$      & $0.45657$   & $-5.25866$  & $-11.3806$\\
	 $\textbf{RG}$         &$\alpha_1$     & $-13.97305$     & $2.59249$   & $\sim -10^{-6}$ &  $-11.3806$\\
	             &$\alpha_2$     & $-11.23408$      & $0.45657$   & $-0.60305$  & $-11.3806$\\
	 $\textbf{LWLY}$         &$\alpha_1$     & $-14.16106$     & $2.59249$   & $-0.69534$  & $-12.2639$\\
	             &$\alpha_2$     & $-5.65005$      & $0.45657$   & $-7.07043$  & $-12.2639$\\
	\hline\end{tabular}  \label{tab:numerical}
	\end{center}
	\end{table}

	\begin{figure}[htbp]
	\begin{center}
	\includegraphics[width=1.\columnwidth]{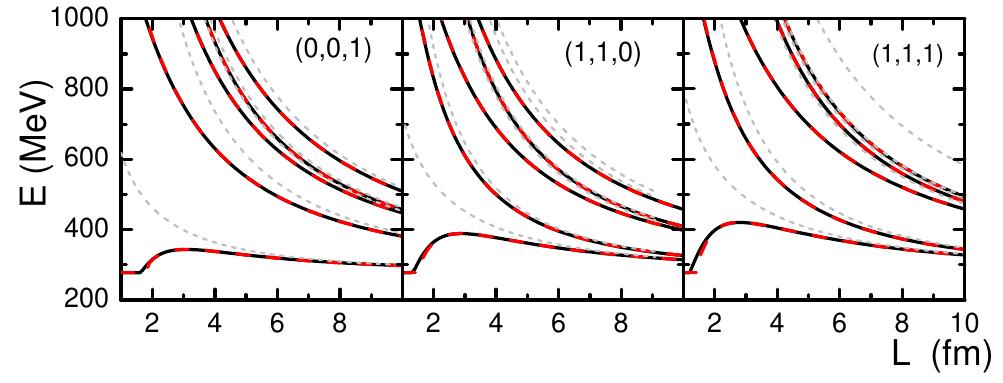}
	\caption{Spectra for systems with total momentums $\frac{2\pi}{L}(0,0,1)$, $\frac{2\pi}{L}(1,1,0)$ and $\frac{2\pi}{L}(1,1,1)$ solved in the scheme \textbf{KSS}/\textbf{RG} (red dashed) and the scheme \textbf{LWLY} (black solid) with the pure S-wave phase shift. Gray short dotted lines represent non-interacting energies.}
	\label{fg:gS}
	\end{center}
	\end{figure}

	\begin{figure}[htbp]
	\begin{center}
	\includegraphics[width=1.\columnwidth]{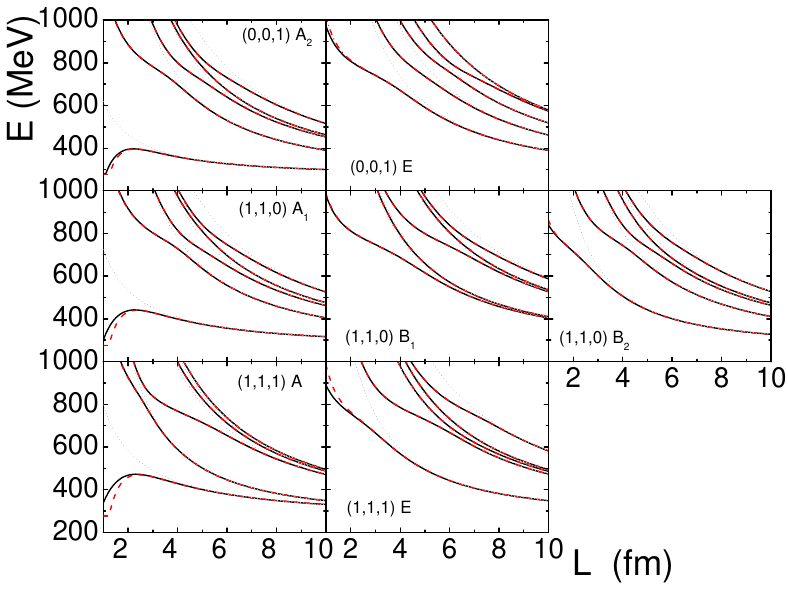}
	\caption{Spectra for systems with total momentums $\frac{2\pi}{L}(0,0,1)$, $\frac{2\pi}{L}(1,1,0)$ and $\frac{2\pi}{L}(1,1,1)$ solved in the scheme \textbf{KSS}/\textbf{RG} (red dashed) and the scheme \textbf{LWLY} (black solid) with the pure P-wave phase shift. Gray short dotted lines represent non-interacting energies.}
	\label{fg:gP}
	\end{center}
	\end{figure}
	
As the detailed derivation given in \cref{app:deriM}, we will introduce a new parameter $\alpha$ which just help to make numerical calculation. 
Now we check the $\alpha$-independence of $\bar{M}_{00}$ with $L=1$ fm, $E^*(q) = 500$ MeV and $\mathbf{P} = \frac{2\pi}{L}(1,1,1)$.
We use two different values of $\alpha$, $\alpha_1 =$ $0.01/(100)^2$ MeV$^{-2}$ and $\alpha_2 =$ $0.1/(100)^2$ MeV$^{-2}$.
The results are shown in \cref{tab:numerical}.
It confirms that these quantization conditions \cref{eq:MMA,eq:MMB,eq:MMC} are indeed independent of $\alpha$.
Furthermore, to our surprise, $\bar{M}_{00}$ for Approaches \textbf{KSS} and \textbf{RG} are exactly the same even though their values of $\bar{M}_b$ and $\bar{M}_c$ are different.
This suggests the exact equivalence between schemes \textbf{KSS} and \textbf{RG}.
In fact, it is true even for any partial waves.
We provide a proof in \cref{app:ABmodel}.

With the solved phase shifts, we can provide the spectra calculated from the quantization conditions of the three schemes in the pure S-wave (shown in \cref{fg:gS}) and the pure P-wave (shown in \cref{fg:gP}) cases.
When $L$ is larger than $2.5$ fm or $m_\pi L \ge 2$, the results of the three schemes are indistinguishable as expected in this case.

\section{Summary and Discussions}
\label{sec:Sum}

In this work, a general formalism of momentum transformations in a moving finite volume is proposed.
It is natural to start by considering starting from the perspective of the rest-frame system, a suitable boost transformation is applied to consider a system at finite total momenta, and it is this set of momenta that one can apply the finite-box quantisation conditions.
With suitable transformations, the discretization scheme produces eigenenergies that are equivalent up to exponentially suppressed corrections of the lattice size.
The suitable three-momentum transformations in \cref{eq:repa2} are constructed as an imitation of the four-momentum Lorentz boost.
To implement this four-momentum boost, the total energy and the energy of one particle must be explicitly specified --- however, there is some arbitrariness in this specification, up to satisfying the appropriate on-shell limits.

Amongst these transformations, we consider three particular cases \cref{eq:SA,eq:SB,eq:SC} in detail.
Two of these correspond to the common implementations used in the modern literature, \textbf{KSS} and \textbf{RG}.
Importantly, we introduce a new approach (\textbf{LWLY}) which has several unique benefits in certain applications.
If the schemes of \textbf{KSS} and \textbf{RG} were to be applied within a Hamiltonian formulation, they necessitate the introduction of an energy-dependence to the Hamiltonian at finite momentum.
Then the equation $\det(H(E)-EI)=0$ is no longer a standard eigenvalue equation since the matrix here depends on its own eigenvalues.
The new scheme \textbf{LWLY} we introduced, that is equivalent to the two schemes up to exponentially-suppressed corrections, does not have this problem.
Furthermore, the eigenvectors of an energy-independent Hamiltonian form a complete orthonormal basis of the Hilbert space of the Hamiltonian.
This enables a probabilistic interpretation that have been used in many previous studies \cite{Hall:2013qba, Wu:2014vma, Wu:2016ixr, Li:2019qvh}. 
In addition, in the hadron physics, the widely used compositeness proposed in Ref.~\cite{Weinberg:1965zz} also requires such a probabilistic interpretation.
The new scheme can also be used in the three-body system to avoid the negative energy or the velocity being larger than the light speed~\cite{Blanton:2020gha}.
This consistency shows our general formalism is successfully producing specific quantization conditions in the moving frame.

\section*{Acknowledgements}
It is a pleasure to thank Derek Leinweber, Stephen Sharpe, Anthony Thomas and James Zanotti for very helpful discussions; we are also grateful to Fernando Romero-L\'opez and Akaki Rusetsky for a critical reading and feedback on a draft version of this manuscript.
This work was partly supported by the National Natural Science Foundation of China (NSFC) under Grants Nos. 12175239 and 12221005 (J.J.W), 
and by the National Key R\&D Program of China under Contract No. 2020YFA0406400
(J.J.W), 
and by the Chinese Academy of Sciences under Grant No. YSBR-101 (J.J.W),
and by the ARC Grant DP220103098 (R.D.Y),
and by the Excellence Hub project ``Unraveling the 3D parton structure of the nucleon with lattice QCD (3D-nucleon)" id EXCELLENCE/0421/0043 co-financed by the European Regional Development Fund and the Republic of Cyprus through the Research and Innovation Foundation (Y.L). 
Y. L. further acknowledges computing time granted on Piz Daint at Centro Svizzero di Calcolo Scientifico (CSCS) via the project with id s1174, JUWELS Booster at the J\"{u}lich Supercomputing Centre (JSC) via the project with id pines, and Cyclone at the Cyprus institute (CYI) via the project with ids P061, P146 and pro22a10951.

\appendix


\section{The proof of equivalence between the quantization conditions of various three-momentum transformations}
\label{app:proof}

In the \textbf{RG} approach, the transformation: 
\begin{align}
	\mathbf{k}^*_\textbf{RG} \to \mathbf{k}_\textbf{RG} = \frac{E(q)}{E^*(q)}\mathbf{k}^*_{\textbf{RG},\parallel} + \frac{\omega_1(q)}{E^*(q)}\mathbf{P} + \mathbf{k}^*_{\textbf{RG},\perp} \,,
\end{align}
is linear (i.e., the parameters $a^*$ and $b^*$ are both independent on $\mathbf{k}$ or $\mathbf{k}^*$), so it is easy to find the inverse transformation:
\begin{align}
	\mathbf{k}^*_\textbf{RG} = \frac{E^*(q)}{E(q)} \mathbf{k}_{\textbf{RG},\parallel} - \frac{\omega_1(q)}{E(q)} \mathbf{P} + \mathbf{k}_{\textbf{RG},\perp}\,.\label{eq:modelBk}
\end{align}
If we can prove that the \textbf{RG} approach is equivalent to any approach X considered in this work with the following general form:
\begin{align}
        \mathbf{k}_X = \mathcal{A}_X(|\mathbf{k}^*_X|)\mathbf{k}^*_{X,\parallel} + \mathcal{B}_X(|\mathbf{k}^*_X|)\mathbf{P} + \mathbf{k}^*_{X,\perp}\,,\label{eq:modelXk}
\end{align}
the equivalence between all approaches is then proved.
In other words, we need to prove that \cref{eq:Mlmlm} for the \textbf{RG} approach and the X approach are equivalent.

Following a similar discussion as made in \cref{subsec:qd}, we need to prove that the subtraction of kernel functions in \cref{eq:Mlmlm} for the \textbf{RG} and X approaches:
\begin{align}
	H(\mathbf{k}^*_X)&\equiv \frac{\left|\left|\frac{\partial \mathbf{k}^*_X}{\partial \mathbf{k}}\right|\right|}{4\omega_1(\mathbf{k}^*_X)\,\omega_2(\mathbf{k}^*_X)}
   \frac{4\pi Y^*_{lm}(\hat{\mathbf{k}}^*_X)Y_{l'm'}(\hat{\mathbf{k}}^*_X)}
   {P_0^*-\left(\omega_1(\mathbf{k}^*_X)+\omega_2(\mathbf{k}^*_X)\right)}
   \left(\frac{|\mathbf{k}^*_X|}{q}\right)^{l+l'},\nonumber\\ 
	H(\mathbf{k}^*_\textbf{RG})&\equiv\frac{\left|\left|\frac{\partial \mathbf{k}^*_\textbf{RG}}{\partial \mathbf{k}}\right|\right|}{4\omega_1(\mathbf{k}^*_\textbf{RG})\,\omega_2(\mathbf{k}^*_\textbf{RG})} 
   \frac{4\pi Y^*_{lm}(\hat{\mathbf{k}}^*_\textbf{RG})Y_{l'm'}(\hat{\mathbf{k}}^*_\textbf{RG})}
   {P_0^*-\left(\omega_1(\mathbf{k}^*_\textbf{RG})+\omega_2(\mathbf{k}^*_\textbf{RG})\right)}
   \left(\frac{|\mathbf{k}^*_\textbf{RG}|}{q}\right)^{l+l'}\,.
\end{align}
is a regular term, i.e., free of pole at $|\mathbf{k}^*_X|=|\mathbf{k}^*_\textbf{RG}|=q$.
Here the notation $\left|\left|\frac{\partial \mathbf{k}^*_X}{\partial \mathbf{k}}\right|\right|$ denote the absolute value of the Jacobian determinant.
The subtraction is then
 \begin{align}
& \lim_{|\mathbf{k}^*_X|\to q}(H(\mathbf{k}^*_X)-H(\mathbf{k}^*_\textbf{RG}))
 =\frac{4\pi}{4\omega_1(q)\omega_2(q)}\Delta,\nonumber\\
& \Delta \equiv
 \lim_{|\mathbf{k}^*_X|\to q}
 \left(
 \left|\left|\frac{\partial \mathbf{k}^*_X}{\partial \mathbf{k}}\right|\right|
 \frac{ Y^*_{lm}(\hat{\mathbf{k}}^*_X)Y_{l'm'}(\hat{\mathbf{k}}^*_X)}
 {P_0^*-\left(\omega_1(\mathbf{k}^*_X)+\omega_2(\mathbf{k}^*_X)\right)}
 -
  \left|\left|\frac{\partial \mathbf{k}^*_\textbf{RG}}{\partial \mathbf{k}}\right|\right|
 \frac{ Y^*_{lm}(\hat{\mathbf{k}}^*_\textbf{RG})Y_{l'm'}(\hat{\mathbf{k}}^*_\textbf{RG})}
 {P_0^*-\left(\omega_1(\mathbf{k}^*_\textbf{RG})+\omega_2(\mathbf{k}^*_\textbf{RG})\right)}
 \right).
\end{align}

Using \cref{eq:modelXk,eq:modelBk}, one can find
\begin{align}\label{eq:com1}
\left|\left|\frac{\partial \mathbf{k}^*_\textbf{RG}}{\partial \mathbf{k}}\right|\right|_{|\mathbf{k}^*_\textbf{RG}|=q}
=
\det\left|\begin{bmatrix}
			\frac{E^*(q)}{E(q)} & 0 & 0 \\
			0 & 1 & 0 \\
			0 & 0 & 1 \\
		\end{bmatrix}  \right|_{|\mathbf{k}^*_\textbf{RG}|=q}=\frac{E^*(q)}{E(q)}\,,
\end{align}
and
\begin{align}\label{eq:com2}
\left|\left|\frac{\partial \mathbf{k}^*_X}{\partial \mathbf{k}}\right|\right|_{|\mathbf{k}^*_X|=q}
&=
\left|\begin{bmatrix}
             \frac{\partial \mathbf{k}_{X,\parallel}}{\partial \mathbf{k}^*_{X\,\parallel} }
           & \frac{\partial \mathbf{k}_{X,\parallel}}{\partial \mathbf{k}^*_{X\,\perp_1} }
           & \frac{\partial \mathbf{k}_{X,\parallel}}{\partial \mathbf{k}^*_{X\,\perp_2} } \\
            0 & 1 & 0 \\
            0 & 0 & 1 \\
        \end{bmatrix}  \right|^{-1}_{|\mathbf{k}^*_X|=q}
        =\left(\frac{\partial \mathbf{k}_{X,\parallel}}{\partial \mathbf{k}^*_{X\,\parallel} }\right)^{-1},\nonumber\\
        &=\left(\mathcal{A}_X(|\mathbf{k}^*_X|)
        +\frac{\mathbf{k}^*_{X,\parallel}}{|\mathbf{k}^*_X|}
      \left(\mathbf{k}^*_{X,\parallel}\frac{\partial \mathcal{A}_X}{\partial |\mathbf{k}^*_X|}
      +\frac{\partial \mathcal{B}_X}{\partial |\mathbf{k}^*_X|}\mathbf{P}\right)\right)^{-1}\,.
\end{align}

The next step is to notice that the moving-frame momentum is a common integral variable for different approaches, it bridges $\mathbf{k}^*_X$ with $\mathbf{k}^*_\textbf{RG}$ as follows:
\begin{align}
        \mathbf{k}^*_\textbf{RG} &= \frac{E^*(q)}{E(q)}\left(\mathcal{A}_X(|\mathbf{k}^*_X|)\mathbf{k}^*_{X,\parallel} + \mathcal{B}_X(|\mathbf{k}^*_X|)\mathbf{P}\right)  - \frac{\omega_1(q)}{E(q)}\mathbf{P}+ \mathbf{k}_{X,\perp} \nonumber\\
        &=\left( \frac{E^*(q)}{E(q)} \mathcal{A}_X(|\mathbf{k}^*_X|)  \right) \mathbf{k}^*_{X,\parallel} +  \left( \frac{E^*(q)}{E(q)}\mathcal{B}_X(|\mathbf{k}^*_X|) - \frac{\omega_1(q)}{E(q)} \right) \mathbf{P} + \mathbf{k}^*_{X,\perp}\,.\label{eq:kks}
\end{align}
This relation will give us
\begin{align}
	&\quad\lim_{|\mathbf{k}^*_X|\to q} \frac{P_0^*-\left(\omega_1(\mathbf{k}^*_\textbf{RG})+\omega_2(\mathbf{k}^*_\textbf{RG})\right)}{P_0^*-\left(\omega_1(\mathbf{k}^*_X)+\omega_2(\mathbf{k}^*_X)\right)} \nonumber\\
	&= \lim_{|\mathbf{k}^*_X|\to q}\frac{-\left(\frac{q}{\omega_1(q)}+\frac{q}{\omega_2(q)}\right)(|\mathbf{k}^*_\textbf{RG}|-q)}{-\left(\frac{q}{\omega_1(q)}+\frac{q}{\omega_2(q)}\right)(|\mathbf{k}^*_X|-q)}=\lim_{|\mathbf{k}^*_X|\to q}\frac{|\mathbf{k}^*_\textbf{RG}|-q}{|\mathbf{k}^*_X|-q} \nonumber\\
	&= \left.\frac{\partial |\mathbf{k}^*_\textbf{RG}|(|\mathbf{k}^*_X|,\theta_{k^*_X},\phi_{k^*_X})}{\partial |\mathbf{k}^*_X|}\right|_{|\mathbf{k}^*_X|=q}\nonumber\\
       &=\left( \frac{E^*(q)}{E(q)}\right)
       \left.\left(\mathcal{A}_X(|\mathbf{k}^*_X|)
        +\frac{\mathbf{k}^*_{X\,\parallel}}{|\mathbf{k}^*_X|}
      \left(\mathbf{k}^*_{X\,\parallel}\frac{\partial \mathcal{A}_X}{\partial |\mathbf{k}^*_X|}
      +\frac{\partial \mathcal{B}_X}{\partial |\mathbf{k}^*_X|}\mathbf{P}\right)\right)\right|_{|\mathbf{k}^*_X|=q} \,,\nonumber\\
      &=
\left|\left|\frac{\partial \mathbf{k}^*_\textbf{RG}}{\partial \mathbf{k}}\right|\right|_{|\mathbf{k}^*_\textbf{RG}|=q} /
\left|\left|\frac{\partial \mathbf{k}^*_X}{\partial \mathbf{k}}\right|\right|_{|\mathbf{k}^*_X|=q}\,,
      \label{eq:bb}
\end{align}
Combining \cref{eq:com1,eq:com2,eq:bb}, we obtain,
\begin{align}
\Delta =& -\left(\frac{q}{\omega_1(q)}+\frac{q}{\omega_2(q)}\right) 
\left|\left|\frac{\partial \mathbf{k}^*_X}{\partial \mathbf{k}}\right|\right|_{|\mathbf{k}^*_X|= q}
\nonumber\\&\times
 \lim_{|\mathbf{k}^*_X|\to q}
 \frac{ Y^*_{lm}(\hat{\mathbf{k}}^*_X)Y_{l'm'}(\hat{\mathbf{k}}^*_X)-Y^*_{lm}(\hat{\mathbf{k}}^*_\textbf{RG})Y_{l'm'}(\hat{\mathbf{k}}^*_\textbf{RG})}
 {|\mathbf{k}^*_X|-q} \,.
\end{align}

In \cref{eq:kks}, since $\mathbf{k}^*_\textbf{RG}$ is expressed as a function of $\mathbf{k}^*_\textbf{X}$, we can consider the spherical harmonic $Y_{l'm'}(\hat{\mathbf{k}}^*_\textbf{RG})$ as a function of the components of $\mathbf{k}^*_\textbf{X}$ in spherical coordinate system: $(|\mathbf{k}^*_\textbf{X}|, \theta_\textbf{X}, \phi_\textbf{X})$.
We also notice $\mathbf{k}^*_X= \mathbf{k}^*_\textbf{RG}$ when $|\mathbf{k}^*_X|=q$, then we have
\begin{align}
    \Delta =& -\left(\frac{q}{\omega_1(q)}+\frac{q}{\omega_2(q)}\right) 
    \left|\left|\frac{\partial \mathbf{k}^*_X}{\partial \mathbf{k}}\right|\right|_{|\mathbf{k}^*_X|= q} \frac{\partial Y^*_{lm}(\hat{\mathbf{k}}^*_\textbf{RG})Y_{l'm'}(\hat{\mathbf{k}}^*_\textbf{RG})}{\partial |\mathbf{k}^*_\textbf{X}|}\Big{|}_{|\mathbf{k}^*_\textbf{X}|=q} \,.
\end{align}
So $\Delta$ is free of poles, and this proves the equivalence.
%


\section{Derivation of $M$ Matrix}\label{app:deriM} 
	In this section, we give the detailed derivation of the $M$ matrix \cref{eq:Mlmlm2}.
	At first one introduces the following equations,
	\begin{align}\label{eq:yyc}
		Y^*_{l_1m_1}Y_{l_2m_2}&=\sum_{l=|l_1-l_2|}^{|l_1+l_2|}\sum_{m=-l}^{+l}\frac{1}{\sqrt{4\pi}}c_{lm,\,l_1m_1,\,l_2m_2}Y_{lm} \,,\nn\\
		c_{lm,\,l_1m_1,\,l_2m_2}&=(-1)^{m+m_1}\sqrt{(2l_1+1)(2l_2+1)(2l+1)}
			\left( \begin{array}{ccc}
			l_1 & l_2 & l   \\
			0 & 0 &0   \\
			\end{array} \right)
			\left( \begin{array}{ccc}
			l_1 & l_2 & l   \\
			-m_1 & m_2 & -m   \\
			\end{array} \right) \,,
	\end{align}
	where the Wigner 3-j symbols are used.
	Then one can rewrite \cref{eq:Mlmlm2} as follows:
	\begin{align}\label{eq:mclm}
	M_{lm,\,l'm'}(q,\mathbf{P})&= \sum_{\tilde{l}=|l-l'|}^{|l+l'|}\sum_{\tilde{m}=-l}^{+l}
	c_{\tilde{l}\tilde{m},\,lm,\,l'm'}
	\bar{M}_{\tilde{l}\tilde{m}}(q,\mathbf{P})
	\nn\\
	\bar{M}_{lm}(q,\mathbf{P})
	&=
	\frac{4\pi}{q}
	\left(\frac{1}{L^3}\sum_{\mathbf{k}=\frac{2\pi}{L}\mathbf{n},\,\mathbf{n} \in \mathbb{Z}^3}
	-\mathcal{P} \int \frac{d^3k}{(2\pi)^3}\right)
	\mathcal{J}
	\frac{\sqrt{4\pi} Y_{lm}(\hat{\mathbf{k}}^*)(|\mathbf{k}^*|/q)^{l}} 
	{q^2-\mathbf{k}^{*\,2}} \,.
	\end{align}
	When $l>0$, the integral vanishes because of the property of $Y_{lm}$, so one has
	\begin{align}
		\bar{M}_{lm}(q,\mathbf{P})=\frac{4\pi}{qL^3}\sum_{\mathbf{k}=\frac{2\pi}{L}\mathbf{n},\,\mathbf{n} \in \mathbb{Z}^3} \mathcal{J} \frac{\sqrt{4\pi} Y_{lm}(\hat{\mathbf{k}}^*)(|\mathbf{k}^*|/q)^{l}} {q^2-\mathbf{k}^{*\,2}} \qquad l>0\,.
	\end{align}
	For $\bar{M}_{00}(q,\mathbf{P})$, one introduces the factor $e^{\alpha\left(q^2-\mathbf{k}^{*\,2}\right)}$ to divide the integrand as a singular part and a non-singular part, and using the Poisson summation formula, the non-singular part can be transformed as follows:
	\begin{align}
		&\bar{M}_{00}(q,\mathbf{P})
		\nonumber\\
		&=
		\frac{4\pi}{q}
		\left(\frac{1}{L^3}\sum_{\mathbf{k}=\frac{2\pi}{L}\mathbf{n},\,\mathbf{n} \in \mathbb{Z}^3}
		-\mathcal{P} \int \frac{d^3k}{(2\pi)^3}\right)
		\frac{\mathcal{J}e^{\alpha\left(q^2-\mathbf{k}^{*\,2}\right)}} 
		{q^2-\mathbf{k}^{*\,2}}
		+\frac{4\pi}{q}\sum_{\mathbf{n}\in\mathbb{Z}^3,\mathbf{n}\neq 0}
		\int \frac{\mathcal{J}d^3k}{(2\pi)^3} e^{iL\nb\cdot\kb}
		\frac{1-e^{\alpha\left(q^2-\mathbf{k}^{*\,2}\right)}}{q^2-\mathbf{k}^{*\,2}}
		\nonumber\\
		&\equiv\bar{M}_a(q,\,\mathbf{P},\,\alpha)+\bar{M}_b(q,\,\mathbf{P},\,\alpha)+\bar{M}_c(q,\,\mathbf{P},\,\alpha),
	\end{align}
	with
	\begin{align}
		\bar{M}_a(q,\,\mathbf{P},\,\alpha)
		&\equiv
		\frac{4\pi}{q}
		\frac{1}{L^3}\sum_{\mathbf{k}=\frac{2\pi}{L}\mathbf{n},\,\mathbf{n} \in \mathbb{Z}^3}\frac{\mathcal{J}e^{\alpha\left(q^2-\mathbf{k}^{*\,2}\right)}} 
		{q^2-\mathbf{k}^{*\,2}}\,,
		\nn\\
		\bar{M}_b(q,\,\alpha)
		&\equiv
		-\frac{4\pi}{q}
		\mathcal{P} \dkv{k^*}
		\frac{e^{\alpha\left(q^2-\mathbf{k}^{*\,2}\right)}} 
		{q^2-\mathbf{k}^{*\,2}}\,,
		\nn\\
		\bar{M}_c(q,\,\mathbf{P},\,\alpha)
		&\equiv
		\frac{4\pi}{q}
		\sum_{\mathbf{n}\in\mathbb{Z}^3, \mathbf{n}\neq 0}
		\dkv{k^*} e^{iL\nb\cdot\kb}
		\frac{1-e^{\alpha\left(q^2-\mathbf{k}^{*\,2}\right)}}{q^2-\mathbf{k}^{*\,2}}\,.
	\end{align}

	Here, although each $\bar{M}_{(a,\,b,\,c)}$ is dependent on the new parameter $\alpha$, their summation is independent of $\alpha$.
	Also, except $\bar{M}_b$, the other two $\bar{M}_i$ both depend on the three-momentum transformation.
	While $\bar{M}_a$ can be calculated numerically, $\bar{M}_b$ can be calculated as follows:
	\begin{align}\label{eq:barMb3}
	\bar{M}_b(q,\,\alpha)
	&\equiv
	-\frac{4\pi}{q}\mathcal{P} \dkv{k^*}
	\frac{e^{\alpha\left(q^2-\mathbf{k}^{*\,2}\right)}} 
	{q^2-\mathbf{k}^{*\,2}} 
	\nn\\
	&=-\frac{4\pi}{q}\frac{1}{2\pi^2}\int d |\mathbf{k}^*|
	\frac{\mathbf{k}^{*\,2}e^{\alpha(q^2-\mathbf{k}^{*\,2})}-q^2}{q^2-\mathbf{k}^{*\,2}}
	\nonumber\\
	&=-\frac{4}{q L}\int d \tilde{k}^*
	\frac{\tilde{k}^{*\,2} e^{\alpha\left(\frac{2\pi}{L}\right)^2\left[\left(\frac{qL}{2\pi}\right)^2-\tilde{k}^{*\,2}\right]}
	-\left(\frac{qL}{2\pi}\right)^2}
	{\left(\frac{qL}{2\pi}\right)^2-\tilde{k}^{*\,2}}\nonumber\\
	&=\frac{1}{\pi qL}\left[\sqrt{\frac{\pi L^2}{\alpha}}-
	\int_0^{\alpha\left(\frac{2\pi}{L}\right)^2}dt \left(e^{t\left(\frac{qL}{2\pi}\right)^2}-1\right)
	\left(\frac{\pi}{t}\right)^{\frac{3}{2}}\right]\,,
	\end{align}
	in the third line of which $|\mathbf{k}^*|\to\tilde{k}^*=|\mathbf{k}^*|L/2\pi$ is used, and the last step is done via the consideration of 
	\begin{align}
		\tilde{k}^2 e^{\alpha\left(\frac{2\pi}{L}\right)^2\left[\left(\frac{qL}{2\pi}\right)^2-\tilde{k}^{*\,2}\right]}
	-\left(\frac{qL}{2\pi}\right)^2 &=
		\tilde{k}^2 \left[e^{\alpha\left(\frac{2\pi}{L}\right)^2\left[\left(\frac{qL}{2\pi}\right)^2-\tilde{k}^{*\,2}\right]}-1\right]\nn\\
		&+\left[\left(\frac{qL}{2\pi}\right)^2-\tilde{k}^{*\,2}\right]\left[e^{-\alpha\left(\frac{2\pi}{L}\right)^2\tilde{k}^{*\,2}}-1\right]\nn\\
		&-\left[\left(\frac{qL}{2\pi}\right)^2-\tilde{k}^{*\,2}\right]e^{-\alpha\left(\frac{2\pi}{L}\right)^2\tilde{k}^{*\,2}}
	\end{align}
	and
	\begin{align}
		\int_0^A dt\,e^{t\,B} = \frac{e^{A\,B}-1}{B} \,.
	\end{align}

	At last one calculates $\bar{M}_c$ with $\mathbf{k} = \mathcal{A}\mathbf{k}^*_{\parallel} + \mathcal{B}\mathbf{P} + \mathbf{k}^*_{\perp}$ as follows:
	\begin{align}\label{eq:barMc3}
	\bar{M}_c(q,\,\mathbf{P},\,\alpha)
	&\equiv
	\frac{4\pi}{q}
	\sum_{\mathbf{n}\in\mathbb{Z}^3, \mathbf{n}\neq 0}
	\dkv{k^*} e^{iL\nb\cdot\kb}
	\frac{1-e^{\alpha\left(q^2-\mathbf{k}^{*\,2}\right)}}{q^2-\mathbf{k}^{*\,2}}
	\nonumber\\
	&=
	-\frac{4\pi}{q}\sum_{\mathbf{n} \in \mathbb{Z}^3, \mathbf{n}\neq 0}
	\int \frac{d^3k^*}{(2\pi)^3}
	e^{iL\nb\cdot(\mathcal{A}\,\mathbf{k}^*_{\parallel}+\mathcal{B}\,\mathbf{P}+\mathbf{k}^*_{\perp})}
	\int^{\alpha}_{0} dt e^{t(q^2-\mathbf{k}^{*\,2})}\nonumber\\
	&=-\frac{1}{2\pi^2 q}\sum_{\mathbf{n} \in \mathbb{Z}^3, \mathbf{n}\neq 0}
	\int^{\alpha}_{0} dt
	\int d^3k^*\,
	e^{iL\mathcal{B}\,\mathbf{n}\cdot\Pb}
	e^{iL(\mathcal{A}\,\mathbf{n}_{\parallel}+\mathbf{n}_{\perp})\cdot\kb^*}
	e^{t(q^2-\mathbf{k}^{*\,2})}\nonumber\\
	&=-\frac{1}{\pi q L}\sum_{\mathbf{n} \in \mathbb{Z}^3, \mathbf{n}\neq 0}
	\int^{\alpha}_{0} 
	dt\,e^{tq^2}
	\int dk^*\,  e^{-tk^{*\,2}}
	\cos [ L\mathcal{B}\,\mathbf{n}\cdot\Pb ]
	\frac{2k^*\sin [L\,D\,k^*]}{D}\,,
	\end{align}
	where we have used the fact that $\bar{M}_c$ is real, and
	\begin{align}
	D&=\sqrt{\mathbf{n}^2 + (\mathcal{A}^2-1)\frac{(\mathbf{n}\cdot\Pb)^2}{\mathbf{P}^2}}\,.
	\end{align}

	To calculate $\bar{M}_c$ numerically, one can first use the Gauss quadrature formula to transform the $t$-integral into a weighted sum, then for each $t$ and $\mathbf{n}$, one splits the range of $k^*$-integration into segments according to the zeros of the trigonometric functions, calculates each segment by using Gauss quadrature formula and re-sums them to get the $k^*$-integral.

In summary, $M(q;\mathbf{P})$ can be further simplified as follows, 
\begin{align}
[M(q;\mathbf{P})]_{lm,l'm'}
&=\sum_{\tilde{l}=|l-l'|}^{|l+l'|}\sum_{\tilde{m}=-l}^{+l}
c_{\tilde{l}\tilde{m},\,lm,\,l'm'}
\frac{4\pi}{q}\bar{M}_{\tilde{l}\tilde{m}}(q,\mathbf{P})\,,\label{eq:mbar}
\\
\bar{M}_{lm}(q,\mathbf{P})&=
\frac{4\pi}{qL^3}\sum_{\mathbf{k}=\frac{2\pi}{L}\mathbf{n},\,\mathbf{n} \in \mathbb{Z}^3} \mathcal{J}^r \frac{\sqrt{4\pi} Y_{lm}(\hat{\mathbf{k}}^*)(|\mathbf{k}^*|/q)^{l}} {q^2-\mathbf{k}^{*\,2}} \qquad l>0 \,,\label{eq:mbarl}\\
\bar{M}_{00}(q,\mathbf{P})
&=\bar{M}_a(q,\,\mathbf{P},\,\alpha)+\bar{M}_b(q,\,\alpha)+\bar{M}_c(q,\,\mathbf{P},\,\alpha)\,,\label{eq:mbar0}\\
\bar{M}_a(q,\,\mathbf{P},\,\alpha)&=
\frac{4\pi}{qL^3}\sum_{\mathbf{k}=\frac{2\pi}{L}\mathbf{n},\,\mathbf{n} \in \mathbb{Z}^3}\frac{\mathcal{J}^r e^{\alpha\left(q^2-\mathbf{k}^{*\,2}\right)}} 
		{q^2-\mathbf{k}^{*\,2}}\,,
\label{eq:mbara}\\
\bar{M}_b(q,\,\alpha)&=
\frac{1}{\pi qL}\left[\sqrt{\frac{\pi L^2}{\alpha}}-
\int_0^{\alpha\left(\frac{2\pi}{L}\right)^2}dt \left(e^{t\left(\frac{qL}{2\pi}\right)^2}-1\right)
\left(\frac{\pi}{t}\right)^{\frac{3}{2}}\right]\,,
\label{eq:mbarb}\\
\bar{M}_c(q,\,\mathbf{P},\,\alpha)&=
-\frac{1}{\pi q L}\sum_{\mathbf{n} \in \mathbb{Z}^3, \mathbf{n}\neq 0}
\int^{\alpha}_{0} dt\, e^{tq^2}
		\int_0^\infty dk^*\,  e^{-tk^{*\,2}}
		\cos [ L\mathcal{B}\,\mathbf{n}\cdot\Pb ]
		\frac{2k^*\sin [L\,D\,k^*]}{D}\,,
\label{eq:mbarc}\\
		D&=\sqrt{\mathbf{n}^2 + (\mathcal{A}^2-1)\frac{(\mathbf{n}\cdot\Pb)^2}{\mathbf{P}^2}} = \sqrt{\mathbf{n}^2 + \left(\frac{\mathbf{n}\cdot\Pb}{a^*}\right)^2} \,.
	\end{align}

For example, the $\bar{M}_{00}(q,\mathbf{P})$ for \textbf{LWLY} can be written as,
\begin{align}\label{eq:MMC}
		\bar{M}^{\textbf{LYLW}}_{00}(q,\,\mathbf{P}) &=
		\frac{4\pi}{q}
		\frac{1}{L^3}
		\sum_{\mathbf{k}=\frac{2\pi}{L}\mathbf{n},\,\mathbf{n} \in \mathbb{Z}^3}
		\frac{\omega_1(\mathbf{k}^*)\omega_2(\mathbf{k}^*)}{\omega_1(\mathbf{k}^*)+\omega_2(\mathbf{k}^*)}\frac{\omega_1(\mathbf{k})+\omega_2(\mathbf{P}-\mathbf{k})}{\omega_1(\mathbf{k})\omega_2(\mathbf{P}-\mathbf{k})}
		\frac{e^{\alpha(q^2-\mathbf{k}^{*\,2})}}{q^2-\mathbf{k}^{*\,2}}\nn\\
		&-\frac{4\pi}{q} 
   	\mathcal{P} \int \frac{d^3 k^*}{(2\pi)^3}
		\frac{e^{\alpha(q^2-\mathbf{k}^{*\,2})}}{q^2-\mathbf{k}^{*\,2}}
		-\frac{1}{\pi qL}\sum_{\mathbf{n} \in \mathbb{Z}^3, \mathbf{n}\neq 0}
		\int^{\alpha}_{0} dt\, e^{tq^2}\int dk^* \times\nn\\
		&\quad e^{-tk^{*\,2}} \cos \left[ L\,\mathbf{n}\cdot\Pb\frac{\omega_1(k^*)}{\omega_1(k^*)+\omega_2(k^*)}  \right]
		\frac{2k^*\sin \left[L\,D_\textbf{LYLW}\,k^* \right]}{D_\textbf{LYLW}} \,,\nn\\
		D_\textbf{LYLW}&=\sqrt{\mathbf{n}^2  +\left(\frac{\mathbf{n}\cdot \mathbf{P}}{\sqrt{k^{*\,2}+m_1^2}+\sqrt{k^{*\,2}+m_2^2}}\right)^2} \,
\end{align}
where in the summation, $\mathbf{k}^*$ is calculated from Eq.~(\ref{eq:kskSC}).
 
\section{Quantization Conditions for Pure S- and Pure P-Waves}\label{app:phes}

\subsection{Pure S-Wave}
	For the pure S-wave, from \cref{eq:qc-mx-2,eq:mbar0}, one has
	\begin{align}\label{eq:phES}
		\cot\delta_0(q)=-\bar{M}_a(q,\mathbf{P},\alpha)-\bar{M}_b(q,\alpha)-\bar{M}_c(q,\mathbf{P},\alpha) \,.
	\end{align}

\subsection{Pure P-Wave}
	For the pure P-wave, \cref{eq:qc-mx-2} is a $3\times3$ matrix equation as follows:
	\begin{align}\label{eq:pdet}
	0 &= 
	\det
	\left( \begin{array}{ccc}
	M_{1-1,1-1}+\cot\delta_1 &M_{1-1,10}   & M_{1-1,1+1}   \\
	M_{10,1-1}      & M_{10,10}+\cot\delta_1 & M_{10,1+1}   \\
	M_{1+1,1-1}   & M_{1+1,10} & M_{1+1,1+1}+\cot\delta_1   \\
	\end{array} \right) \,.
	\end{align}
	Applying \cref{eq:yyc} and \cref{eq:mclm}, it then becomes
	\begin{align}\label{eq:pdetidep}
	0=
		\det
		\left( \begin{array}{ccc}
		\bar{M}_{00}-\sqrt{\frac{1}{5}}\bar{M}_{20}+\cot\delta_1
		& -\sqrt{\frac{3}{5}}\bar{M}_{21}
		&-\sqrt{\frac{6}{5}}\bar{M}_{22}   \\
		 \sqrt{\frac{3}{5}}\bar{M}_{2-1}
		&\bar{M}_{00}+\sqrt{\frac{4}{5}}\bar{M}_{20}+\cot\delta_1
		&\sqrt{\frac{3}{5}}\bar{M}_{21}   \\
		 -\sqrt{\frac{6}{5}}\bar{M}_{2-2}
		& -\sqrt{\frac{3}{5}}\bar{M}_{2-1}
		&\bar{M}_{00}-\sqrt{\frac{1}{5}}\bar{M}_{20}+\cot\delta_1
		\\
		\end{array} \right) \,.
	\end{align}

	Symmetry groups in the finite volume are different for different $\mathbf{P}$.
	In the rest case, i.e. $\mathbf{P}=(0,0,0)$, the group is the cubic group $O_h$.
	In the moving case, however, only elements in $O_h$ that keep $\mathbf{P}$ fixed will survive, and the resulting group is just the little group of $O_h$.
	Next, we will discuss three different $\mathbf{P}$ in detail.

	\begin{itemize}
	 	 \item $\mathbf{P} = \frac{2\pi}{L}(0,0,1)$: we have $\bar{M}_{2\pm1} = \bar{M}_{2\pm2} = 0$, then \cref{eq:pdetidep} can be simplified to be
			  \begin{align}
			0 &= \det
			\left( \begin{array}{ccc}
			\bar{M}_{00}-\sqrt{\frac{1}{5}}\bar{M}_{20}+\cot\delta_1
			& 0
			& 0   \\
			  0
			&\bar{M}_{00}+\sqrt{\frac{4}{5}}\bar{M}_{20}+\cot\delta_1
			& 0   \\
			  0
			& 0
			&\bar{M}_{00}-\sqrt{\frac{1}{5}}\bar{M}_{20}+\cot\delta_1
			\\
			\end{array} \right),\nonumber\\
			\end{align}
			so we have
			  \begin{align}
			\cot\delta_1&=-\bar{M}_{00}-\sqrt{\frac{4}{5}}\bar{M}_{20}\,,\label{eq:p001A}\\
			\cot\delta_1&=-\bar{M}_{00}+\sqrt{\frac{1}{5}}\bar{M}_{20}\,.\label{eq:p001E}
			\end{align}
			The energies solved from \cref{eq:p001A} and \cref{eq:p001E} will correspond to the irreducible representations $A_1$ and $E$ respectively of the little group for $(0,0,1)$.

	 	\item $\mathbf{P} = \frac{2\pi}{L}(1,1,0)$: we have $\bar{M}_{2\pm1} = 0$ and $\bar{M}_{2-2} = -\bar{M}_{22}$, then \cref{eq:pdetidep} can be simplified to be
			  \begin{align}
			0 &= \det
			\left( \begin{array}{ccc}
			\bar{M}_{00}-\sqrt{\frac{1}{5}}\bar{M}_{20}+\cot\delta_1
			& 0
			& -\sqrt{\frac{6}{5}}\bar{M}_{22}   \\
			  0
			&\bar{M}_{00}+\sqrt{\frac{4}{5}}\bar{M}_{20}+\cot\delta_1
			& 0   \\
			  \sqrt{\frac{6}{5}}\bar{M}_{22}
			& 0
			&\bar{M}_{00}-\sqrt{\frac{1}{5}}\bar{M}_{20}+\cot\delta_1
			\\
			\end{array} \right),\nonumber\\
			\end{align}
			then we have
			  \begin{align}
			\cot\delta_1&=-\bar{M}_{00}+\sqrt{\frac{1}{5}}\bar{M}_{20}
			+i\sqrt{\frac{6}{5}}\bar{M}_{22}\,,\label{eq:p110A}\\
			\cot\delta_1&=-\bar{M}_{00}-\sqrt{\frac{4}{5}}\bar{M}_{20}\,,\label{eq:p110B1}\\
			\cot\delta_1&=-\bar{M}_{00}+\sqrt{\frac{1}{5}}\bar{M}_{20}
			-i\sqrt{\frac{6}{5}}\bar{M}_{22}\,.\label{eq:p110B2}
			\end{align}
			  The energies solved from \cref{eq:p110A}, \cref{eq:p110B1} and \cref{eq:p110B2} will correspond to the irreducible representations $A_1$, $B_1$ and $B_2$ in the little group for $(1,1,0)$.

	    \item $\mathbf{P} = \frac{2\pi}{L}(1,1,1)$:  we have $\bar{M}_{2\pm1} = (-1\pm i)\bar{M}_{22}$,  $\bar{M}_{2-2} = -\bar{M}_{22}$, and $\bar{M}_{20} = 0$, then \cref{eq:pdetidep} can be simplified to be
			  \begin{align}
			0 &= \det
			\left( \begin{array}{ccc}
			\bar{M}_{00}+\cot\delta_1
			& -\sqrt{\frac{6}{5}}e^{i\frac{3\pi}{4}}\bar{M}_{22}
			& -\sqrt{\frac{6}{5}}\bar{M}_{22}   \\
			  \sqrt{\frac{6}{5}}e^{i\frac{5\pi}{4}}\bar{M}_{22}
			&\bar{M}_{00}+\cot\delta_1
			& \sqrt{\frac{6}{5}}e^{i\frac{3\pi}{4}}\bar{M}_{22}
			   \\
			  \sqrt{\frac{6}{5}}\bar{M}_{22}
			& -\sqrt{\frac{6}{5}}e^{i\frac{5\pi}{4}}\bar{M}_{22}
			&\bar{M}_{00}+\cot\delta_1
			\\
			\end{array} \right),
			\end{align}
			and
			  \begin{align}
			\cot\delta_1&=-\bar{M}_{00}-i\sqrt{\frac{6}{5}}\bar{M}_{22}\,,\label{eq:p111A}\\
			\cot\delta_1&=-\bar{M}_{00}+i2\sqrt{\frac{6}{5}}\bar{M}_{22}\,.\label{eq:p111E}
			\end{align}
			  The energies solved from \cref{eq:p111A} and \cref{eq:p111E} will correspond to the irreducible representations $A_1$ and $E$ in the little group for $(1,1,1)$.
			\end{itemize}

\section{Proof of Exact Equivalence between Approaches KSS and RG}\label{app:ABmodel}
	In the numerical analysis in \cref{sec:NCC}, we found $\bar{M}^{\textbf{KSS}}_{00}=\bar{M}^{\textbf{RG}}_{00}$.
	Here we provide a proof for it.
	Then the proof will be generalized to any partial-wave cases. 
	It means that Approaches \textbf{KSS} and \textbf{RG} are exactly equivalent.
	
	From the definition \cref{eq:mclm}, it is equivalent to prove that
	\begin{align}\label{eq:mamb}
		\left( \skv{\mathbf{k}} - \dkv{k} \right) \left[ \frac{\mathcal{J}^\textbf{KSS}}{q^2-\mathbf{k}^{*\,2}_\textbf{KSS}} - \frac{\mathcal{J}^\textbf{RG}}{q^2-\mathbf{k}^{*\,2}_\textbf{RG}} \right] 
	\end{align}
	vanishes exactly, where we have removed the principal value integration as the integrand is free of poles.
	Using the Poisson summation formula, it becomes
	\begin{align}
		\sum_{\mathbf{n}\neq0,\,\mathbf{n} \in \mathbb{Z}^3}\dkv{k} e^{iL\nb\cdot\kb} \left[ \frac{\mathcal{J}^\textbf{KSS}}{q^2-\mathbf{k}^{*\,2}_\textbf{KSS}} - \frac{\mathcal{J}^\textbf{RG}}{q^2-\mathbf{k}^{*\,2}_\textbf{RG}} \right] \,,
	\end{align}
	then one transforms the moving three-momentums back to the rest three-momentums to get
	\begin{align}
		&\quad\sum_{\mathbf{n}\neq0,\,\mathbf{n} \in \mathbb{Z}^3}\dkv{k^*} \left[e^{iL\nb\cdot\kb_\textbf{KSS}} - e^{iL\nb\cdot\kb_\textbf{RG}}\right]\frac{1}{q^2-\mathbf{k}^{*\,2}} \nn\\
		&= \sum_{\mathbf{n}\neq0,\,\mathbf{n} \in \mathbb{Z}^3}\dkv{k^*} e^{iL\nb\cdot(\frac{E(q)}{E^*(q)}\mathbf{k}^*_{\parallel}+\mathbf{k}^*_{\perp})} \left[e^{iL\nb\cdot \mathbf{P} \frac{\sqrt{\mathbf{k}^{*\,2}+m_1^2}}{E^*(q)}} - e^{iL\nb\cdot \mathbf{P}\left(\frac{1}{2}+\frac{m^2_1-m^2_2}{2E^{*\,2}(q)}\right)}\right]\frac{1}{q^2-\mathbf{k}^{*\,2}} \nn\\
		&= \sum_{\mathbf{n}\neq0,\,\mathbf{n} \in \mathbb{Z}^3}\dkv{k^*} e^{iL\kb^*\cdot(\frac{E(q)}{E^*(q)}\mathbf{n}_{\parallel}+\mathbf{n}_{\perp})} \left[e^{iL\nb\cdot \mathbf{P} \frac{\sqrt{\mathbf{k}^{*\,2}+m_1^2}}{E^*(q)}} - e^{iL\nb\cdot \mathbf{P}\left(\frac{1}{2}+\frac{m^2_1-m^2_2}{2E^{*\,2}(q)}\right)}\right]\frac{1}{q^2-\mathbf{k}^{*\,2}} \nn\\
		&= \sum_{\mathbf{n}\neq0,\,\mathbf{n} \in \mathbb{Z}^3}\dkv{k^*} e^{iL\kb^*\cdot\nb_\gamma} \left[\cos\left({2\alpha_\textbf{KSS}\pi\,\mathbf{n}\cdot \mathbf{d}}\right) - \cos\left({2\alpha_\textbf{RG}\pi\nb\cdot\db}\right)\right] \frac{1}{q^2-\mathbf{k}^{*\,2}} \,,
	\end{align}
	where in the last step we have used the invariance under $\mathbf{k}^*\to-\mathbf{k}^*$ and $\mathbf{n}\to-\mathbf{n}$, and 
	\begin{align}
		\mathbf{P}=\frac{2\pi}{L}\mathbf{d}\,,\quad 
		\alpha_\textbf{KSS}=\frac{\sqrt{\mathbf{k}^{*\,2}+m_1^2}}{E^*(q)} \,,\quad
		\alpha_\textbf{RG}=\frac{1}{2}+\frac{m^2_1-m^2_2}{2E^{*\,2}(q)} \,,\quad
		\mathbf{n}_\gamma=\frac{E(q)}{E^*(q)}\mathbf{n}^*_{\parallel}+\mathbf{n}^*_{\perp}\,.
	\end{align}
	Then performing the integration over solid angle, one gets
	\begin{align}
		\frac{1}{4\pi^2L}\sum_{\mathbf{n}\neq0,\,\mathbf{n} \in \mathbb{Z}^3} \frac{1}{n_\gamma} \int d k^* \left[\cos\left({2\alpha_\textbf{KSS}\pi\,\mathbf{n}\cdot \mathbf{d}}\right) - \cos\left({2\alpha_\textbf{RG}\pi\nb\cdot\db}\right)\right] \frac{2k^{*}\sin(L\,n_\gamma\,k^*)}{q^2-k^{*\,2}} \,.
	\end{align}

	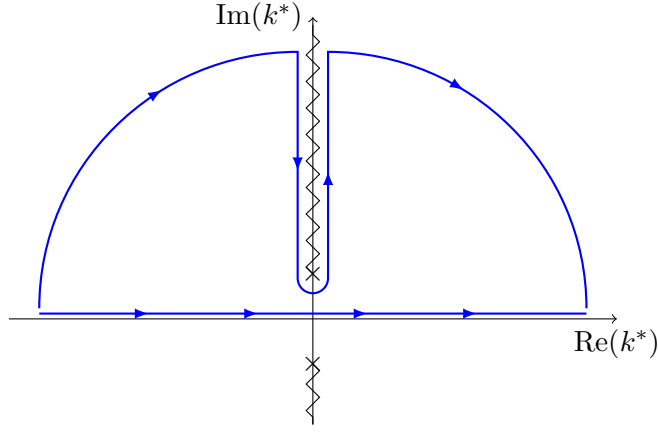
\begin{figure}[h]
    \centerline{	  
    \begin{tikzpicture}[xscale=2,yscale=2]
	    \draw[->] (0,-.7) --(0,2) node [left] {$\mathrm{Im}(k^*)$};
	    \draw[->] (-2,0) --(2,0) node [below] {$\mathrm{Re}(k^*)$};
	    \node at (0,.3) {$\times$};
	    \draw[decorate, decoration = {zigzag}] (0,.3) -- (0,1.95);
	    \node at (0,-.3) {$\times$};
	    \draw[decorate, decoration = {zigzag}] (0,-.3) -- (0,-.7);
	    \draw[thick,blue, yshift=1,
			decoration={ markings,
			      mark=at position 0.2 with {\arrow{latex}}, 
			      mark=at position 0.4 with {\arrow{latex}},
			      mark=at position 0.6 with {\arrow{latex}}, 
			      mark=at position 0.8 with {\arrow{latex}}}, 
			      postaction={decorate}]
			 (-1.8,0) -- (1.8,0);
	    \draw[thick,blue, yshift=2,
			decoration={ markings,
			      mark=at position 0.2 with {\arrow{latex}}, 
			      mark=at position 0.4 with {\arrow{latex}},
			      mark=at position 0.6 with {\arrow{latex}}, 
			      mark=at position 0.8 with {\arrow{latex}}
			      }, 
			      postaction={decorate}]
			 (-1.8,0) arc (180:90:1.7) -- (-.1,.2) arc (-180:0:.1) -- (.1,1.7) arc (90:0:1.7) ;
	  \end{tikzpicture}	
	  }
	\caption{The contour at the complex $k^*$-plane}\label{fig:contour}
	\end{figure}

	Now, let us focus on the integration part:
	\begin{align}\label{eq:extI}
		I&\equiv\int_0^{\infty} d k^* \left[\cos\left({2\alpha_\textbf{KSS}\pi\,\mathbf{n}\cdot \mathbf{d}}\right) - \cos\left({2\alpha_\textbf{RG}\pi\nb\cdot\db}\right)\right] \frac{2k^{*}\sin(L\,n_\gamma\,k^*)}{q^2-k^{*\,2}} \nn\\
		&=\frac{1}{2}\int_{-\infty}^{+\infty} d k^* \left[\cos\left({2\alpha_\textbf{KSS}\pi\,\mathbf{n}\cdot \mathbf{d}}\right) - \cos\left({2\alpha_\textbf{RG}\pi\nb\cdot\db}\right)\right] \frac{2k^{*}\sin(L\,n_\gamma\,k^*)}{q^2-k^{*\,2}} \nn\\
		&=\mathrm{Im} \int_{-\infty}^{+\infty} d k^* \left[\cos\left({2\alpha_\textbf{KSS}\pi\,\mathbf{n}\cdot \mathbf{d}}\right) - \cos\left({2\alpha_\textbf{RG}\pi\nb\cdot\db}\right)\right] \frac{k^{*}e^{i\,L\,n_\gamma\,k^*}}{q^2-k^{*\,2}} \,.
	\end{align}
	As the poles on the real axis of $k^*$ cancel out, the only non-analyticities come from the branch points $k^*\to\pm im$ of $\alpha_\textbf{KSS}$.  
	Noting that $\mathrm{Im} (L\,n_\gamma\,k^*\pm2\alpha_\textbf{KSS}\pi\,\mathbf{n}\cdot \mathbf{d})\to+\infty$ when $|k^*|\to\infty$ in the upper $k^*$-complex plane for any $\mathbf{n}\neq 0$, one can perform the contour integration shown in \cref{fig:contour} to get
	\begin{align}
		I = \mathrm{Im} \left(\int_{-\varepsilon+i\,\infty}^{-\varepsilon+im} dk^* - \int_{+\varepsilon+i\,\infty}^{+\varepsilon+im} dk^* \right)\left[\cos\left({2\alpha_\textbf{KSS}\pi\,\mathbf{n}\cdot \mathbf{d}}\right) - \cos\left({2\alpha_\textbf{RG}\pi\nb\cdot\db}\right)\right] \frac{k^{*}e^{i\,L\,n_\gamma\,k^*}}{q^2-k^{*\,2}} \,.
   \end{align}
	The only obstruction to cancel the two integrals comes from $\cos(2\alpha_\textbf{KSS}\pi\nb\cdot\db)$ because $\alpha_\textbf{KSS}$ takes opposite values when crossing the branch cut. However the evenness of the cosine function makes it safe for the cancellation. So one can conclude that $I=0$.

	To generalize the discussion to any partial waves, it is enough to prove that
	\begin{align}
		\dkv{k^*} e^{iL\kb^*\cdot\nb_\gamma} \left[\cos\left({2\alpha_\textbf{KSS}\pi\,\mathbf{n}\cdot \mathbf{d}}\right) - \cos\left({2\alpha_\textbf{RG}\pi\nb\cdot\db}\right)\right] \frac{Y_{lm}(\hat{\mathbf{k}}^*)(k^*/q)^{l}}{q^2-\mathbf{k}^{*\,2}}
	\end{align}
	vanishes for any $(l,m)$.
	Although the $z$-direction of $Y_{lm}$ is in general not the direction of $\mathbf{n}_\gamma$, the rotational property of $Y_{lm}$ allows it to be linearly combined by $Y_{lm}^{\mathbf{z}=\hat{\mathbf{n}}_\gamma}$, so one only needs to deal with
	\begin{align}
		&\quad\dkv{k^*} e^{iL \mathbf{k}^*\cdot\nb_\gamma} \left[\cos\left({2\alpha_\textbf{KSS}\pi\,\mathbf{n}\cdot \mathbf{d}}\right) - \cos\left({2\alpha_\textbf{RG}\pi\nb\cdot\db}\right)\right] \frac{Y_{lm}^{\mathbf{z}=\hat{\mathbf{n}}_\gamma}(\hat{\mathbf{k}}^*)(k^*/q)^{l}}{q^2-\mathbf{k}^{*\,2}} \nn\\
		&=\int \frac{k^{*\,2}d k^*}{(2\pi)^3} \frac{\cos\left({2\alpha_\textbf{KSS}\pi\,\mathbf{n}\cdot \mathbf{d}}\right) - \cos\left({2\alpha_\textbf{RG}\pi\nb\cdot\db}\right)}{q^2-k^{*\,2}} (k^*/q)^{l}\int d\Omega \, e^{iLk^*n_\gamma \cos\theta}\,Y_{lm}^{\mathbf{z}=\hat{\mathbf{n}}_\gamma}(\theta,\phi) \,,
	\end{align}
	where $\theta$ is the angle between $\mathbf{k}^*$ and $\mathbf{n}_\gamma$. This expression vanishes for $m\neq0$, and $Y_{l0}$ are just polynomials of $\cos\theta$, so one needs to work with expressions like
	\begin{align}
		\int_{-1}^{1} [d\cos\theta] \, e^{iL k^* n_\gamma\cos\theta}\,\cos^j\theta = \left[\left(\frac{1}{i}\frac{\partial }{\partial \lambda}\right)^j \frac{2\sin\lambda}{\lambda}\right]_{\lambda=L k^* n_\gamma} \,.
	\end{align}
	Noting that $Y_{lm}$ only contains $j$-even terms when $l$ even and $j$-odd terms when $l$ odd, and combining with the factor $(k^*/q)^{l}$, the resulting integrands are always even under $k^*\to-k^*$.
	This allows a similar process as the first step in \cref{eq:extI} to extend the range of the integration from $\int_0^{\infty} dk^*$ to $\int_{-\infty}^{+\infty} dk^*$, and for the second step in \cref{eq:extI}, one needs to treat $\sin(L k^* n_\gamma)$ and $\cos(L k^* n_\gamma)$ as $\mathrm{Im}\,[e^{iL k^* n_\gamma}]$ and $\mathrm{Re}\,[e^{iL k^* n_\gamma}]$ respectively. 
	One will find the resulting analytical structure is the same as before, so the same contour integration and discussion follow, and one can conclude that Approaches \textbf{KSS} and \textbf{RG} are equivalent exactly.


\bibliography{refs}

\begin{thebibliography}{73}%
\makeatletter
\providecommand \@ifxundefined [1]{%
 \@ifx{#1\undefined}
}%
\providecommand \@ifnum [1]{%
 \ifnum #1\expandafter \@firstoftwo
 \else \expandafter \@secondoftwo
 \fi
}%
\providecommand \@ifx [1]{%
 \ifx #1\expandafter \@firstoftwo
 \else \expandafter \@secondoftwo
 \fi
}%
\providecommand \natexlab [1]{#1}%
\providecommand \enquote  [1]{``#1''}%
\providecommand \bibnamefont  [1]{#1}%
\providecommand \bibfnamefont [1]{#1}%
\providecommand \citenamefont [1]{#1}%
\providecommand \href@noop [0]{\@secondoftwo}%
\providecommand \href [0]{\begingroup \@sanitize@url \@href}%
\providecommand \@href[1]{\@@startlink{#1}\@@href}%
\providecommand \@@href[1]{\endgroup#1\@@endlink}%
\providecommand \@sanitize@url [0]{\catcode `\\12\catcode `\$12\catcode
  `\&12\catcode `\#12\catcode `\^12\catcode `\_12\catcode `\%12\relax}%
\providecommand \@@startlink[1]{}%
\providecommand \@@endlink[0]{}%
\providecommand \url  [0]{\begingroup\@sanitize@url \@url }%
\providecommand \@url [1]{\endgroup\@href {#1}{\urlprefix }}%
\providecommand \urlprefix  [0]{URL }%
\providecommand \Eprint [0]{\href }%
\providecommand \doibase [0]{http://dx.doi.org/}%
\providecommand \selectlanguage [0]{\@gobble}%
\providecommand \bibinfo  [0]{\@secondoftwo}%
\providecommand \bibfield  [0]{\@secondoftwo}%
\providecommand \translation [1]{[#1]}%
\providecommand \BibitemOpen [0]{}%
\providecommand \bibitemStop [0]{}%
\providecommand \bibitemNoStop [0]{.\EOS\space}%
\providecommand \EOS [0]{\spacefactor3000\relax}%
\providecommand \BibitemShut  [1]{\csname bibitem#1\endcsname}%
\let\auto@bib@innerbib\@empty
\bibitem [{\citenamefont {Shepherd}\ \emph {et~al.}(2016)\citenamefont
  {Shepherd}, \citenamefont {Dudek},\ and\ \citenamefont
  {Mitchell}}]{Shepherd:2016dni}%
  \BibitemOpen
  \bibfield  {author} {\bibinfo {author} {\bibfnamefont {M.~R.}\ \bibnamefont
  {Shepherd}}, \bibinfo {author} {\bibfnamefont {J.~J.}\ \bibnamefont {Dudek}},
  \ and\ \bibinfo {author} {\bibfnamefont {R.~E.}\ \bibnamefont {Mitchell}},\
  }\href {\doibase 10.1038/nature18011} {\bibfield  {journal} {\bibinfo
  {journal} {Nature}\ }\textbf {\bibinfo {volume} {534}},\ \bibinfo {pages}
  {487} (\bibinfo {year} {2016})},\ \Eprint {http://arxiv.org/abs/1802.08131}
  {arXiv:1802.08131 [hep-ph]} \BibitemShut {NoStop}%
\bibitem [{\citenamefont {Guo}\ \emph {et~al.}(2018)\citenamefont {Guo},
  \citenamefont {Hanhart}, \citenamefont {Meißner}, \citenamefont {Wang},
  \citenamefont {Zhao},\ and\ \citenamefont {Zou}}]{Guo:2017jvc}%
  \BibitemOpen
  \bibfield  {author} {\bibinfo {author} {\bibfnamefont {F.-K.}\ \bibnamefont
  {Guo}}, \bibinfo {author} {\bibfnamefont {C.}~\bibnamefont {Hanhart}},
  \bibinfo {author} {\bibfnamefont {U.-G.}\ \bibnamefont {Meißner}}, \bibinfo
  {author} {\bibfnamefont {Q.}~\bibnamefont {Wang}}, \bibinfo {author}
  {\bibfnamefont {Q.}~\bibnamefont {Zhao}}, \ and\ \bibinfo {author}
  {\bibfnamefont {B.-S.}\ \bibnamefont {Zou}},\ }\href {\doibase
  10.1103/RevModPhys.90.015004} {\bibfield  {journal} {\bibinfo  {journal}
  {Rev. Mod. Phys.}\ }\textbf {\bibinfo {volume} {90}},\ \bibinfo {pages}
  {015004} (\bibinfo {year} {2018})},\ \Eprint
  {http://arxiv.org/abs/1705.00141} {arXiv:1705.00141} \BibitemShut {NoStop}%
\bibitem [{\citenamefont {Liu}(2017)}]{Liu:2016kbb}%
  \BibitemOpen
  \bibfield  {author} {\bibinfo {author} {\bibfnamefont {C.}~\bibnamefont
  {Liu}},\ }\href {\doibase 10.22323/1.256.0006} {\bibfield  {journal}
  {\bibinfo  {journal} {PoS}\ }\textbf {\bibinfo {volume} {LATTICE2016}},\
  \bibinfo {pages} {006} (\bibinfo {year} {2017})},\ \Eprint
  {http://arxiv.org/abs/1612.00103} {arXiv:1612.00103 [hep-lat]} \BibitemShut
  {NoStop}%
\bibitem [{\citenamefont {Briceno}\ \emph {et~al.}(2018)\citenamefont
  {Briceno}, \citenamefont {Dudek},\ and\ \citenamefont
  {Young}}]{Briceno:2017max}%
  \BibitemOpen
  \bibfield  {author} {\bibinfo {author} {\bibfnamefont {R.~A.}\ \bibnamefont
  {Briceno}}, \bibinfo {author} {\bibfnamefont {J.~J.}\ \bibnamefont {Dudek}},
  \ and\ \bibinfo {author} {\bibfnamefont {R.~D.}\ \bibnamefont {Young}},\
  }\href {\doibase 10.1103/RevModPhys.90.025001} {\bibfield  {journal}
  {\bibinfo  {journal} {Rev. Mod. Phys.}\ }\textbf {\bibinfo {volume} {90}},\
  \bibinfo {pages} {025001} (\bibinfo {year} {2018})},\ \Eprint
  {http://arxiv.org/abs/1706.06223} {arXiv:1706.06223 [hep-lat]} \BibitemShut
  {NoStop}%
\bibitem [{\citenamefont {Padmanath}(2019)}]{Padmanath:2019wid}%
  \BibitemOpen
  \bibfield  {author} {\bibinfo {author} {\bibfnamefont {M.}~\bibnamefont
  {Padmanath}}\ }(\bibinfo {year} {2019})\ \Eprint
  {http://arxiv.org/abs/1905.09651} {arXiv:1905.09651 [hep-ex, physics:hep-lat,
  physics:hep-ph]} \BibitemShut {NoStop}%
\bibitem [{\citenamefont {Detmold}\ and\ \citenamefont
  {Savage}(2008)}]{Detmold:2008gh}%
  \BibitemOpen
  \bibfield  {author} {\bibinfo {author} {\bibfnamefont {W.}~\bibnamefont
  {Detmold}}\ and\ \bibinfo {author} {\bibfnamefont {M.~J.}\ \bibnamefont
  {Savage}},\ }\href {\doibase 10.1103/PhysRevD.77.057502} {\bibfield
  {journal} {\bibinfo  {journal} {Phys. Rev. D}\ }\textbf {\bibinfo {volume}
  {77}},\ \bibinfo {pages} {057502} (\bibinfo {year} {2008})},\ \Eprint
  {http://arxiv.org/abs/0801.0763} {arXiv:0801.0763 [hep-lat]} \BibitemShut
  {NoStop}%
\bibitem [{\citenamefont {Detmold}\ and\ \citenamefont
  {Smigielski}(2011)}]{Detmold:2011kw}%
  \BibitemOpen
  \bibfield  {author} {\bibinfo {author} {\bibfnamefont {W.}~\bibnamefont
  {Detmold}}\ and\ \bibinfo {author} {\bibfnamefont {B.}~\bibnamefont
  {Smigielski}},\ }\href {\doibase 10.1103/PhysRevD.84.014508} {\bibfield
  {journal} {\bibinfo  {journal} {Phys. Rev. D}\ }\textbf {\bibinfo {volume}
  {84}},\ \bibinfo {pages} {014508} (\bibinfo {year} {2011})}\BibitemShut
  {NoStop}%
\bibitem [{\citenamefont {Briceño}\ and\ \citenamefont
  {Davoudi}(2013)}]{Briceno:2012rv}%
  \BibitemOpen
  \bibfield  {author} {\bibinfo {author} {\bibfnamefont {R.~A.}\ \bibnamefont
  {Briceño}}\ and\ \bibinfo {author} {\bibfnamefont {Z.}~\bibnamefont
  {Davoudi}},\ }\href {\doibase 10.1103/PhysRevD.87.094507} {\bibfield
  {journal} {\bibinfo  {journal} {Phys. Rev. D}\ }\textbf {\bibinfo {volume}
  {87}},\ \bibinfo {pages} {094507} (\bibinfo {year} {2013})},\ \Eprint
  {http://arxiv.org/abs/1212.3398} {arXiv:1212.3398 [hep-lat]} \BibitemShut
  {NoStop}%
\bibitem [{\citenamefont {Polejaeva}\ and\ \citenamefont
  {Rusetsky}(2012)}]{Polejaeva:2012ut}%
  \BibitemOpen
  \bibfield  {author} {\bibinfo {author} {\bibfnamefont {K.}~\bibnamefont
  {Polejaeva}}\ and\ \bibinfo {author} {\bibfnamefont {A.}~\bibnamefont
  {Rusetsky}},\ }\href {\doibase 10.1140/epja/i2012-12067-8} {\bibfield
  {journal} {\bibinfo  {journal} {Eur. Phys. J. A}\ }\textbf {\bibinfo {volume}
  {48}},\ \bibinfo {pages} {67} (\bibinfo {year} {2012})},\ \Eprint
  {http://arxiv.org/abs/1203.1241} {arXiv:1203.1241 [hep-lat]} \BibitemShut
  {NoStop}%
\bibitem [{\citenamefont {Hansen}\ and\ \citenamefont
  {Sharpe}(2014)}]{Hansen:2014eka}%
  \BibitemOpen
  \bibfield  {author} {\bibinfo {author} {\bibfnamefont {M.~T.}\ \bibnamefont
  {Hansen}}\ and\ \bibinfo {author} {\bibfnamefont {S.~R.}\ \bibnamefont
  {Sharpe}},\ }\href {\doibase 10.1103/PhysRevD.90.116003} {\bibfield
  {journal} {\bibinfo  {journal} {Phys. Rev. D}\ }\textbf {\bibinfo {volume}
  {90}},\ \bibinfo {pages} {116003} (\bibinfo {year} {2014})},\ \Eprint
  {http://arxiv.org/abs/1408.5933} {arXiv:1408.5933 [hep-lat]} \BibitemShut
  {NoStop}%
\bibitem [{\citenamefont {Hansen}\ and\ \citenamefont
  {Sharpe}(2015)}]{Hansen:2015zga}%
  \BibitemOpen
  \bibfield  {author} {\bibinfo {author} {\bibfnamefont {M.~T.}\ \bibnamefont
  {Hansen}}\ and\ \bibinfo {author} {\bibfnamefont {S.~R.}\ \bibnamefont
  {Sharpe}},\ }\href {\doibase 10.1103/PhysRevD.92.114509} {\bibfield
  {journal} {\bibinfo  {journal} {Phys. Rev. D}\ }\textbf {\bibinfo {volume}
  {92}},\ \bibinfo {pages} {114509} (\bibinfo {year} {2015})},\ \Eprint
  {http://arxiv.org/abs/1504.04248} {arXiv:1504.04248 [hep-lat]} \BibitemShut
  {NoStop}%
\bibitem [{\citenamefont {Briceño}\ \emph {et~al.}(2017)\citenamefont
  {Briceño}, \citenamefont {Hansen},\ and\ \citenamefont
  {Sharpe}}]{Briceno:2017tce}%
  \BibitemOpen
  \bibfield  {author} {\bibinfo {author} {\bibfnamefont {R.~A.}\ \bibnamefont
  {Briceño}}, \bibinfo {author} {\bibfnamefont {M.~T.}\ \bibnamefont
  {Hansen}}, \ and\ \bibinfo {author} {\bibfnamefont {S.~R.}\ \bibnamefont
  {Sharpe}},\ }\href {\doibase 10.1103/PhysRevD.95.074510} {\bibfield
  {journal} {\bibinfo  {journal} {Phys. Rev. D}\ }\textbf {\bibinfo {volume}
  {95}},\ \bibinfo {pages} {074510} (\bibinfo {year} {2017})},\ \Eprint
  {http://arxiv.org/abs/1701.07465} {arXiv:1701.07465 [hep-lat]} \BibitemShut
  {NoStop}%
\bibitem [{\citenamefont {Hammer}\ \emph
  {et~al.}(2017{\natexlab{a}})\citenamefont {Hammer}, \citenamefont {Pang},\
  and\ \citenamefont {Rusetsky}}]{Hammer:2017uqm}%
  \BibitemOpen
  \bibfield  {author} {\bibinfo {author} {\bibfnamefont {H.-W.}\ \bibnamefont
  {Hammer}}, \bibinfo {author} {\bibfnamefont {J.-Y.}\ \bibnamefont {Pang}}, \
  and\ \bibinfo {author} {\bibfnamefont {A.}~\bibnamefont {Rusetsky}},\ }\href
  {\doibase 10.1007/JHEP09(2017)109} {\bibfield  {journal} {\bibinfo  {journal}
  {JHEP}\ }\textbf {\bibinfo {volume} {09}},\ \bibinfo {pages} {109} (\bibinfo
  {year} {2017}{\natexlab{a}})},\ \Eprint {http://arxiv.org/abs/1706.07700}
  {arXiv:1706.07700 [hep-lat]} \BibitemShut {NoStop}%
\bibitem [{\citenamefont {Hammer}\ \emph
  {et~al.}(2017{\natexlab{b}})\citenamefont {Hammer}, \citenamefont {Pang},\
  and\ \citenamefont {Rusetsky}}]{Hammer:2017kms}%
  \BibitemOpen
  \bibfield  {author} {\bibinfo {author} {\bibfnamefont {H.~W.}\ \bibnamefont
  {Hammer}}, \bibinfo {author} {\bibfnamefont {J.~Y.}\ \bibnamefont {Pang}}, \
  and\ \bibinfo {author} {\bibfnamefont {A.}~\bibnamefont {Rusetsky}},\ }\href
  {\doibase 10.1007/JHEP10(2017)115} {\bibfield  {journal} {\bibinfo  {journal}
  {JHEP}\ }\textbf {\bibinfo {volume} {10}},\ \bibinfo {pages} {115} (\bibinfo
  {year} {2017}{\natexlab{b}})},\ \Eprint {http://arxiv.org/abs/1707.02176}
  {arXiv:1707.02176 [hep-lat]} \BibitemShut {NoStop}%
\bibitem [{\citenamefont {Mai}\ and\ \citenamefont
  {D\"oring}(2017)}]{Mai:2017bge}%
  \BibitemOpen
  \bibfield  {author} {\bibinfo {author} {\bibfnamefont {M.}~\bibnamefont
  {Mai}}\ and\ \bibinfo {author} {\bibfnamefont {M.}~\bibnamefont {D\"oring}},\
  }\href {\doibase 10.1140/epja/i2017-12440-1} {\bibfield  {journal} {\bibinfo
  {journal} {Eur. Phys. J. A}\ }\textbf {\bibinfo {volume} {53}},\ \bibinfo
  {pages} {240} (\bibinfo {year} {2017})},\ \Eprint
  {http://arxiv.org/abs/1709.08222} {arXiv:1709.08222 [hep-lat]} \BibitemShut
  {NoStop}%
\bibitem [{\citenamefont {Mai}\ and\ \citenamefont
  {Doring}(2019)}]{Mai:2018djl}%
  \BibitemOpen
  \bibfield  {author} {\bibinfo {author} {\bibfnamefont {M.}~\bibnamefont
  {Mai}}\ and\ \bibinfo {author} {\bibfnamefont {M.}~\bibnamefont {Doring}},\
  }\href {\doibase 10.1103/PhysRevLett.122.062503} {\bibfield  {journal}
  {\bibinfo  {journal} {Phys. Rev. Lett.}\ }\textbf {\bibinfo {volume} {122}},\
  \bibinfo {pages} {062503} (\bibinfo {year} {2019})},\ \Eprint
  {http://arxiv.org/abs/1807.04746} {arXiv:1807.04746 [hep-lat]} \BibitemShut
  {NoStop}%
\bibitem [{\citenamefont {Brice\~no}\ \emph {et~al.}(2019)\citenamefont
  {Brice\~no}, \citenamefont {Hansen},\ and\ \citenamefont
  {Sharpe}}]{Briceno:2018aml}%
  \BibitemOpen
  \bibfield  {author} {\bibinfo {author} {\bibfnamefont {R.~A.}\ \bibnamefont
  {Brice\~no}}, \bibinfo {author} {\bibfnamefont {M.~T.}\ \bibnamefont
  {Hansen}}, \ and\ \bibinfo {author} {\bibfnamefont {S.~R.}\ \bibnamefont
  {Sharpe}},\ }\href {\doibase 10.1103/PhysRevD.99.014516} {\bibfield
  {journal} {\bibinfo  {journal} {Phys. Rev. D}\ }\textbf {\bibinfo {volume}
  {99}},\ \bibinfo {pages} {014516} (\bibinfo {year} {2019})},\ \Eprint
  {http://arxiv.org/abs/1810.01429} {arXiv:1810.01429 [hep-lat]} \BibitemShut
  {NoStop}%
\bibitem [{\citenamefont {Jackura}\ \emph {et~al.}(2019)\citenamefont
  {Jackura}, \citenamefont {Dawid}, \citenamefont {Fernández-Ramírez},
  \citenamefont {Mathieu}, \citenamefont {Mikhasenko}, \citenamefont {Pilloni},
  \citenamefont {Sharpe},\ and\ \citenamefont {Szczepaniak}}]{Jackura:2019bmu}%
  \BibitemOpen
  \bibfield  {author} {\bibinfo {author} {\bibfnamefont {A.~W.}\ \bibnamefont
  {Jackura}}, \bibinfo {author} {\bibfnamefont {S.~M.}\ \bibnamefont {Dawid}},
  \bibinfo {author} {\bibfnamefont {C.}~\bibnamefont {Fernández-Ramírez}},
  \bibinfo {author} {\bibfnamefont {V.}~\bibnamefont {Mathieu}}, \bibinfo
  {author} {\bibfnamefont {M.}~\bibnamefont {Mikhasenko}}, \bibinfo {author}
  {\bibfnamefont {A.}~\bibnamefont {Pilloni}}, \bibinfo {author} {\bibfnamefont
  {S.~R.}\ \bibnamefont {Sharpe}}, \ and\ \bibinfo {author} {\bibfnamefont
  {A.~P.}\ \bibnamefont {Szczepaniak}},\ }\href {\doibase
  10.1103/PhysRevD.100.034508} {\bibfield  {journal} {\bibinfo  {journal}
  {Phys. Rev. D}\ }\textbf {\bibinfo {volume} {100}},\ \bibinfo {pages}
  {034508} (\bibinfo {year} {2019})}\BibitemShut {NoStop}%
\bibitem [{\citenamefont {H\"orz}\ and\ \citenamefont
  {Hanlon}(2019)}]{Horz:2019rrn}%
  \BibitemOpen
  \bibfield  {author} {\bibinfo {author} {\bibfnamefont {B.}~\bibnamefont
  {H\"orz}}\ and\ \bibinfo {author} {\bibfnamefont {A.}~\bibnamefont
  {Hanlon}},\ }\href {\doibase 10.1103/PhysRevLett.123.142002} {\bibfield
  {journal} {\bibinfo  {journal} {Phys. Rev. Lett.}\ }\textbf {\bibinfo
  {volume} {123}},\ \bibinfo {pages} {142002} (\bibinfo {year} {2019})},\
  \Eprint {http://arxiv.org/abs/1905.04277} {arXiv:1905.04277 [hep-lat]}
  \BibitemShut {NoStop}%
\bibitem [{\citenamefont {Blanton}\ \emph {et~al.}(2020)\citenamefont
  {Blanton}, \citenamefont {Romero-López},\ and\ \citenamefont
  {Sharpe}}]{Blanton:2019vdk}%
  \BibitemOpen
  \bibfield  {author} {\bibinfo {author} {\bibfnamefont {T.~D.}\ \bibnamefont
  {Blanton}}, \bibinfo {author} {\bibfnamefont {F.}~\bibnamefont
  {Romero-López}}, \ and\ \bibinfo {author} {\bibfnamefont {S.~R.}\
  \bibnamefont {Sharpe}},\ }\href {\doibase 10.1103/PhysRevLett.124.032001}
  {\bibfield  {journal} {\bibinfo  {journal} {Phys. Rev. Lett.}\ }\textbf
  {\bibinfo {volume} {124}},\ \bibinfo {pages} {032001} (\bibinfo {year}
  {2020})},\ \Eprint {http://arxiv.org/abs/1909.02973} {arXiv:1909.02973}
  \BibitemShut {NoStop}%
\bibitem [{\citenamefont {Blanton}\ \emph {et~al.}(2019)\citenamefont
  {Blanton}, \citenamefont {Romero-L\'opez},\ and\ \citenamefont
  {Sharpe}}]{Blanton:2019igq}%
  \BibitemOpen
  \bibfield  {author} {\bibinfo {author} {\bibfnamefont {T.~D.}\ \bibnamefont
  {Blanton}}, \bibinfo {author} {\bibfnamefont {F.}~\bibnamefont
  {Romero-L\'opez}}, \ and\ \bibinfo {author} {\bibfnamefont {S.~R.}\
  \bibnamefont {Sharpe}},\ }\href {\doibase 10.1007/JHEP03(2019)106} {\bibfield
   {journal} {\bibinfo  {journal} {JHEP}\ }\textbf {\bibinfo {volume} {03}},\
  \bibinfo {pages} {106} (\bibinfo {year} {2019})},\ \Eprint
  {http://arxiv.org/abs/1901.07095} {arXiv:1901.07095 [hep-lat]} \BibitemShut
  {NoStop}%
\bibitem [{\citenamefont {Mai}\ \emph {et~al.}(2020)\citenamefont {Mai},
  \citenamefont {Döring}, \citenamefont {Culver},\ and\ \citenamefont
  {Alexandru}}]{Mai:2019fba}%
  \BibitemOpen
  \bibfield  {author} {\bibinfo {author} {\bibfnamefont {M.}~\bibnamefont
  {Mai}}, \bibinfo {author} {\bibfnamefont {M.}~\bibnamefont {Döring}},
  \bibinfo {author} {\bibfnamefont {C.}~\bibnamefont {Culver}}, \ and\ \bibinfo
  {author} {\bibfnamefont {A.}~\bibnamefont {Alexandru}},\ }\href {\doibase
  10.1103/PhysRevD.101.054510} {\bibfield  {journal} {\bibinfo  {journal}
  {Phys. Rev. D}\ }\textbf {\bibinfo {volume} {101}},\ \bibinfo {pages}
  {054510} (\bibinfo {year} {2020})}\BibitemShut {NoStop}%
\bibitem [{\citenamefont {Culver}\ \emph {et~al.}(2020)\citenamefont {Culver},
  \citenamefont {Mai}, \citenamefont {Brett}, \citenamefont {Alexandru},\ and\
  \citenamefont {Döring}}]{Culver:2019vvu}%
  \BibitemOpen
  \bibfield  {author} {\bibinfo {author} {\bibfnamefont {C.}~\bibnamefont
  {Culver}}, \bibinfo {author} {\bibfnamefont {M.}~\bibnamefont {Mai}},
  \bibinfo {author} {\bibfnamefont {R.}~\bibnamefont {Brett}}, \bibinfo
  {author} {\bibfnamefont {A.}~\bibnamefont {Alexandru}}, \ and\ \bibinfo
  {author} {\bibfnamefont {M.}~\bibnamefont {Döring}},\ }\href {\doibase
  10.1103/PhysRevD.101.114507} {\bibfield  {journal} {\bibinfo  {journal}
  {Phys. Rev. D}\ }\textbf {\bibinfo {volume} {101}},\ \bibinfo {pages}
  {114507} (\bibinfo {year} {2020})}\BibitemShut {NoStop}%
\bibitem [{\citenamefont {Pang}\ \emph {et~al.}(2019)\citenamefont {Pang},
  \citenamefont {Wu}, \citenamefont {Hammer}, \citenamefont {Meißner},\ and\
  \citenamefont {Rusetsky}}]{Pang:2019dfe}%
  \BibitemOpen
  \bibfield  {author} {\bibinfo {author} {\bibfnamefont {J.-Y.}\ \bibnamefont
  {Pang}}, \bibinfo {author} {\bibfnamefont {J.-J.}\ \bibnamefont {Wu}},
  \bibinfo {author} {\bibfnamefont {H.-W.}\ \bibnamefont {Hammer}}, \bibinfo
  {author} {\bibfnamefont {U.-G.}\ \bibnamefont {Meißner}}, \ and\ \bibinfo
  {author} {\bibfnamefont {A.}~\bibnamefont {Rusetsky}},\ }\href {\doibase
  10.1103/PhysRevD.99.074513} {\bibfield  {journal} {\bibinfo  {journal} {Phys.
  Rev. D}\ }\textbf {\bibinfo {volume} {99}},\ \bibinfo {pages} {074513}
  (\bibinfo {year} {2019})},\ \Eprint {http://arxiv.org/abs/1902.01111}
  {arXiv:1902.01111 [hep-lat]} \BibitemShut {NoStop}%
\bibitem [{\citenamefont {Romero-López}\ \emph {et~al.}(2019)\citenamefont
  {Romero-López}, \citenamefont {Sharpe}, \citenamefont {Blanton},
  \citenamefont {Briceño},\ and\ \citenamefont
  {Hansen}}]{Romero-Lopez:2019qrt}%
  \BibitemOpen
  \bibfield  {author} {\bibinfo {author} {\bibfnamefont {F.}~\bibnamefont
  {Romero-López}}, \bibinfo {author} {\bibfnamefont {S.~R.}\ \bibnamefont
  {Sharpe}}, \bibinfo {author} {\bibfnamefont {T.~D.}\ \bibnamefont {Blanton}},
  \bibinfo {author} {\bibfnamefont {R.~A.}\ \bibnamefont {Briceño}}, \ and\
  \bibinfo {author} {\bibfnamefont {M.~T.}\ \bibnamefont {Hansen}},\ }\href
  {\doibase 10.1007/JHEP10(2019)007} {\bibfield  {journal} {\bibinfo  {journal}
  {JHEP}\ }\textbf {\bibinfo {volume} {10}},\ \bibinfo {pages} {007} (\bibinfo
  {year} {2019})},\ \Eprint {http://arxiv.org/abs/1908.02411} {arXiv:1908.02411
  [hep-lat]} \BibitemShut {NoStop}%
\bibitem [{\citenamefont {Blanton}\ and\ \citenamefont
  {Sharpe}(2021)}]{Blanton:2020gmf}%
  \BibitemOpen
  \bibfield  {author} {\bibinfo {author} {\bibfnamefont {T.~D.}\ \bibnamefont
  {Blanton}}\ and\ \bibinfo {author} {\bibfnamefont {S.~R.}\ \bibnamefont
  {Sharpe}},\ }\href {\doibase 10.1103/PhysRevD.103.054503} {\bibfield
  {journal} {\bibinfo  {journal} {Phys. Rev. D}\ }\textbf {\bibinfo {volume}
  {103}},\ \bibinfo {pages} {054503} (\bibinfo {year} {2021})},\ \Eprint
  {http://arxiv.org/abs/2011.05520} {arXiv:2011.05520 [hep-lat]} \BibitemShut
  {NoStop}%
\bibitem [{\citenamefont {Blanton}\ and\ \citenamefont
  {Sharpe}(2020{\natexlab{a}})}]{Blanton:2020gha}%
  \BibitemOpen
  \bibfield  {author} {\bibinfo {author} {\bibfnamefont {T.~D.}\ \bibnamefont
  {Blanton}}\ and\ \bibinfo {author} {\bibfnamefont {S.~R.}\ \bibnamefont
  {Sharpe}},\ }\href {\doibase 10.1103/PhysRevD.102.054520} {\bibfield
  {journal} {\bibinfo  {journal} {Phys. Rev. D}\ }\textbf {\bibinfo {volume}
  {102}},\ \bibinfo {pages} {054520} (\bibinfo {year} {2020}{\natexlab{a}})},\
  \Eprint {http://arxiv.org/abs/2007.16188} {arXiv:2007.16188 [hep-lat]}
  \BibitemShut {NoStop}%
\bibitem [{\citenamefont {Blanton}\ and\ \citenamefont
  {Sharpe}(2020{\natexlab{b}})}]{Blanton:2020jnm}%
  \BibitemOpen
  \bibfield  {author} {\bibinfo {author} {\bibfnamefont {T.~D.}\ \bibnamefont
  {Blanton}}\ and\ \bibinfo {author} {\bibfnamefont {S.~R.}\ \bibnamefont
  {Sharpe}},\ }\href {\doibase 10.1103/PhysRevD.102.054515} {\bibfield
  {journal} {\bibinfo  {journal} {Phys. Rev. D}\ }\textbf {\bibinfo {volume}
  {102}},\ \bibinfo {pages} {054515} (\bibinfo {year} {2020}{\natexlab{b}})},\
  \Eprint {http://arxiv.org/abs/2007.16190} {arXiv:2007.16190 [hep-lat]}
  \BibitemShut {NoStop}%
\bibitem [{\citenamefont {M\"uller}\ and\ \citenamefont
  {Rusetsky}(2021)}]{Muller:2020wjo}%
  \BibitemOpen
  \bibfield  {author} {\bibinfo {author} {\bibfnamefont {F.}~\bibnamefont
  {M\"uller}}\ and\ \bibinfo {author} {\bibfnamefont {A.}~\bibnamefont
  {Rusetsky}},\ }\href {\doibase 10.1007/JHEP03(2021)152} {\bibfield  {journal}
  {\bibinfo  {journal} {JHEP}\ }\textbf {\bibinfo {volume} {03}},\ \bibinfo
  {pages} {152} (\bibinfo {year} {2021})},\ \Eprint
  {http://arxiv.org/abs/2012.13957} {arXiv:2012.13957 [hep-lat]} \BibitemShut
  {NoStop}%
\bibitem [{\citenamefont {Fischer}\ \emph {et~al.}(2021)\citenamefont
  {Fischer}, \citenamefont {Kostrzewa}, \citenamefont {Liu}, \citenamefont
  {Romero-L\'opez}, \citenamefont {Ueding},\ and\ \citenamefont
  {Urbach}}]{Fischer:2020jzp}%
  \BibitemOpen
  \bibfield  {author} {\bibinfo {author} {\bibfnamefont {M.}~\bibnamefont
  {Fischer}}, \bibinfo {author} {\bibfnamefont {B.}~\bibnamefont {Kostrzewa}},
  \bibinfo {author} {\bibfnamefont {L.}~\bibnamefont {Liu}}, \bibinfo {author}
  {\bibfnamefont {F.}~\bibnamefont {Romero-L\'opez}}, \bibinfo {author}
  {\bibfnamefont {M.}~\bibnamefont {Ueding}}, \ and\ \bibinfo {author}
  {\bibfnamefont {C.}~\bibnamefont {Urbach}},\ }\href {\doibase
  10.1140/epjc/s10052-021-09206-5} {\bibfield  {journal} {\bibinfo  {journal}
  {Eur. Phys. J. C}\ }\textbf {\bibinfo {volume} {81}},\ \bibinfo {pages} {436}
  (\bibinfo {year} {2021})},\ \Eprint {http://arxiv.org/abs/2008.03035}
  {arXiv:2008.03035 [hep-lat]} \BibitemShut {NoStop}%
\bibitem [{\citenamefont {Hansen}\ \emph {et~al.}(2021)\citenamefont {Hansen},
  \citenamefont {Brice\~no}, \citenamefont {Edwards}, \citenamefont {Thomas},\
  and\ \citenamefont {Wilson}}]{Hansen:2020otl}%
  \BibitemOpen
  \bibfield  {author} {\bibinfo {author} {\bibfnamefont {M.~T.}\ \bibnamefont
  {Hansen}}, \bibinfo {author} {\bibfnamefont {R.~A.}\ \bibnamefont
  {Brice\~no}}, \bibinfo {author} {\bibfnamefont {R.~G.}\ \bibnamefont
  {Edwards}}, \bibinfo {author} {\bibfnamefont {C.~E.}\ \bibnamefont {Thomas}},
  \ and\ \bibinfo {author} {\bibfnamefont {D.~J.}\ \bibnamefont {Wilson}}
  (\bibinfo {collaboration} {Hadron Spectrum}),\ }\href {\doibase
  10.1103/PhysRevLett.126.012001} {\bibfield  {journal} {\bibinfo  {journal}
  {Phys. Rev. Lett.}\ }\textbf {\bibinfo {volume} {126}},\ \bibinfo {pages}
  {012001} (\bibinfo {year} {2021})},\ \Eprint
  {http://arxiv.org/abs/2009.04931} {arXiv:2009.04931 [hep-lat]} \BibitemShut
  {NoStop}%
\bibitem [{\citenamefont {Hansen}\ \emph {et~al.}(2020)\citenamefont {Hansen},
  \citenamefont {Romero-L\'opez},\ and\ \citenamefont
  {Sharpe}}]{Hansen:2020zhy}%
  \BibitemOpen
  \bibfield  {author} {\bibinfo {author} {\bibfnamefont {M.~T.}\ \bibnamefont
  {Hansen}}, \bibinfo {author} {\bibfnamefont {F.}~\bibnamefont
  {Romero-L\'opez}}, \ and\ \bibinfo {author} {\bibfnamefont {S.~R.}\
  \bibnamefont {Sharpe}},\ }\href {\doibase 10.1007/JHEP07(2020)047} {\bibfield
   {journal} {\bibinfo  {journal} {JHEP}\ }\textbf {\bibinfo {volume} {07}},\
  \bibinfo {pages} {047} (\bibinfo {year} {2020})},\ \bibinfo {note} {[Erratum:
  JHEP 02, 014 (2021)]},\ \Eprint {http://arxiv.org/abs/2003.10974}
  {arXiv:2003.10974 [hep-lat]} \BibitemShut {NoStop}%
\bibitem [{\citenamefont {Alexandru}\ \emph {et~al.}(2020)\citenamefont
  {Alexandru}, \citenamefont {Brett}, \citenamefont {Culver}, \citenamefont
  {D\"oring}, \citenamefont {Guo}, \citenamefont {Lee},\ and\ \citenamefont
  {Mai}}]{Alexandru:2020xqf}%
  \BibitemOpen
  \bibfield  {author} {\bibinfo {author} {\bibfnamefont {A.}~\bibnamefont
  {Alexandru}}, \bibinfo {author} {\bibfnamefont {R.}~\bibnamefont {Brett}},
  \bibinfo {author} {\bibfnamefont {C.}~\bibnamefont {Culver}}, \bibinfo
  {author} {\bibfnamefont {M.}~\bibnamefont {D\"oring}}, \bibinfo {author}
  {\bibfnamefont {D.}~\bibnamefont {Guo}}, \bibinfo {author} {\bibfnamefont
  {F.~X.}\ \bibnamefont {Lee}}, \ and\ \bibinfo {author} {\bibfnamefont
  {M.}~\bibnamefont {Mai}},\ }\href {\doibase 10.1103/PhysRevD.102.114523}
  {\bibfield  {journal} {\bibinfo  {journal} {Phys. Rev. D}\ }\textbf {\bibinfo
  {volume} {102}},\ \bibinfo {pages} {114523} (\bibinfo {year} {2020})},\
  \Eprint {http://arxiv.org/abs/2009.12358} {arXiv:2009.12358 [hep-lat]}
  \BibitemShut {NoStop}%
\bibitem [{\citenamefont {M\"uller}\ \emph {et~al.}(2022)\citenamefont
  {M\"uller}, \citenamefont {Pang}, \citenamefont {Rusetsky},\ and\
  \citenamefont {Wu}}]{Muller:2021uur}%
  \BibitemOpen
  \bibfield  {author} {\bibinfo {author} {\bibfnamefont {F.}~\bibnamefont
  {M\"uller}}, \bibinfo {author} {\bibfnamefont {J.-Y.}\ \bibnamefont {Pang}},
  \bibinfo {author} {\bibfnamefont {A.}~\bibnamefont {Rusetsky}}, \ and\
  \bibinfo {author} {\bibfnamefont {J.-J.}\ \bibnamefont {Wu}},\ }\href
  {\doibase 10.1007/JHEP02(2022)158} {\bibfield  {journal} {\bibinfo  {journal}
  {JHEP}\ }\textbf {\bibinfo {volume} {02}},\ \bibinfo {pages} {158} (\bibinfo
  {year} {2022})},\ \Eprint {http://arxiv.org/abs/2110.09351} {arXiv:2110.09351
  [hep-lat]} \BibitemShut {NoStop}%
\bibitem [{\citenamefont {Brett}\ \emph {et~al.}(2021)\citenamefont {Brett},
  \citenamefont {Culver}, \citenamefont {Mai}, \citenamefont {Alexandru},
  \citenamefont {D\"oring},\ and\ \citenamefont {Lee}}]{Brett:2021wyd}%
  \BibitemOpen
  \bibfield  {author} {\bibinfo {author} {\bibfnamefont {R.}~\bibnamefont
  {Brett}}, \bibinfo {author} {\bibfnamefont {C.}~\bibnamefont {Culver}},
  \bibinfo {author} {\bibfnamefont {M.}~\bibnamefont {Mai}}, \bibinfo {author}
  {\bibfnamefont {A.}~\bibnamefont {Alexandru}}, \bibinfo {author}
  {\bibfnamefont {M.}~\bibnamefont {D\"oring}}, \ and\ \bibinfo {author}
  {\bibfnamefont {F.~X.}\ \bibnamefont {Lee}},\ }\href {\doibase
  10.1103/PhysRevD.104.014501} {\bibfield  {journal} {\bibinfo  {journal}
  {Phys. Rev. D}\ }\textbf {\bibinfo {volume} {104}},\ \bibinfo {pages}
  {014501} (\bibinfo {year} {2021})},\ \Eprint
  {http://arxiv.org/abs/2101.06144} {arXiv:2101.06144 [hep-lat]} \BibitemShut
  {NoStop}%
\bibitem [{\citenamefont {Mai}\ \emph {et~al.}(2021{\natexlab{a}})\citenamefont
  {Mai}, \citenamefont {D\"oring},\ and\ \citenamefont
  {Rusetsky}}]{Mai:2021lwb}%
  \BibitemOpen
  \bibfield  {author} {\bibinfo {author} {\bibfnamefont {M.}~\bibnamefont
  {Mai}}, \bibinfo {author} {\bibfnamefont {M.}~\bibnamefont {D\"oring}}, \
  and\ \bibinfo {author} {\bibfnamefont {A.}~\bibnamefont {Rusetsky}},\ }\href
  {\doibase 10.1140/epjs/s11734-021-00146-5} {\bibfield  {journal} {\bibinfo
  {journal} {Eur. Phys. J. ST}\ }\textbf {\bibinfo {volume} {230}},\ \bibinfo
  {pages} {1623} (\bibinfo {year} {2021}{\natexlab{a}})},\ \Eprint
  {http://arxiv.org/abs/2103.00577} {arXiv:2103.00577 [hep-lat]} \BibitemShut
  {NoStop}%
\bibitem [{\citenamefont {Blanton}\ \emph {et~al.}(2021)\citenamefont
  {Blanton}, \citenamefont {Hanlon}, \citenamefont {H\"orz}, \citenamefont
  {Morningstar}, \citenamefont {Romero-L\'opez},\ and\ \citenamefont
  {Sharpe}}]{Blanton:2021llb}%
  \BibitemOpen
  \bibfield  {author} {\bibinfo {author} {\bibfnamefont {T.~D.}\ \bibnamefont
  {Blanton}}, \bibinfo {author} {\bibfnamefont {A.~D.}\ \bibnamefont {Hanlon}},
  \bibinfo {author} {\bibfnamefont {B.}~\bibnamefont {H\"orz}}, \bibinfo
  {author} {\bibfnamefont {C.}~\bibnamefont {Morningstar}}, \bibinfo {author}
  {\bibfnamefont {F.}~\bibnamefont {Romero-L\'opez}}, \ and\ \bibinfo {author}
  {\bibfnamefont {S.~R.}\ \bibnamefont {Sharpe}},\ }\href {\doibase
  10.1007/JHEP10(2021)023} {\bibfield  {journal} {\bibinfo  {journal} {JHEP}\
  }\textbf {\bibinfo {volume} {10}},\ \bibinfo {pages} {023} (\bibinfo {year}
  {2021})},\ \Eprint {http://arxiv.org/abs/2106.05590} {arXiv:2106.05590
  [hep-lat]} \BibitemShut {NoStop}%
\bibitem [{\citenamefont {Mai}\ \emph {et~al.}(2021{\natexlab{b}})\citenamefont
  {Mai}, \citenamefont {Alexandru}, \citenamefont {Brett}, \citenamefont
  {Culver}, \citenamefont {D\"oring}, \citenamefont {Lee},\ and\ \citenamefont
  {Sadasivan}}]{Mai:2021nul}%
  \BibitemOpen
  \bibfield  {author} {\bibinfo {author} {\bibfnamefont {M.}~\bibnamefont
  {Mai}}, \bibinfo {author} {\bibfnamefont {A.}~\bibnamefont {Alexandru}},
  \bibinfo {author} {\bibfnamefont {R.}~\bibnamefont {Brett}}, \bibinfo
  {author} {\bibfnamefont {C.}~\bibnamefont {Culver}}, \bibinfo {author}
  {\bibfnamefont {M.}~\bibnamefont {D\"oring}}, \bibinfo {author}
  {\bibfnamefont {F.~X.}\ \bibnamefont {Lee}}, \ and\ \bibinfo {author}
  {\bibfnamefont {D.}~\bibnamefont {Sadasivan}} (\bibinfo {collaboration}
  {GWQCD}),\ }\href {\doibase 10.1103/PhysRevLett.127.222001} {\bibfield
  {journal} {\bibinfo  {journal} {Phys. Rev. Lett.}\ }\textbf {\bibinfo
  {volume} {127}},\ \bibinfo {pages} {222001} (\bibinfo {year}
  {2021}{\natexlab{b}})},\ \Eprint {http://arxiv.org/abs/2107.03973}
  {arXiv:2107.03973 [hep-lat]} \BibitemShut {NoStop}%
\bibitem [{\citenamefont {Hansen}\ and\ \citenamefont
  {Sharpe}(2019)}]{Hansen:2019nir}%
  \BibitemOpen
  \bibfield  {author} {\bibinfo {author} {\bibfnamefont {M.~T.}\ \bibnamefont
  {Hansen}}\ and\ \bibinfo {author} {\bibfnamefont {S.~R.}\ \bibnamefont
  {Sharpe}},\ }\href {\doibase 10.1146/annurev-nucl-101918-023723} {\bibfield
  {journal} {\bibinfo  {journal} {Ann. Rev. Nucl. Part. Sci.}\ }\textbf
  {\bibinfo {volume} {69}},\ \bibinfo {pages} {65} (\bibinfo {year} {2019})},\
  \Eprint {http://arxiv.org/abs/1901.00483} {arXiv:1901.00483 [hep-lat]}
  \BibitemShut {NoStop}%
\bibitem [{\citenamefont {Lüscher}(1986)}]{Luscher:1986pf}%
  \BibitemOpen
  \bibfield  {author} {\bibinfo {author} {\bibfnamefont {M.}~\bibnamefont
  {Lüscher}},\ }\href {\doibase 10.1007/BF01211097} {\bibfield  {journal}
  {\bibinfo  {journal} {Commun.Math. Phys.}\ }\textbf {\bibinfo {volume}
  {105}},\ \bibinfo {pages} {153} (\bibinfo {year} {1986})}\BibitemShut
  {NoStop}%
\bibitem [{\citenamefont {Lüscher}(1991)}]{Luscher:1990ux}%
  \BibitemOpen
  \bibfield  {author} {\bibinfo {author} {\bibfnamefont {M.}~\bibnamefont
  {Lüscher}},\ }\href {\doibase 10.1016/0550-3213(91)90366-6} {\bibfield
  {journal} {\bibinfo  {journal} {Nuclear Physics B}\ }\textbf {\bibinfo
  {volume} {354}},\ \bibinfo {pages} {531} (\bibinfo {year}
  {1991})}\BibitemShut {NoStop}%
\bibitem [{\citenamefont {Rummukainen}\ and\ \citenamefont
  {Gottlieb}(1995)}]{Rummukainen:1995vs}%
  \BibitemOpen
  \bibfield  {author} {\bibinfo {author} {\bibfnamefont {K.}~\bibnamefont
  {Rummukainen}}\ and\ \bibinfo {author} {\bibfnamefont {S.}~\bibnamefont
  {Gottlieb}},\ }\href {\doibase 10.1016/0550-3213(95)00313-H} {\bibfield
  {journal} {\bibinfo  {journal} {Nuclear Physics B}\ }\textbf {\bibinfo
  {volume} {450}},\ \bibinfo {pages} {397} (\bibinfo {year} {1995})},\ \Eprint
  {http://arxiv.org/abs/hep-lat/9503028} {arXiv:hep-lat/9503028} \BibitemShut
  {NoStop}%
\bibitem [{\citenamefont {Dudek}\ \emph {et~al.}(2013)\citenamefont {Dudek},
  \citenamefont {Edwards},\ and\ \citenamefont {Thomas}}]{Dudek:2012xn}%
  \BibitemOpen
  \bibfield  {author} {\bibinfo {author} {\bibfnamefont {J.~J.}\ \bibnamefont
  {Dudek}}, \bibinfo {author} {\bibfnamefont {R.~G.}\ \bibnamefont {Edwards}},
  \ and\ \bibinfo {author} {\bibfnamefont {C.~E.}\ \bibnamefont {Thomas}}
  (\bibinfo {collaboration} {Hadron Spectrum}),\ }\href {\doibase
  10.1103/PhysRevD.87.034505} {\bibfield  {journal} {\bibinfo  {journal} {Phys.
  Rev. D}\ }\textbf {\bibinfo {volume} {87}},\ \bibinfo {pages} {034505}
  (\bibinfo {year} {2013})},\ \bibinfo {note} {[Erratum: Phys.Rev.D 90, 099902
  (2014)]},\ \Eprint {http://arxiv.org/abs/1212.0830} {arXiv:1212.0830
  [hep-ph]} \BibitemShut {NoStop}%
\bibitem [{\citenamefont {Dudek}\ \emph {et~al.}(2014)\citenamefont {Dudek},
  \citenamefont {Edwards}, \citenamefont {Thomas},\ and\ \citenamefont
  {Wilson}}]{Dudek:2014qha}%
  \BibitemOpen
  \bibfield  {author} {\bibinfo {author} {\bibfnamefont {J.~J.}\ \bibnamefont
  {Dudek}}, \bibinfo {author} {\bibfnamefont {R.~G.}\ \bibnamefont {Edwards}},
  \bibinfo {author} {\bibfnamefont {C.~E.}\ \bibnamefont {Thomas}}, \ and\
  \bibinfo {author} {\bibfnamefont {D.~J.}\ \bibnamefont {Wilson}} (\bibinfo
  {collaboration} {Hadron Spectrum}),\ }\href {\doibase
  10.1103/PhysRevLett.113.182001} {\bibfield  {journal} {\bibinfo  {journal}
  {Phys. Rev. Lett.}\ }\textbf {\bibinfo {volume} {113}},\ \bibinfo {pages}
  {182001} (\bibinfo {year} {2014})},\ \Eprint {http://arxiv.org/abs/1406.4158}
  {arXiv:1406.4158 [hep-ph]} \BibitemShut {NoStop}%
\bibitem [{\citenamefont {Kim}\ \emph {et~al.}(2005)\citenamefont {Kim},
  \citenamefont {Sachrajda},\ and\ \citenamefont {Sharpe}}]{Kim:2005gf}%
  \BibitemOpen
  \bibfield  {author} {\bibinfo {author} {\bibfnamefont {C.}~\bibnamefont
  {Kim}}, \bibinfo {author} {\bibfnamefont {C.}~\bibnamefont {Sachrajda}}, \
  and\ \bibinfo {author} {\bibfnamefont {S.~R.}\ \bibnamefont {Sharpe}},\
  }\href {\doibase 10.1016/j.nuclphysb.2005.08.029} {\bibfield  {journal}
  {\bibinfo  {journal} {Nuclear Physics B}\ }\textbf {\bibinfo {volume}
  {727}},\ \bibinfo {pages} {218} (\bibinfo {year} {2005})},\ \Eprint
  {http://arxiv.org/abs/hep-lat/0507006} {arXiv:hep-lat/0507006} \BibitemShut
  {NoStop}%
\bibitem [{\citenamefont {Hall}\ \emph {et~al.}(2013)\citenamefont {Hall},
  \citenamefont {Hsu}, \citenamefont {Leinweber}, \citenamefont {Thomas},\ and\
  \citenamefont {Young}}]{Hall:2013qba}%
  \BibitemOpen
  \bibfield  {author} {\bibinfo {author} {\bibfnamefont {J.~M.~M.}\
  \bibnamefont {Hall}}, \bibinfo {author} {\bibfnamefont {A.~C.-P.}\
  \bibnamefont {Hsu}}, \bibinfo {author} {\bibfnamefont {D.~B.}\ \bibnamefont
  {Leinweber}}, \bibinfo {author} {\bibfnamefont {A.~W.}\ \bibnamefont
  {Thomas}}, \ and\ \bibinfo {author} {\bibfnamefont {R.~D.}\ \bibnamefont
  {Young}},\ }\href {\doibase 10.1103/PhysRevD.87.094510} {\bibfield  {journal}
  {\bibinfo  {journal} {Phys. Rev. D}\ }\textbf {\bibinfo {volume} {87}},\
  \bibinfo {pages} {094510} (\bibinfo {year} {2013})},\ \Eprint
  {http://arxiv.org/abs/1303.4157} {arXiv:1303.4157 [hep-lat]} \BibitemShut
  {NoStop}%
\bibitem [{\citenamefont {Wu}\ \emph {et~al.}(2014)\citenamefont {Wu},
  \citenamefont {Lee}, \citenamefont {Thomas},\ and\ \citenamefont
  {Young}}]{Wu:2014vma}%
  \BibitemOpen
  \bibfield  {author} {\bibinfo {author} {\bibfnamefont {J.-J.}\ \bibnamefont
  {Wu}}, \bibinfo {author} {\bibfnamefont {T.~S.~H.}\ \bibnamefont {Lee}},
  \bibinfo {author} {\bibfnamefont {A.~W.}\ \bibnamefont {Thomas}}, \ and\
  \bibinfo {author} {\bibfnamefont {R.~D.}\ \bibnamefont {Young}},\ }\href
  {\doibase 10.1103/PhysRevC.90.055206} {\bibfield  {journal} {\bibinfo
  {journal} {Phys. Rev. C}\ }\textbf {\bibinfo {volume} {90}},\ \bibinfo
  {pages} {055206} (\bibinfo {year} {2014})},\ \Eprint
  {http://arxiv.org/abs/1402.4868} {arXiv:1402.4868 [hep-lat]} \BibitemShut
  {NoStop}%
\bibitem [{\citenamefont {Abell}\ \emph {et~al.}(2022)\citenamefont {Abell},
  \citenamefont {Leinweber}, \citenamefont {Thomas},\ and\ \citenamefont
  {Wu}}]{Abell:2021awi}%
  \BibitemOpen
  \bibfield  {author} {\bibinfo {author} {\bibfnamefont {C.~D.}\ \bibnamefont
  {Abell}}, \bibinfo {author} {\bibfnamefont {D.~B.}\ \bibnamefont
  {Leinweber}}, \bibinfo {author} {\bibfnamefont {A.~W.}\ \bibnamefont
  {Thomas}}, \ and\ \bibinfo {author} {\bibfnamefont {J.-J.}\ \bibnamefont
  {Wu}},\ }\href {\doibase 10.1103/PhysRevD.106.034506} {\bibfield  {journal}
  {\bibinfo  {journal} {Phys. Rev. D}\ }\textbf {\bibinfo {volume} {106}},\
  \bibinfo {pages} {034506} (\bibinfo {year} {2022})},\ \Eprint
  {http://arxiv.org/abs/2110.14113} {arXiv:2110.14113 [hep-lat]} \BibitemShut
  {NoStop}%
\bibitem [{\citenamefont {Li}\ \emph {et~al.}(2021)\citenamefont {Li},
  \citenamefont {Wu}, \citenamefont {Leinweber},\ and\ \citenamefont
  {Thomas}}]{Li:2021mob}%
  \BibitemOpen
  \bibfield  {author} {\bibinfo {author} {\bibfnamefont {Y.}~\bibnamefont
  {Li}}, \bibinfo {author} {\bibfnamefont {J.-J.}\ \bibnamefont {Wu}}, \bibinfo
  {author} {\bibfnamefont {D.~B.}\ \bibnamefont {Leinweber}}, \ and\ \bibinfo
  {author} {\bibfnamefont {A.~W.}\ \bibnamefont {Thomas}},\ }\href {\doibase
  10.1103/PhysRevD.103.094518} {\bibfield  {journal} {\bibinfo  {journal}
  {Phys. Rev. D}\ }\textbf {\bibinfo {volume} {103}},\ \bibinfo {pages}
  {094518} (\bibinfo {year} {2021})},\ \Eprint
  {http://arxiv.org/abs/2103.12260} {arXiv:2103.12260 [hep-lat]} \BibitemShut
  {NoStop}%
\bibitem [{\citenamefont {Wu}\ \emph {et~al.}(2017)\citenamefont {Wu},
  \citenamefont {Kamano}, \citenamefont {Lee}, \citenamefont {Leinweber},\ and\
  \citenamefont {Thomas}}]{Wu:2016ixr}%
  \BibitemOpen
  \bibfield  {author} {\bibinfo {author} {\bibfnamefont {J.-J.}\ \bibnamefont
  {Wu}}, \bibinfo {author} {\bibfnamefont {H.}~\bibnamefont {Kamano}}, \bibinfo
  {author} {\bibfnamefont {T.-S.~H.}\ \bibnamefont {Lee}}, \bibinfo {author}
  {\bibfnamefont {D.~B.}\ \bibnamefont {Leinweber}}, \ and\ \bibinfo {author}
  {\bibfnamefont {A.~W.}\ \bibnamefont {Thomas}},\ }\href {\doibase
  10.1103/PhysRevD.95.114507} {\bibfield  {journal} {\bibinfo  {journal} {Phys.
  Rev. D}\ }\textbf {\bibinfo {volume} {95}},\ \bibinfo {pages} {114507}
  (\bibinfo {year} {2017})},\ \Eprint {http://arxiv.org/abs/1611.05970}
  {arXiv:1611.05970 [hep-lat]} \BibitemShut {NoStop}%
\bibitem [{\citenamefont {Li}\ \emph {et~al.}(2020)\citenamefont {Li},
  \citenamefont {Wu}, \citenamefont {Abell}, \citenamefont {Leinweber},\ and\
  \citenamefont {Thomas}}]{Li:2019qvh}%
  \BibitemOpen
  \bibfield  {author} {\bibinfo {author} {\bibfnamefont {Y.}~\bibnamefont
  {Li}}, \bibinfo {author} {\bibfnamefont {J.-J.}\ \bibnamefont {Wu}}, \bibinfo
  {author} {\bibfnamefont {C.~D.}\ \bibnamefont {Abell}}, \bibinfo {author}
  {\bibfnamefont {D.~B.}\ \bibnamefont {Leinweber}}, \ and\ \bibinfo {author}
  {\bibfnamefont {A.~W.}\ \bibnamefont {Thomas}},\ }\href {\doibase
  10.1103/PhysRevD.101.114501} {\bibfield  {journal} {\bibinfo  {journal}
  {Phys. Rev. D}\ }\textbf {\bibinfo {volume} {101}},\ \bibinfo {pages}
  {114501} (\bibinfo {year} {2020})},\ \Eprint
  {http://arxiv.org/abs/1910.04973} {arXiv:1910.04973 [hep-lat]} \BibitemShut
  {NoStop}%
\bibitem [{\citenamefont {Bloch}\ and\ \citenamefont
  {Horowitz}(1958)}]{Bloch:1958determination}%
  \BibitemOpen
  \bibfield  {author} {\bibinfo {author} {\bibfnamefont {C.}~\bibnamefont
  {Bloch}}\ and\ \bibinfo {author} {\bibfnamefont {J.}~\bibnamefont
  {Horowitz}},\ }\href {\doibase 10.1016/0029-5582(58)90136-6} {\bibfield
  {journal} {\bibinfo  {journal} {Nuclear Physics}\ }\textbf {\bibinfo {volume}
  {8}},\ \bibinfo {pages} {91} (\bibinfo {year} {1958})}\BibitemShut {NoStop}%
\bibitem [{\citenamefont {Luu}(2005)}]{Luu:2005qr}%
  \BibitemOpen
  \bibfield  {author} {\bibinfo {author} {\bibfnamefont {T.~C.}\ \bibnamefont
  {Luu}},\ }\href {\doibase 10.1088/0954-3899/31/8/009} {\bibfield  {journal}
  {\bibinfo  {journal} {Journal of Physics G: Nuclear and Particle Physics}\
  }\textbf {\bibinfo {volume} {31}},\ \bibinfo {pages} {S1311} (\bibinfo {year}
  {2005})}\BibitemShut {NoStop}%
\bibitem [{\citenamefont {Li}\ and\ \citenamefont {Wu}(2022)}]{Li:2022aru}%
  \BibitemOpen
  \bibfield  {author} {\bibinfo {author} {\bibfnamefont {Y.}~\bibnamefont
  {Li}}\ and\ \bibinfo {author} {\bibfnamefont {J.-J.}\ \bibnamefont {Wu}},\
  }\href {\doibase 10.1103/PhysRevD.105.116024} {\bibfield  {journal} {\bibinfo
   {journal} {Phys. Rev.D}\ }\textbf {\bibinfo {volume} {105}},\ \bibinfo
  {pages} {116024} (\bibinfo {year} {2022})},\ \Eprint
  {http://arxiv.org/abs/2204.05510} {arxiv:2204.05510 [hep-ph]} \BibitemShut
  {NoStop}%
\bibitem [{\citenamefont {Doring}\ \emph {et~al.}(2011)\citenamefont {Doring},
  \citenamefont {Meissner}, \citenamefont {Oset},\ and\ \citenamefont
  {Rusetsky}}]{Doring:2011vk}%
  \BibitemOpen
  \bibfield  {author} {\bibinfo {author} {\bibfnamefont {M.}~\bibnamefont
  {Doring}}, \bibinfo {author} {\bibfnamefont {U.-G.}\ \bibnamefont
  {Meissner}}, \bibinfo {author} {\bibfnamefont {E.}~\bibnamefont {Oset}}, \
  and\ \bibinfo {author} {\bibfnamefont {A.}~\bibnamefont {Rusetsky}},\ }\href
  {\doibase 10.1140/epja/i2011-11139-7} {\bibfield  {journal} {\bibinfo
  {journal} {Eur. Phys. J.}\ }\textbf {\bibinfo {volume} {A47}},\ \bibinfo
  {pages} {139} (\bibinfo {year} {2011})},\ \Eprint
  {http://arxiv.org/abs/1107.3988} {arXiv:1107.3988 [hep-lat]} \BibitemShut
  {NoStop}%
\bibitem [{\citenamefont {Bernard}\ \emph {et~al.}(2008)\citenamefont
  {Bernard}, \citenamefont {Lage}, \citenamefont {Meissner},\ and\
  \citenamefont {Rusetsky}}]{Bernard:2008ax}%
  \BibitemOpen
  \bibfield  {author} {\bibinfo {author} {\bibfnamefont {V.}~\bibnamefont
  {Bernard}}, \bibinfo {author} {\bibfnamefont {M.}~\bibnamefont {Lage}},
  \bibinfo {author} {\bibfnamefont {U.-G.}\ \bibnamefont {Meissner}}, \ and\
  \bibinfo {author} {\bibfnamefont {A.}~\bibnamefont {Rusetsky}},\ }\href
  {\doibase 10.1088/1126-6708/2008/08/024} {\bibfield  {journal} {\bibinfo
  {journal} {JHEP}\ }\textbf {\bibinfo {volume} {08}},\ \bibinfo {pages} {024}
  (\bibinfo {year} {2008})},\ \Eprint {http://arxiv.org/abs/0806.4495}
  {arXiv:0806.4495 [hep-lat]} \BibitemShut {NoStop}%
\bibitem [{\citenamefont {Dirac}(1936)}]{Dirac:1936tg}%
  \BibitemOpen
  \bibfield  {author} {\bibinfo {author} {\bibfnamefont {P.~A.~M.}\
  \bibnamefont {Dirac}},\ }\href {\doibase 10.1098/rspa.1936.0111} {\bibfield
  {journal} {\bibinfo  {journal} {Proc. Roy. Soc. Lond. A}\ }\textbf {\bibinfo
  {volume} {155}},\ \bibinfo {pages} {447} (\bibinfo {year}
  {1936})}\BibitemShut {NoStop}%
\bibitem [{\citenamefont {Klein}\ and\ \citenamefont
  {Lee}(1974)}]{Klein:1974aa}%
  \BibitemOpen
  \bibfield  {author} {\bibinfo {author} {\bibfnamefont {A.}~\bibnamefont
  {Klein}}\ and\ \bibinfo {author} {\bibfnamefont {T.-S.~H.}\ \bibnamefont
  {Lee}},\ }\href {\doibase 10.1103/PhysRevD.10.4308} {\bibfield  {journal}
  {\bibinfo  {journal} {Phys. Rev. D}\ }\textbf {\bibinfo {volume} {10}},\
  \bibinfo {pages} {4308} (\bibinfo {year} {1974})}\BibitemShut {NoStop}%
\bibitem [{\citenamefont {Beane}(2004)}]{Beane:2004tw}%
  \BibitemOpen
  \bibfield  {author} {\bibinfo {author} {\bibfnamefont {S.~R.}\ \bibnamefont
  {Beane}},\ }\href {\doibase 10.1103/PhysRevD.70.034507} {\bibfield  {journal}
  {\bibinfo  {journal} {Phys. Rev. D}\ }\textbf {\bibinfo {volume} {70}},\
  \bibinfo {pages} {034507} (\bibinfo {year} {2004})},\ \Eprint
  {http://arxiv.org/abs/hep-lat/0403015} {arXiv:hep-lat/0403015} \BibitemShut
  {NoStop}%
\bibitem [{\citenamefont {Hansen}\ \emph {et~al.}(2024)\citenamefont {Hansen},
  \citenamefont {Romero-L\'opez},\ and\ \citenamefont
  {Sharpe}}]{Hansen:2024ffk}%
  \BibitemOpen
  \bibfield  {author} {\bibinfo {author} {\bibfnamefont {M.~T.}\ \bibnamefont
  {Hansen}}, \bibinfo {author} {\bibfnamefont {F.}~\bibnamefont
  {Romero-L\'opez}}, \ and\ \bibinfo {author} {\bibfnamefont {S.~R.}\
  \bibnamefont {Sharpe}},\ }\href@noop {} {\  (\bibinfo {year} {2024})},\
  \Eprint {http://arxiv.org/abs/2401.06609} {arXiv:2401.06609 [hep-lat]}
  \BibitemShut {NoStop}%
\bibitem [{\citenamefont {Raposo}\ and\ \citenamefont
  {Hansen}(2023)}]{Raposo:2023oru}%
  \BibitemOpen
  \bibfield  {author} {\bibinfo {author} {\bibfnamefont {A.~B.~a.}\
  \bibnamefont {Raposo}}\ and\ \bibinfo {author} {\bibfnamefont {M.~T.}\
  \bibnamefont {Hansen}},\ }\href@noop {} {\  (\bibinfo {year} {2023})},\
  \Eprint {http://arxiv.org/abs/2311.18793} {arXiv:2311.18793 [hep-lat]}
  \BibitemShut {NoStop}%
\bibitem [{\citenamefont {Meng}\ \emph {et~al.}(2023)\citenamefont {Meng},
  \citenamefont {Baru}, \citenamefont {Epelbaum}, \citenamefont {Filin},\ and\
  \citenamefont {Gasparyan}}]{Meng:2023bmz}%
  \BibitemOpen
  \bibfield  {author} {\bibinfo {author} {\bibfnamefont {L.}~\bibnamefont
  {Meng}}, \bibinfo {author} {\bibfnamefont {V.}~\bibnamefont {Baru}}, \bibinfo
  {author} {\bibfnamefont {E.}~\bibnamefont {Epelbaum}}, \bibinfo {author}
  {\bibfnamefont {A.~A.}\ \bibnamefont {Filin}}, \ and\ \bibinfo {author}
  {\bibfnamefont {A.~M.}\ \bibnamefont {Gasparyan}},\ }\href@noop {} {\
  (\bibinfo {year} {2023})},\ \Eprint {http://arxiv.org/abs/2312.01930}
  {arXiv:2312.01930 [hep-lat]} \BibitemShut {NoStop}%
\bibitem [{\citenamefont {Bubna}\ \emph {et~al.}(2024)\citenamefont {Bubna},
  \citenamefont {Hammer}, \citenamefont {M\"uller}, \citenamefont {Pang},
  \citenamefont {Rusetsky},\ and\ \citenamefont {Wu}}]{Bubna:2024izx}%
  \BibitemOpen
  \bibfield  {author} {\bibinfo {author} {\bibfnamefont {R.}~\bibnamefont
  {Bubna}}, \bibinfo {author} {\bibfnamefont {H.-W.}\ \bibnamefont {Hammer}},
  \bibinfo {author} {\bibfnamefont {F.}~\bibnamefont {M\"uller}}, \bibinfo
  {author} {\bibfnamefont {J.-Y.}\ \bibnamefont {Pang}}, \bibinfo {author}
  {\bibfnamefont {A.}~\bibnamefont {Rusetsky}}, \ and\ \bibinfo {author}
  {\bibfnamefont {J.-J.}\ \bibnamefont {Wu}},\ }\href@noop {} {\  (\bibinfo
  {year} {2024})},\ \Eprint {http://arxiv.org/abs/2402.12985} {arXiv:2402.12985
  [hep-lat]} \BibitemShut {NoStop}%
\bibitem [{\citenamefont {Göckeler}\ \emph {et~al.}(2012)\citenamefont
  {Göckeler}, \citenamefont {Horsley}, \citenamefont {Lage}, \citenamefont
  {Meißner}, \citenamefont {Rakow}, \citenamefont {Rusetsky}, \citenamefont
  {Schierholz},\ and\ \citenamefont {Zanotti}}]{Gockeler:2012yj}%
  \BibitemOpen
  \bibfield  {author} {\bibinfo {author} {\bibfnamefont {M.}~\bibnamefont
  {Göckeler}}, \bibinfo {author} {\bibfnamefont {R.}~\bibnamefont {Horsley}},
  \bibinfo {author} {\bibfnamefont {M.}~\bibnamefont {Lage}}, \bibinfo {author}
  {\bibfnamefont {U.-G.}\ \bibnamefont {Meißner}}, \bibinfo {author}
  {\bibfnamefont {P.~E.~L.}\ \bibnamefont {Rakow}}, \bibinfo {author}
  {\bibfnamefont {A.}~\bibnamefont {Rusetsky}}, \bibinfo {author}
  {\bibfnamefont {G.}~\bibnamefont {Schierholz}}, \ and\ \bibinfo {author}
  {\bibfnamefont {J.~M.}\ \bibnamefont {Zanotti}},\ }\href {\doibase
  10.1103/PhysRevD.86.094513} {\bibfield  {journal} {\bibinfo  {journal} {Phys.
  Rev. D}\ }\textbf {\bibinfo {volume} {86}},\ \bibinfo {pages} {094513}
  (\bibinfo {year} {2012})},\ \Eprint {http://arxiv.org/abs/1206.4141}
  {arXiv:1206.4141 [hep-lat]} \BibitemShut {NoStop}%
\bibitem [{\citenamefont {Leskovec}\ and\ \citenamefont
  {Prelovsek}(2012)}]{Leskovec:2012gb}%
  \BibitemOpen
  \bibfield  {author} {\bibinfo {author} {\bibfnamefont {L.}~\bibnamefont
  {Leskovec}}\ and\ \bibinfo {author} {\bibfnamefont {S.}~\bibnamefont
  {Prelovsek}},\ }\href {\doibase 10.1103/PhysRevD.85.114507} {\bibfield
  {journal} {\bibinfo  {journal} {Phys. Rev. D}\ }\textbf {\bibinfo {volume}
  {85}},\ \bibinfo {pages} {114507} (\bibinfo {year} {2012})}\BibitemShut
  {NoStop}%
\bibitem [{\citenamefont {Batley}\ \emph {et~al.}(2008)\citenamefont {Batley}
  \emph {et~al.}}]{Batley:2007zz}%
  \BibitemOpen
  \bibfield  {author} {\bibinfo {author} {\bibfnamefont {J.~R.}\ \bibnamefont
  {Batley}} \emph {et~al.} (\bibinfo {collaboration} {NA48/2}),\ }\href
  {\doibase 10.1140/epjc/s10052-008-0547-0} {\bibfield  {journal} {\bibinfo
  {journal} {Eur. Phys. J.}\ }\textbf {\bibinfo {volume} {C54}},\ \bibinfo
  {pages} {411} (\bibinfo {year} {2008})}\BibitemShut {NoStop}%
\bibitem [{\citenamefont {Hyams}\ \emph {et~al.}(1973)\citenamefont {Hyams},
  \citenamefont {Jones}, \citenamefont {Weilhammer}, \citenamefont {Blum},
  \citenamefont {Dietl}, \citenamefont {Grayer}, \citenamefont {Koch},
  \citenamefont {Lorenz}, \citenamefont {Lütjens}, \citenamefont {Männer},
  \citenamefont {Meissburger}, \citenamefont {Ochs}, \citenamefont {Stierlin},\
  and\ \citenamefont {Wagner}}]{Hyams:1973zf}%
  \BibitemOpen
  \bibfield  {author} {\bibinfo {author} {\bibfnamefont {B.}~\bibnamefont
  {Hyams}}, \bibinfo {author} {\bibfnamefont {C.}~\bibnamefont {Jones}},
  \bibinfo {author} {\bibfnamefont {P.}~\bibnamefont {Weilhammer}}, \bibinfo
  {author} {\bibfnamefont {W.}~\bibnamefont {Blum}}, \bibinfo {author}
  {\bibfnamefont {H.}~\bibnamefont {Dietl}}, \bibinfo {author} {\bibfnamefont
  {G.}~\bibnamefont {Grayer}}, \bibinfo {author} {\bibfnamefont
  {W.}~\bibnamefont {Koch}}, \bibinfo {author} {\bibfnamefont {E.}~\bibnamefont
  {Lorenz}}, \bibinfo {author} {\bibfnamefont {G.}~\bibnamefont {Lütjens}},
  \bibinfo {author} {\bibfnamefont {W.}~\bibnamefont {Männer}}, \bibinfo
  {author} {\bibfnamefont {J.}~\bibnamefont {Meissburger}}, \bibinfo {author}
  {\bibfnamefont {W.}~\bibnamefont {Ochs}}, \bibinfo {author} {\bibfnamefont
  {U.}~\bibnamefont {Stierlin}}, \ and\ \bibinfo {author} {\bibfnamefont
  {F.}~\bibnamefont {Wagner}},\ }\href {\doibase 10.1016/0550-3213(73)90618-4}
  {\bibfield  {journal} {\bibinfo  {journal} {Nuclear Physics B}\ }\textbf
  {\bibinfo {volume} {64}},\ \bibinfo {pages} {134} (\bibinfo {year}
  {1973})}\BibitemShut {NoStop}%
\bibitem [{\citenamefont {Estabrooks}\ \emph {et~al.}(1973)\citenamefont
  {Estabrooks}, \citenamefont {Martin}, \citenamefont {Grayer}, \citenamefont
  {Hyams}, \citenamefont {Jones}, \citenamefont {Weilhammer}, \citenamefont
  {Blum}, \citenamefont {Dietl}, \citenamefont {Koch}, \citenamefont {Lorenz},
  \citenamefont {Lütjens}, \citenamefont {Männer}, \citenamefont
  {Meissburger},\ and\ \citenamefont {Stierlin}}]{Estabrooks:1973dya}%
  \BibitemOpen
  \bibfield  {author} {\bibinfo {author} {\bibfnamefont {P.}~\bibnamefont
  {Estabrooks}}, \bibinfo {author} {\bibfnamefont {A.~D.}\ \bibnamefont
  {Martin}}, \bibinfo {author} {\bibfnamefont {G.}~\bibnamefont {Grayer}},
  \bibinfo {author} {\bibfnamefont {B.}~\bibnamefont {Hyams}}, \bibinfo
  {author} {\bibfnamefont {C.}~\bibnamefont {Jones}}, \bibinfo {author}
  {\bibfnamefont {P.}~\bibnamefont {Weilhammer}}, \bibinfo {author}
  {\bibfnamefont {W.}~\bibnamefont {Blum}}, \bibinfo {author} {\bibfnamefont
  {H.}~\bibnamefont {Dietl}}, \bibinfo {author} {\bibfnamefont
  {W.}~\bibnamefont {Koch}}, \bibinfo {author} {\bibfnamefont {E.}~\bibnamefont
  {Lorenz}}, \bibinfo {author} {\bibfnamefont {G.}~\bibnamefont {Lütjens}},
  \bibinfo {author} {\bibfnamefont {W.}~\bibnamefont {Männer}}, \bibinfo
  {author} {\bibfnamefont {J.}~\bibnamefont {Meissburger}}, \ and\ \bibinfo
  {author} {\bibfnamefont {U.}~\bibnamefont {Stierlin}},\ }in\ \href {\doibase
  10.1063/1.2947126} {\emph {\bibinfo {booktitle} {AIP Conference
  Proceedings}}},\ Vol.~\bibinfo {volume} {13}\ (\bibinfo  {publisher}
  {American Institute of Physics},\ \bibinfo {year} {1973})\ pp.\ \bibinfo
  {pages} {37--79}\BibitemShut {NoStop}%
\bibitem [{\citenamefont {Protopopescu}\ \emph {et~al.}(1973)\citenamefont
  {Protopopescu}, \citenamefont {Alston-Garnjost}, \citenamefont
  {Barbaro-Galtieri}, \citenamefont {Flatté}, \citenamefont {Friedman},
  \citenamefont {Lasinski}, \citenamefont {Lynch}, \citenamefont {Rabin},\ and\
  \citenamefont {Solmitz}}]{Protopopescu:1973sh}%
  \BibitemOpen
  \bibfield  {author} {\bibinfo {author} {\bibfnamefont {S.~D.}\ \bibnamefont
  {Protopopescu}}, \bibinfo {author} {\bibfnamefont {M.}~\bibnamefont
  {Alston-Garnjost}}, \bibinfo {author} {\bibfnamefont {A.}~\bibnamefont
  {Barbaro-Galtieri}}, \bibinfo {author} {\bibfnamefont {S.~M.}\ \bibnamefont
  {Flatté}}, \bibinfo {author} {\bibfnamefont {J.~H.}\ \bibnamefont
  {Friedman}}, \bibinfo {author} {\bibfnamefont {T.~A.}\ \bibnamefont
  {Lasinski}}, \bibinfo {author} {\bibfnamefont {G.~R.}\ \bibnamefont {Lynch}},
  \bibinfo {author} {\bibfnamefont {M.~S.}\ \bibnamefont {Rabin}}, \ and\
  \bibinfo {author} {\bibfnamefont {F.~T.}\ \bibnamefont {Solmitz}},\ }\href
  {\doibase 10.1103/PhysRevD.7.1279} {\bibfield  {journal} {\bibinfo  {journal}
  {Phys. Rev. D}\ }\textbf {\bibinfo {volume} {7}},\ \bibinfo {pages} {1279}
  (\bibinfo {year} {1973})}\BibitemShut {NoStop}%
\bibitem [{\citenamefont {Grayer}\ \emph {et~al.}(1972)\citenamefont {Grayer},
  \citenamefont {Hyams}, \citenamefont {Jones}, \citenamefont {Schlein},
  \citenamefont {Blum}, \citenamefont {Dietl}, \citenamefont {Koch},
  \citenamefont {Lorenz}, \citenamefont {Lütjens}, \citenamefont {Männer},
  \citenamefont {Meissburger}, \citenamefont {Ochs}, \citenamefont {Stierlin},\
  and\ \citenamefont {Weilhammer}}]{doi:10.1063/1.2948709}%
  \BibitemOpen
  \bibfield  {author} {\bibinfo {author} {\bibfnamefont {G.}~\bibnamefont
  {Grayer}}, \bibinfo {author} {\bibfnamefont {B.}~\bibnamefont {Hyams}},
  \bibinfo {author} {\bibfnamefont {C.}~\bibnamefont {Jones}}, \bibinfo
  {author} {\bibfnamefont {P.}~\bibnamefont {Schlein}}, \bibinfo {author}
  {\bibfnamefont {W.}~\bibnamefont {Blum}}, \bibinfo {author} {\bibfnamefont
  {H.}~\bibnamefont {Dietl}}, \bibinfo {author} {\bibfnamefont
  {W.}~\bibnamefont {Koch}}, \bibinfo {author} {\bibfnamefont {E.}~\bibnamefont
  {Lorenz}}, \bibinfo {author} {\bibfnamefont {G.}~\bibnamefont {Lütjens}},
  \bibinfo {author} {\bibfnamefont {W.}~\bibnamefont {Männer}}, \bibinfo
  {author} {\bibfnamefont {J.}~\bibnamefont {Meissburger}}, \bibinfo {author}
  {\bibfnamefont {W.}~\bibnamefont {Ochs}}, \bibinfo {author} {\bibfnamefont
  {U.}~\bibnamefont {Stierlin}}, \ and\ \bibinfo {author} {\bibfnamefont
  {P.}~\bibnamefont {Weilhammer}},\ }\href {\doibase 10.1063/1.2948709}
  {\bibfield  {journal} {\bibinfo  {journal} {AIP Conference Proceedings}\
  }\textbf {\bibinfo {volume} {8}},\ \bibinfo {pages} {5} (\bibinfo {year}
  {1972})},\ \Eprint
  {http://arxiv.org/abs/https://aip.scitation.org/doi/pdf/10.1063/1.2948709}
  {https://aip.scitation.org/doi/pdf/10.1063/1.2948709} \BibitemShut {NoStop}%
\bibitem [{\citenamefont {Männer}(1974)}]{Manner:1974ak}%
  \BibitemOpen
  \bibfield  {author} {\bibinfo {author} {\bibfnamefont {W.}~\bibnamefont
  {Männer}},\ }in\ \href {\doibase 10.1063/1.2947385} {\emph {\bibinfo
  {booktitle} {AIP Conference Proceedings}}},\ Vol.~\bibinfo {volume} {21}\
  (\bibinfo  {publisher} {American Institute of Physics},\ \bibinfo {year}
  {1974})\ pp.\ \bibinfo {pages} {22--45}\BibitemShut {NoStop}%
\bibitem [{\citenamefont {Froggatt}\ and\ \citenamefont
  {Petersen}(1977)}]{Froggatt:1977hu}%
  \BibitemOpen
  \bibfield  {author} {\bibinfo {author} {\bibfnamefont {C.~D.}\ \bibnamefont
  {Froggatt}}\ and\ \bibinfo {author} {\bibfnamefont {J.~L.}\ \bibnamefont
  {Petersen}},\ }\href {\doibase 10.1016/0550-3213(77)90021-9} {\bibfield
  {journal} {\bibinfo  {journal} {Nucl. Phys.}\ }\textbf {\bibinfo {volume}
  {B129}},\ \bibinfo {pages} {89} (\bibinfo {year} {1977})}\BibitemShut
  {NoStop}%
\bibitem [{\citenamefont {Weinberg}(1965)}]{Weinberg:1965zz}%
  \BibitemOpen
  \bibfield  {author} {\bibinfo {author} {\bibfnamefont {S.}~\bibnamefont
  {Weinberg}},\ }\href {\doibase 10.1103/PhysRev.137.B672} {\bibfield
  {journal} {\bibinfo  {journal} {Phys. Rev.}\ }\textbf {\bibinfo {volume}
  {137}},\ \bibinfo {pages} {B672} (\bibinfo {year} {1965})}\BibitemShut
  {NoStop}%
\end{thebibliography}%

\end{document}